\documentclass[iop]{emulateapj}
\def\LIR{\hbox{$L_{\rm IR}$}}

\def\HII{\hbox{H\,{\sc ii}}}

\def\NeII{\hbox{[Ne\,{\sc ii}]12.81\,\micron}}
\def\NeIIno{\hbox{[Ne\,{\sc ii}]}}

\def\SIV{\hbox{[S\,{\sc iv}]10.51\,\micron}}

\def\SSi{\hbox{$S_{\rm Si\,9.7\,\mu m}$}}

\def\PAHa{\hbox{11.3\,\micron\,PAH}}
\def\PAHas{\hbox{11.3\,\micron\,PAHs}}
\def\PAHb{\hbox{8.6\,\micron\,PAH}}

\def\PAHc{\hbox{7.7\,\micron\,PAH}}
\def\PAHcs{\hbox{7.7\,\micron\,PAHs}}
\def\PAHd{\hbox{6.2\,\micron\,PAH}}
\def\PAHds{\hbox{6.2\,\micron\,PAHs}}
\def\PAHe{\hbox{12.5\,\micron\,PAH}}

\def\Paalpha{\hbox{\rm Pa$\alpha$}}

\def\deg{$^{\circ}$}

\def\Lsun{\hbox{$L_\odot$}}

\def\LIR{\hbox{$L_{\rm IR}$}}

\def\MH2{\hbox{$M_{H_2}$}}

\shorttitle{The Extended MIR Emission in (U)LIRGs. II. Feature Emission}
\shortauthors{D\'{\i}az-Santos et al.}

\begin{document}

\title{The Spatial Extent of (U)LIRG\lowercase{s} in the mid-Infrared. II. Feature Emission}
%\title{Quantifying the Extended mid-Infrared Emission of (U)LIRG\lowercase{s} I: Continuum Emission and Implications for Herschel}


\author{T.~D\'{\i}az-Santos\altaffilmark{1,$\dagger$},
V.~Charmandaris\altaffilmark{1,2},
L.~Armus\altaffilmark{3},
S.~Stierwalt\altaffilmark{3},
S.~Haan\altaffilmark{3},
J.~M.~Mazzarella\altaffilmark{4},
J.~H.~Howell\altaffilmark{3},
S.~Veilleux\altaffilmark{5},
E.~J.~Murphy\altaffilmark{6},
A.~O.~Petric\altaffilmark{3},
P.~Appleton\altaffilmark{7},
A.~S.~Evans\altaffilmark{8,9},
D.~B.~Sanders\altaffilmark{10},
and J.~A.~Surace\altaffilmark{3}
}

\altaffiltext{1}{University of Crete, Department of Physics, GR-71003, Heraklion, Greece: tanio@physics.uoc.gr}
\altaffiltext{$\dagger$}{Now at the Spitzer Science Center, California Institute of Technology, MS 220-6, Pasadena, CA 91125}
\altaffiltext{2}{IESL/FORTH - Department of Physics and Institute of Theoretical \& Computational Physics, 71003, Heraklion, Greece and Chercheur Associ\'e, Observatoire de  Paris, F-75014, Paris, France}
\altaffiltext{3}{Spitzer Science Center, California Institute of Technology, MS 220-6, Pasadena, CA 91125}
\altaffiltext{4}{Infrared Processing \& Analysis Center, MS 100-22, California Institute of Technology, Pasadena, CA 91125}
%\altaffiltext{5}{Department of Physics and Astronomy, University of New York at Stony Brook, NY 11794-3800}
\altaffiltext{5}{Department of Physics, University of Oregon, Eugene, OR 97403}
\altaffiltext{6}{Observatories of the Carnegie Institution for Science, 813 Santa Barbara Street, Pasadena, CA 91101}
\altaffiltext{7}{NASA Herschel Science Center, IPAC, MS 100-22, California Institute of Technology, Pasadena, CA 91125}
\altaffiltext{8}{Department of Astronomy, University of Virginia, P.O. Box 400325, Charlottesville, VA 22904}
\altaffiltext{9}{National Radio Astronomy Observatory, 520 Edgemont Road, Charlottesville, VA 22903}
\altaffiltext{10}{Institute for Astronomy, University of Hawaii, 2680 Woodlawn Drive, Honolulu, HI 96822}

\begin{abstract}

%This is the second paper of a series studying the compactness of the mid-infrared (MIR) emission of (U)LIRGs in the GOALS galaxy sample. While the first paper concentrated on the continuum emission, this is focused on the analysis of the fraction of extended emission (FEE) of the spectral features (polycyclic aromatic hydrocarbons -PAHs-, emission lines and the 9.7$\,\micron$ silicate absorption feature). We find that the $5-14\,\micron$ spectrum of the extended emission component of the galaxies in the sample, irrespective of their \LIR, AGN contribution to the nuclear MIR continuum, the presence of a central compact starburst, or their FEE$_\lambda$ type classification, is similar for all of them. This implies that the properties of the star formation processes taking place at d$\,\gtrsim\,1.5\,$kpc from the nuclei of (U)LIRGs, i.e., in their disks, are the same, and resemble to what is found in local lower-luminous starbursts galaxies. On the other hand, the spectrum of the nuclear component of (U)LIRGs is very diverse. The presence of an AGN or a powerful starburst in their nuclei makes their MIR continumm to become compact, showing larger luminosity surface densities, but does not affect the extent of the PAH emission. In fact, both processes are indistingushible in terms of how they modify the PAH-to-continuum FEE ratio of (U)LIRGs, at least at the spatial scales probed by \textit{Spitzer}.

We present results from the second part of our analysis of the extended mid-infrared (MIR) emission of the Great Observatories All-Sky LIRG Survey (GOALS) sample based on $5-14\,\micron$ low-resolution spectra obtained with the Infrared Spectrograph on \textit{Spitzer}. We calculate the fraction of extended emission as a function of wavelength for all galaxies in the sample, FEE$_{\lambda}$, defined as the fraction of the emission that originates outside of the unresolved central component of a source, and spatially separate the MIR spectrum of a galaxy into its nuclear and extended components.

We find that the \NeII\ emission line is as compact as the hot dust MIR continuum, while the polycyclic aromatic hydrocarbon (PAH) emission is more extended. In addition, the 6.2 and \PAHc\ emission is more compact than that of the \PAHa, which is consistent with the formers being enhanced in a more ionized medium.
%former being a neutral PAH and the later being ionized.
The presence of an AGN or a powerful nuclear starburst increases the compactness and the luminosity surface density of the hot dust MIR continuum, but has a negligible effect on the spatial extent of the PAH emission on kpc-scales. Furthermore, it appears that both processes, AGN and/or nuclear starburst, are indistinguishable in terms of how they modify the integrated PAH-to-continuum ratio of the FEE in (U)LIRGs. Globally, the $5-14\,\micron$ spectra of the extended emission component are homogeneous for all galaxies in the GOALS sample.
%of the GOALS galaxies is similar irrespective of their %\LIR, AGN contribution to the MIR continuum, the presence of a central compact starburst, or theirFEE$_{\lambda}$ type classification.
This suggests that, independently of the spatial distribution of the various MIR features, the physical properties of star formation occurring at distances farther than 1.5\,kpc from the nuclei of (U)LIRGs are very similar, resembling local star-forming galaxies with $\LIR\,<\,10^{11}\,\Lsun$, as well as star formation-dominated ULIRGs at $z\,\sim\,2$. In contrast, the MIR spectra of the nuclear component of local ULIRGs \textit{and} LIRGs are very diverse. These results imply that the observed variety of the integrated MIR properties of local (U)LIRGs arise, on average, only from the processes that are taking place in their cores.

\end{abstract}

\keywords{infrared: galaxies --- galaxies: evolution --- galaxies: interactions --- galaxies: starburst --- galaxies: active}

%________________________________________________________________
\section{Introduction}\label{s:introduction}

The mid-infrared (MIR) spectra of galaxies provide key information that allow us to characterize the global properties of their star formation, as well as to develop continuum and emission line diagnostics in order to infer whether they harbor an active galactic nucleus (AGN), with less confusion due to dust extinction compared to observations at optical wavelengths (\citealt{Lutz1998b}; \citealt{Genzel1998}; \citealt{Charmandaris2004}; \citealt{Brandl2006}; \citealt{Smith2007}; \citealt{Armus2007}; \citealt{Veilleux2009}; \citealt{Wu2009}; \citealt{Sales2010}; \citealt{Petric2011}). However, when studying the integrated emission of galaxies, the spatial distributions of the sources contributing to their observed infrared (IR) spectral energy distributions, such as \HII\ regions, photo-dissociation regions (PDRs), cold molecular clouds, and/or non-thermal emission sources, are averaged together. That is, all the intrinsic properties of the different phases of the inter-stellar medium (ISM) are mixed, despite the fact that they can be substantially different (\citealt{Laurent2000}; \citealt{Smith2004}). Therefore, even a rough decomposition between the nuclear and disk components of a galaxy can yield valuable information on where the IR luminosity originates, as well as details on the processes associated to the different emission components.

The access to this spatial information is even more vital in the study of luminous and ultra-luminous infrared galaxies, (U)LIRGs, for which it is known that the bulk of their energy production is generated within their central regions and emitted in the IR on scales of a few kpc. This class of galaxies, although not very numerous in the nearby Universe \citep{Sanders1996}, are responsible for the bulk of the obscured star formation at $z\,\geq\,1$ (\citealt{LeFloch2005}; \citealt{PG2005}; \citealt{Caputi2007}; \citealt{Magnelli2011}; \citealt{Murphy2011}). Thus, the study of the spectra of local (U)LIRGs, where the angular resolution achieved by space-born observatories such as \textit{Spitzer} is sufficient to spatially-resolve their components, is essential for interpreting the internal mechanisms that govern their IR emission and for helping to understand the properties of high redshift (U)LIRGs.

Several MIR spectroscopic studies have already been carried out using ground-based telescopes which provide, for the closest and brightest sources, spatial resolution of a few tens of pc (\citealt{Soifer2002}; \citealt{Soifer2003}; \citealt{DS2010a}). These works have demonstrated that the MIR continuum emission, ionization lines, and polycyclic aromatic hydrocarbon (PAH) features arise from different regions in (U)LIRGs, and that the sources of emission are arranged in more compact configurations than in galaxies with lower IR luminosities. This is consistent with the merger-induced nature of most local (U)LIRGs (\citealt{Veilleux2002}; \citealt{Armus2009}) and the funneling of large quantities of gas and dust mass towards their nuclei during the interaction. In addition, MIR spectral maps using the IRS instrument on-board \textit{Spitzer} have been obtained for 15 local LIRGs revealing an increase of the electron density in their nuclei (\citealt{PS2010}).

Despite the small number of galaxies studied in these works, some general conclusions about the properties of (U)LIRGs as a whole could be derived. 
Yet, these trends are confounded by systematics caused by intrinsic variations of processes that dominate at sub-kpc scales. Studying these internal processes is therefore essential if we wish to completely understand the global trends. The Great Observatories All-sky LIRG Survey (GOALS; \citealt{Armus2009}), which consist of 291 galaxies (202 systems), is the ideal local LIRG sample for this task because it is large enough to allow us to break it up into sub-samples based, for example, on their interaction stage, or the AGN contribution to their MIR emission (\citealt{Petric2011}). Each sub-sample has enough galaxies in order to perform a robust statistical analysis and examine how the various MIR properties vary from one sub-sample to another and within a given sub-sample. Such an analysis was performed by \cite{DS2010b}, hereafter Paper~I, who analyzed the compactness of the MIR continuum emission at 13.2$\,\micron$ of (U)LIRGs in the GOALS sample and discussed its correlation to several physical properties of the systems such as their \LIR, MIR AGN fraction, and far infrared (FIR) colors.

The wealth of information available in the $5-14\,\micron$ spectral range allow us to further expand our analysis to several other key MIR features. In this second paper we explore how the compactness of the emission due to polycyclic aromatic hydrocarbon (PAH) and atomic lines compares to that of the warm dust continuum and investigate the origin of the observed differences.
%The analysis presented in these papers are based on \textit{Spitzer}/IRS SL staring observations covering the 5 to $\sim\,15\,\micron$ wavelength range with a spectral resolution of R$\sim\,60-130$ and the aim of the study is to separate and quantify the extended emission of (U)LIRGs from the contribution of the unresolved nuclear component.
An analysis regarding the nuclear PAH emission ratios among the GOALS sample will be further addressed in a following study by Stierwalt et al. (2011, in prep.). The paper is structured as follows: in Section 2 we briefly remind the reader of the datasets used and the methodology of our analysis which was described in detail in Paper~I. Our results are reported in Section 3, where we present average MIR spectra of the extended and nuclear emission of the three main types of sources we found in Paper~I. We also explore the role of dust extinction as well as the presence of an active galactic nucleus (AGN) in the observed spatial profiles of the various MIR spectral features. Finally, in Section 4 we present our conclusions.

\begin{figure*}
\epsscale{1.1}
%\plotone{./figures/f2a.ps}\plotone{./figures/f2b.ps}\plotone{./figures/f2c.ps}
\plotone{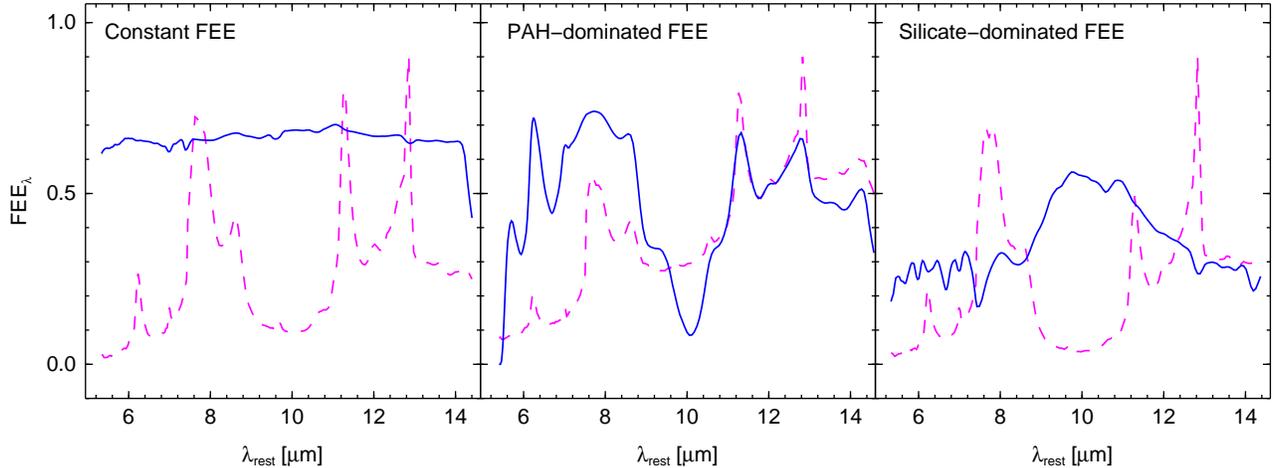}
%\vspace{.5cm}
\caption{\footnotesize The FEE$_\lambda$ function, smoothed with a 4-pixel box to reduce noise, are plotted in blue for 3 galaxies that serve as examples of the main types identified in the sample. The Spitzer/IRS $\sim$5-14$\mu$m spectrum of each galaxy, scaled to arbitrary units, is also plotted as a pink dashed line for reference. Left panel: NGC~3110 with constant/featureless FEE$_\lambda$. Middle panel: NGC~1365 displaying a PAH and line extended emission. Right panel: MCG+08-11-002 with silicate-extended emission.
%The FEE$_\lambda$ functions of each type have been normalized to a value of 0.5, 0.35 and 0.25 at 13.2$\,\micron$, respectively.
}\label{f:feetype}
%\vspace{.25cm}
\end{figure*}

\section{Observations and Data Analysis}\label{s:irsobs}

\subsection{The sample}\label{ss:sample}

The sample we analyze is the Great Observatories All-Sky LIRG Survey (GOALS; \citealt{Armus2009}). GOALS comprises a complete, flux-limited sample of galaxies in the local Universe drawn from the Revised Bright Galaxy Sample (RBGS, \citealt{Sanders2003}) selected to be systems in the (U)LIRG luminosity classes. \cite{Armus2009} describe in detail how the sample was selected as well its global characteristics.
% Additional information related to the presence on an active galactic nucleus (AGN) is discussed by \cite{Petric2011}. More specifically,
In addition, using a number of MIR diagnostics, \cite{Petric2011} estimate the AGN contribution to the MIR luminosity of the systems and, based on their apparent morphology, classify each galaxy into a stage of interaction, ranging from isolated systems to advanced mergers. In \cite{Howell2010} the relation between the UV and MIR emission of the galaxies in the sample is investigated, while \cite{Haan2011} present a thorough analysis of the nuclear structure of the galaxy sample using high spatial resolution NIR and optical images obtained with the \textit{Hubble Space Telescope}. We refer the reader to these papers since we will rely on their findings for the interpretation of our results. Out of the 291 galaxies (202 systems) included in the GOALS sample, a total of 221 are used for this study, 200 LIRGs and 21 ULIRGs. All of them have low-resolution spectroscopic \textit{Spitzer}/IRS observations in the $5-14\,\micron$ rest-frame wavelength range. The angular resolution of IRS ($\sim\,3\farcs6$ at $13.2\,\micron$) allows us to separate regions of physical scales ranging from 0.22\,kpc at the distance of the closest LIRG ($\sim\,12\,$Mpc), to 6.1\,kpc at $\sim\,340\,$Mpc where the farthest ULIRG of the sample is located. The spatial resolution at the median distance of our galaxy sample, 91\,Mpc, is 1.7\,kpc (also at $13.2\,\micron$). A table with the main characteristics of the galaxies such as their distance, \LIR, or FIR colors can be found in Paper~I. We also analyzed separately a compilation of ULIRGs from the works of \cite{Imanishi2007,Imanishi2009,Imanishi2010}, but these sources cannot be included in our present study because they are too distant and, as a consequence, they are unresolved at all wavelengths observed with the IRS.

\subsection{Analysis}\label{s:analysis}

Our analysis is based on the calculation of the fraction of extended emission of galaxies as a function of wavelength, FEE$_\lambda$, which is defined as the fraction of emission in a galaxy that does \textit{not} arise from its spatially unresolved central component. Detailed information on the calculation of the FEE$_\lambda$ functions can be found in Paper~I. For practical purposes, we remind the reader that its formal definition is:

\begin{equation}\label{e:feevsfir}
FEE_\lambda=\frac{EE_\lambda}{E_\lambda(total)}
\end{equation}

\noindent
where FEE$_\lambda$, EE$_\lambda$ and E$_\lambda(total)$ are the fraction extended emission (ranging from zero to unity), the extended emission, and the total emission of a galaxy within the IRS slit at each wavelength respectively. Essentially, the FEE$_\lambda$ of a source can be considered as the complementary part of its compactness, which can be defined as $1-$FEE$_\lambda$. We also define the core size of a galaxy as the full width half-maximum (FWHM) of a Gaussian fitted to the spatial profile of its nuclear emission along the \textit{Spitzer}/IRS slit at a given wavelength. We note that while the core size represents how extended the nuclear emission is, the FEE also accounts for low surface brightness emission that is more extended than the core of the source (beyond the wings of the central Gaussian). In Table~\ref{t:fees} (available in electronic format) we provide the FEE$_\lambda$ measurements at the wavelengths of the MIR spectral features considered in this work.

% We report as the final FEE$_\lambda$ at a given wavelength the average of the values obtained for each one of the two nod positions of the IRS slits. Comparing the difference between the two nods, we calculate the statistical uncertainty of the FEE$_\lambda$. The final uncertainty includes additional sources of error and depends on the details of the target. For reference, the typical uncertainty on the FEE is between 0.05 and 0.15.

\section{Results And Discussion}

In Paper~I we grouped the GOALS galaxies into 3 types of FEE$_\lambda$ functions: constant/featureless, PAH- and line-extended, and silicate-extended. We excluded 8 sources that appear unresolved, thus having an FEE$_{5-14\mu m}\,\simeq\,0$, and 11 that could not be classified in any of the three main FEE$_\lambda$ types. However, for completeness, in the present paper we do take them into account. Figure~\ref{f:feetype} shows examples of the FEE$_\lambda$ function types.

\subsection{The FEE of Continuum, PAHs, and Emission Lines}\label{ss:feerelations}

\begin{figure*}
\epsscale{.45}
\plotone{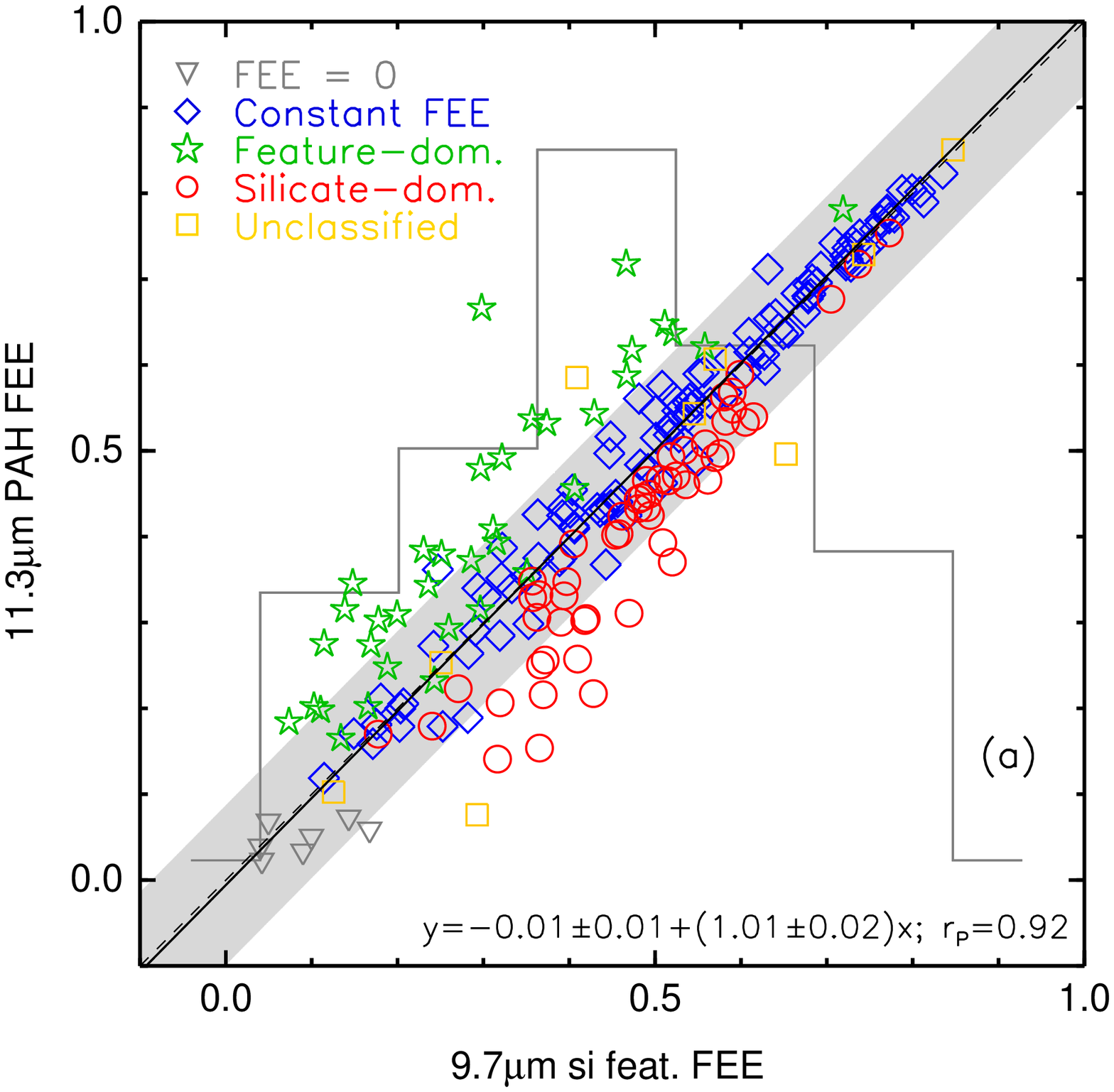}\plotone{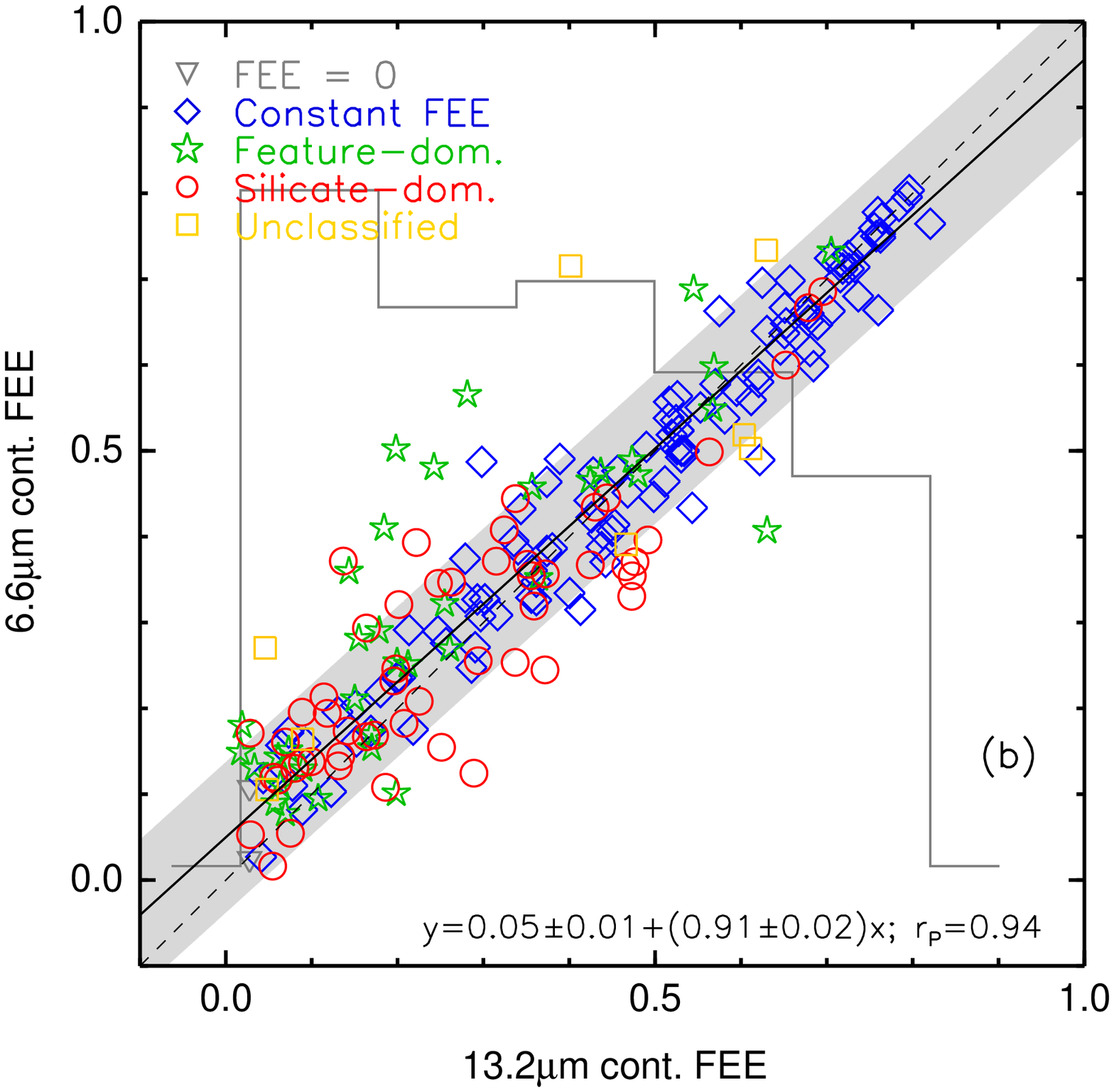}\plotone{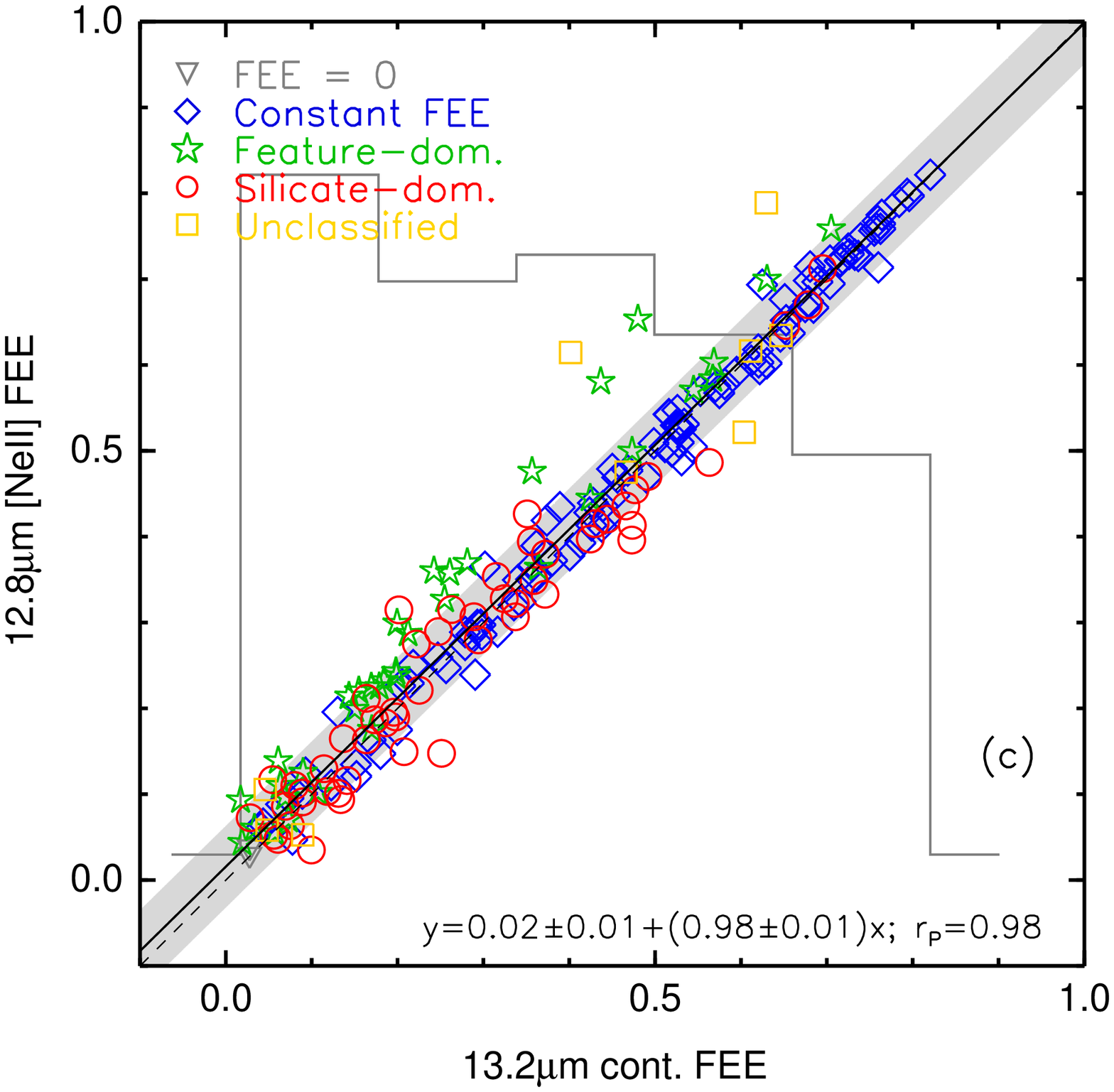}\plotone{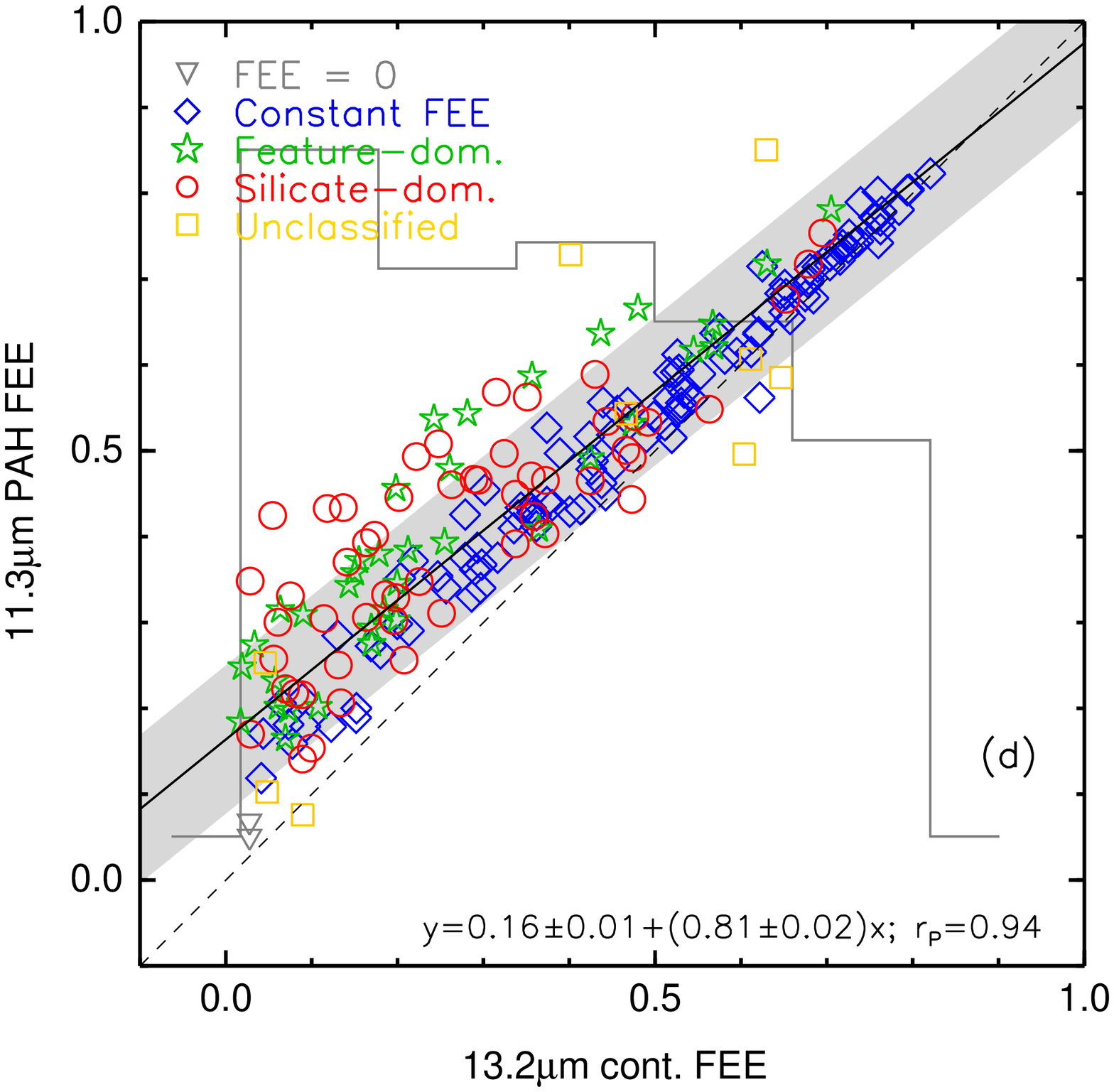}\plotone{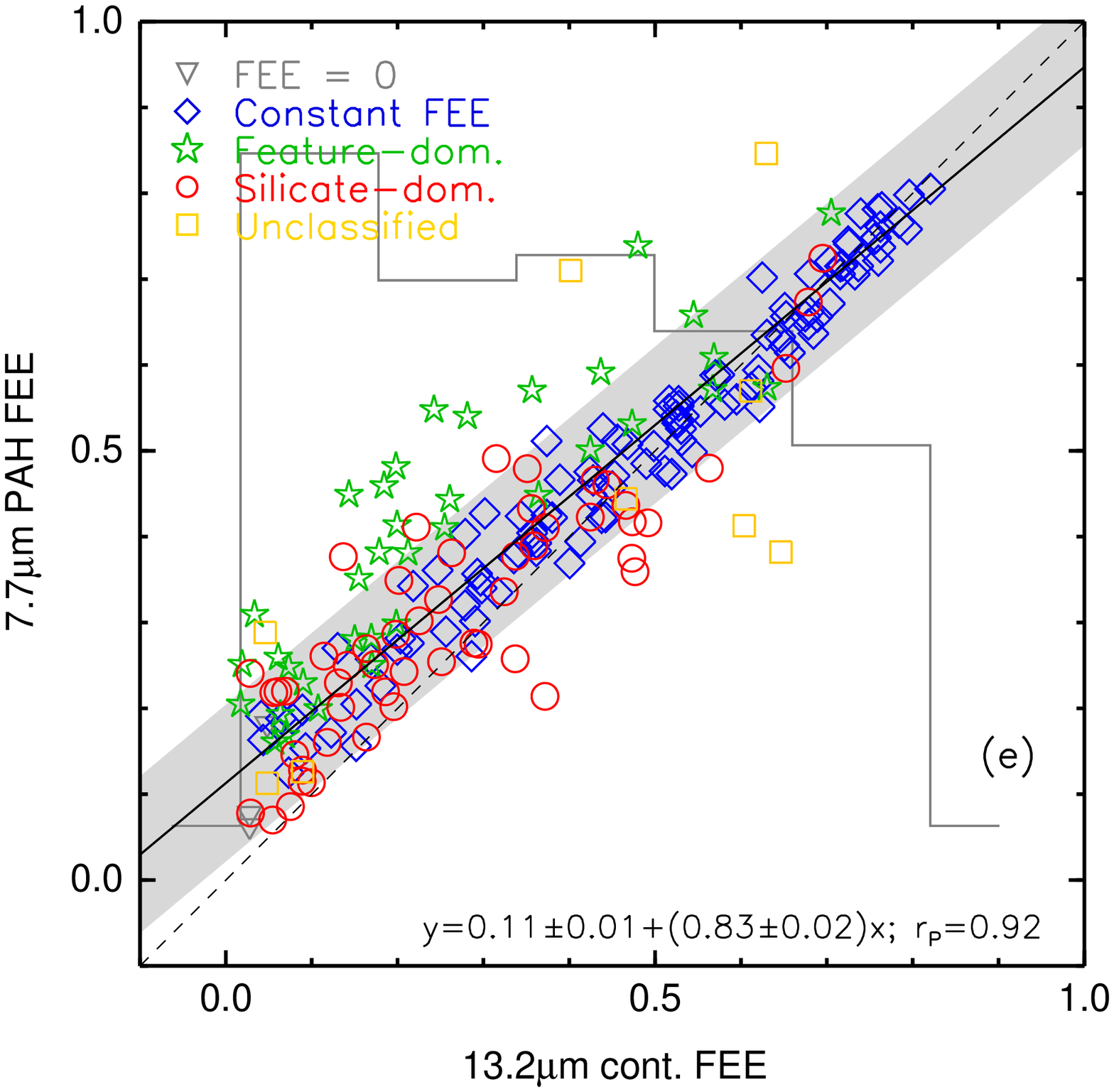}\plotone{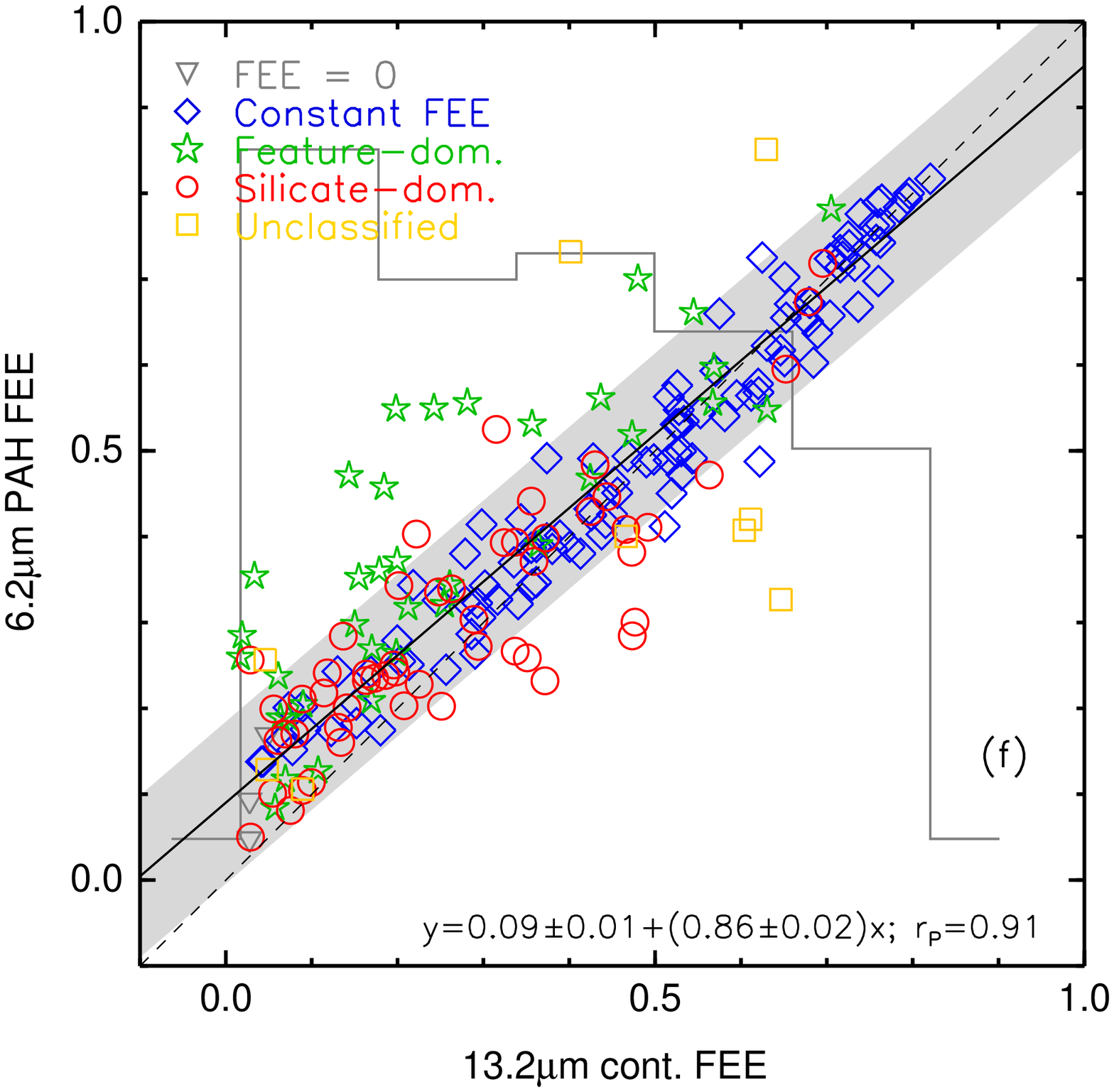}
%\plotone{./figures/f4f.ps}
%\vspace{.5cm}
\caption{\footnotesize From left to right and top to bottom: (a) FEE$_{11.3\mu m}$ (\PAHa) vs. FEE$_{9.7\mu m}$ (peak of the silicate absorption feature). Then, FEE$_{13.2\mu m}$ (continuum) vs.: (b) FEE$_{6.6\mu m}$ (continuum), (c) FEE$_{12.8\mu m}$ (\NeII\ line), (d) FEE$_{11.3\mu m}$, (e) FEE$_{7.7\mu m}$ (\PAHc), and (f) FEE$_{6.2\mu m}$ (\PAHd). The different symbols and colors represent the FEE$_\lambda$ galaxy types (unresolved: dark grey inverted triangles; constant/featureless: blue diamonds; PAH-dominated: green stars; silicate-dominated: red circles; and unclassified: yellow squares). The black dashed line shows the one-to-one relation, while the black solid line is the linear fit to the data, whose parameters can be found at the bottom right of the plots. The shaded region is the 1$\,\sigma$ uncertainty of the fit, while the light gray underlying line is the arbitrarily-scaled histogram of the data distributed in equal bins of FEE$_{9.7\mu m}$ or FEE$_{13.2\mu m}$.}\label{f:contfeevsfees}
\vspace{.25cm}
\end{figure*}

In Figure~\ref{f:contfeevsfees} we examine how the FEE of several MIR spectral features, averaged at the wavelength where they peak $\pm\,0.1\,\micron$, vary as a function of the FEE of the continuum emission at 13.2$\,\micron$, which was studied in Paper~I. From left to right and top to bottom we display: the \PAHa\ FEE as a function of the FEE at the peak of the 9.7$\,\micron$ silicate feature, and the 6.6$\,\micron$ continuum FEE, \NeII\ emission line (+ \PAHe) FEE, \PAHa\ FEE, \PAHc\ FEE, and \PAHd\ FEE as a function of the 13.2$\,\micron$ continuum FEE.

Figure~\ref{f:contfeevsfees}a
%presents the FEE$_{11.3\mu m}$ (\PAHa) as a function of the FEE$_{9.7\mu m}$ (peak of the silicate feature) and
shows, in a quantitative way, the visual classification of the FEE$_\lambda$ functions of galaxies performed in Paper~I, which we used for separating sources with characteristic FEEs$_\lambda$ in three main types/categories (Figure~\ref{f:feetype}). Within the uncertainties, (U)LIRGs showing a constant/featureless FEE$_\lambda$ (blue diamonds) are located on the one-one relation (black dashed line). Sources presenting an excess of extended PAH emission with respect to that of the continuum (green stars) are above the line, while galaxies with the FEE$_\lambda$ peaking at 9.7$\,\micron$ (red circles) are generally located below. We refer the reader to Paper~I for more details about this classification.

The principal reason we plot the 6.6$\,\micron$ continuum in Figure~\ref{f:contfeevsfees}b is to verify that the loss of angular resolution as a function of wavelength that affects the $5-14\,\micron$ SL spectra is not biasing our measurements of the FEE$_\lambda$, such that it decreases as wavelength increases. We can see that this is not the case, as the best fit to the FEE$_{6.6\mu m}$ and FEE$_{13.2\mu m}$ data, both representing continuum emission, is compatible, within the uncertainty (see the gray shadow), with the one-to-one relation.
% The large dispersion found in the data is caused by the larger uncertainty in the FEE measurement at 6.6$\,\micron$ due to the under-sampling effect of the spectra at these wavelengths ($\lambda_{obs}\,\lesssim\,7.8\,\micron$), as explained in Paper~I. Nevertheless, despite of the scatter, there is no systematic tendency for finding lower FEE$_{13.2\mu m}$ values than those measured at 6.6$\,\micron$, which
This implies that the decrease in angular resolution as the wavelength increases does not affect the trends significantly, at least in terms of the FEE measurements, when enough galaxies are considered.

Figure~\ref{f:contfeevsfees}c shows the FEE averaged at the location of the \NeII\ emission line as a function of FEE$_{13.2\mu m}$. Both quantities are well correlated, which is in agreement with previous studies of smaller samples of local LIRGs \citep{PS2010}, and implies that the physical processes (AGN and/or star formation) that are responsible for the \NeII\ line and the MIR continuum emission are equally concentrated and that they are likely the same. On the other hand, the FEE of the \PAHa\ is larger than that of the continuum or the \NeII\ emission in many galaxies (Figure~\ref{f:contfeevsfees}d). Moreover, this is independent of the FEE$_\lambda$ function type and does not only occur for silicate-dominated FEE$_\lambda$ type galaxies (red circles; see related caveat in Section~\ref{ss:extrole}, where we discuss the effect of the extinction on the spatial distribution of the MIR emission of spectral features located within the wavelength range of the $9.7\,\micron$ silicate absorption band). This result is in agreement with works based on high spatial resolution $N$-band ($8-13\,\micron$) spectroscopy of local LIRGs where it has been found that the spatial distribution of the \NeII\ emission and the MIR continuum in star-forming regions follows that of the \Paalpha\ emission \citep{DS2010a}, which is in turn associated to young, compact \HII\ regions \citep[e.g.,][]{AAH2002}. Meanwhile, the \PAHa\ emission can be powered not only by \HII\ regions but also by rather ``evolved'' ($\gtrsim\,8-10\,$Myr) stars, which emit fewer ionizing photons. In other words, the \PAHa/\NeII\ flux ratio depends on the age of the ionizing stellar populations (\citealt{DS2010a}). This diffuse PAH emission would be located in between the circumnuclear young star-forming regions and also farther from the nucleus of (U)LIRGs (\citealt{DS2008}; \citealt{PS2010}; \citealt{Bendo2010}).
% PAH destruction due to the presence of an AGN might also play a role in the spatial distribution of PAH emission in the nuclei of LIRGs at sub-kpc spatial scales \citep{Voit1992} which can be probed only for the most nearby galaxies of our sample using ground-based telescopes (\citealt{DS2010a}).
One might think that, because of the limited spectral resolution of the data, the \NeII\ line could be contaminated by the emission of the \PAHe\ feature. However, this is not the case, because if the \PAHe\ were as extended as the \PAHa, or even as the 7.7 or \PAHds\ (see below), and the \PAHe\ was dominating the emission at 12.8$\,\micron$ then, for a given value of the FEE$_{13.2\mu m}$, the FEE$_{12.8\mu m}$ should appear much more extended and/or with a larger scatter than what is observed in Figure~\ref{f:contfeevsfees}c. Instead, one can identify a tight correlation between both FEEs at 12.8 and 13.2$\,\micron$.

Figure~\ref{f:contfeevsfees}e and \ref{f:contfeevsfees}f show that the 7.7 and \PAHd\ emissions are as compact as, or just slightly more extended than, that of the MIR continuum and hence the \NeII\ emission, but less than the \PAHa.
% The ordinate in the origin is significantly larger for the later.
This is in agreement with the notion that the 7.7 and 6.2$\,\micron$ emission bands are enhanced for ionized PAHs, while the \PAHa\ emission is similar for neutral and ionized molecules (e.g., \citealt{Allamandola1999}; \citealt{Galliano2008}). Since ionized PAH carriers require harder radiation fields in order to be excited than neutral PAHs \citep{Draine2001}, they will be more tightly associated to younger massive star forming regions which, in turn, are also strong \NeII\ emitters. Given the nature of LIRGs and ULIRGs, many of which harbor strong nuclear starbursts, this would result in having both the 7.7 and the \PAHds\ slightly more extended than the \NeII\ line but less than the \PAHa.

\begin{figure}
\epsscale{1.1}
\plotone{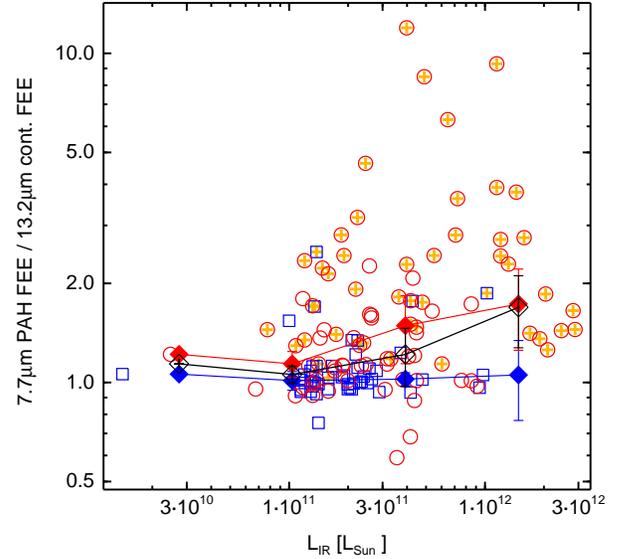}
%\vspace{.5cm}
\caption{\footnotesize FEE$_{7.7\mu m}$/FEE$_{13.2\mu m}$ ratio as a function of the \LIR\ of our (U)LIRG sample.  The symbols and colors depend on the \textit{IRAS} log($f_{60\,\mu m}/f_{100\,\mu m}$) FIR ratio of the galaxies. The blue squares and red circles represent cold and warm sources with log($f_{60\,\mu m}/f_{100\,\mu m}$) ratios lower and larger than $-0.2$ respectively. The blue and red filled diamonds show the median values of the FEE$_{7.7\mu m}$/FEE$_{13.2\mu m}$ ratio for the two groups of sources in equally-distributed IR luminosity bins. The black diamonds represent the same ratio but considering all sources. Galaxies whose core sizes are unresolved at 13.2$\,\micron$ are marked with yellow crosses. }\label{f:feepahcont-lir}
\vspace{.25cm}
\end{figure}

In Paper~I we also examined the variations of the FEE of the continuum at 13.2$\,\micron$ as a function of the \LIR. Here we also explore how the FEE of PAHs in (U)LIRGs depends on the \LIR\ and compare them with their corresponding FEE$_{13.2\mu m}$. Figure~\ref{f:feepahcont-lir} shows that, despite the large dispersion, the median of the FEE$_{7.7\mu m}$/FEE$_{13.2\mu m}$ ratio (black solid line) slightly increases with \LIR. The more IR luminous a galaxy is, the more extended its \PAHa\ emission is compared with the MIR continuum. This is in agreement with our earlier result in Paper I, where we found that the compactness of the MIR continuum increases for galaxies with $\LIR\,\gtrsim\,10^{11.8}\,\Lsun$. However, the 6.2, 7.7, and \PAHas\, do not show such a steep decrease of their FEEs around these IR luminosities. Therefore, it is mainly the MIR continuum that becomes more compact as the \LIR\ of galaxies increases, making the median PAH-to-continuum FEE ratio to increase too. Likewise, the most luminous galaxies are also those having the warmest log($f_{60\,\mu m}/f_{100\,\mu m}$) FIR ratios. This is expected for ULIRGs, which for a long time are known to host nuclear starbursts and AGNs and show on average warm colors \citep{Sanders1996}, but these same properties are also displayed by several sources with lower IR luminosities ($10^{11}\,\Lsun\,\lesssim\,L_{\rm IR}\,\lesssim\,10^{12}\,\Lsun$). Indeed, independently of their \LIR, the warmest galaxies, those with log($f_{60\,\mu m}/f_{100\,\mu m}$)$\,> -0.2$ (identified by red circles in Figure~\ref{f:feepahcont-lir}), are those driving the trend and accounting for the dispersion seen in the data.
%, while those colder display a rather constant FEE$_{7.7\mu m}$/FEE$_{13.2\mu m}$ ratio. 
As we will discuss later, this is in agreement with their compact nature and the general scenario we will propose in Section~\ref{ss:pahcontfir}.

\subsection{The Role of Dust Extinction}\label{ss:extrole}

We also investigated the specific distribution of (U)LIRGs of different FEE$_\lambda$ types in Figure~\ref{f:contfeevsfees}a$-$e. As we mentioned above, Figure~\ref{f:contfeevsfees}c shows that the FEE$_{11.3\mu m}$ is larger than the FEE$_{13.2\mu m}$ in many (U)LIRGs. However, for some of them this may be an extinction effect.
% This can be interpreted as: (1) the \PAHa\ emission is truly spreading over the disks of the galaxies farther out than the MIR continuum and \NeII\ line emission; (2) the continuum is more compact than the \PAHa\ emission; (3) the PAHs are being destroyed in the nuclei of some galaxies; (4) an extinction effect. 

Even in the MIR, dust obscuration must be taken into account in (U)LIRGs since their nuclear extinction is much larger than that of their extended regions \citep[e.g.,][]{GM2009}.
%However, this does not necessarily imply that the \PAHa\ is intrinsically more extended than the continuum emission or, viceversa, that the continuum is more compact than the \PAHa\ emission, but it can also be explained by the fact that the nuclear extinction of some (U)LIRGs can be much larger than that of their extended regions \citep[see][]{GM2009}.
Galaxies with a silicate-dominated FEE$_\lambda$ type are marked as red circles in Figure~\ref{f:contfeevsfees}. We observe that for these sources the \PAHa\ feature, which is located within the 9.7$\,\micron$ silicate absorption feature, appears more extended (i.e., have large FEE$_{11.3\mu m}$ values). This is not because there is more PAH emission in the disks of these galaxies but instead because the emission is suppressed in their nuclei due to the extinction, as we already suggested in Paper~I. Therefore, the differential extinction between the unresolved and extended components plays an important role in the ``apparent'' spatial distribution of the emission of the different MIR features in these galaxies. Sources with extreme differences between the extinction in their nuclei and extended regions (large $\Delta\SSi$, see Section~\ref{ss:resspec}) will show an artificial excess in their FEE$_{11.3\mu m}$/FEE$_{13.2\mu m}$ ratio (see implications in next Sections). Indeed, this also applies in general to all features within the $\sim\,8-12.5\,\micron$ wavelength range affected by the 9.7$\,\micron$ silicate feature, such as the \PAHb\ and the \SIV\ line. As a consequence, in order to interpret correctly the FEE ratios of those galaxies with large $\Delta\SSi$, a careful decomposition of their spectra in different dust components and feature emissions must be done to properly account for extinction effects (Stierwalt et al. 2011, in prep.).

Nevertheless, not all (U)LIRGs have highly obscured nuclei at 9.7$\,\micron$. There are several galaxies in our sample, those with a PAH-dominated FEE$_\lambda$ type, that show high PAH-to-continuum FEE ratios but, unlike galaxies with silicate-dominated FEE$_\lambda$ types, they do not display large differences between the extinction of their unresolved and extended components.

\subsection{The Influence of an AGN}\label{ss:pahagnfrac}

%\begin{figure}
%\epsscale{1.}
%\plotone{./f4.ps}
%%\vspace{.5cm}
%\caption{\footnotesize Nuclear \SSi\ $-$ extended \SSi, $\Delta$\SSi, as a function of the MIR AGN fraction. The symbols, lines, and colors ar as in Figure~\ref{f:fee113-ssinucssiext}.}\label{f:ssinucssiext-agnfrac}
%\vspace{.5cm}
%\end{figure}

\begin{figure}
\epsscale{1.1}
\plotone{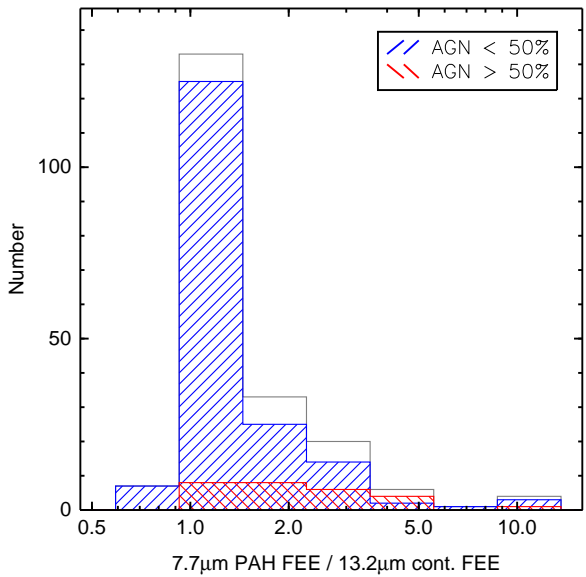}
%\plotone{./figures/f5b.ps}
%\vspace{.5cm}
\caption{\footnotesize Histogram of the FEE$_{7.7\mu m}$/FEE$_{13.2\mu m}$ ratio of our galaxies grouped as a function of the AGN contribution to their MIR continuum. A total of 205 (U)LIRGs for which we have measurements of their FEE$_{7.7\mu m}$/FEE$_{13.2\mu m}$ ratios and AGN-fractions from Petric et al. (2011) were used. The solid lines represent the median value of each distribution, while the dotted lines show the $\pm\,1\,\sigma$ uncertainty with respect to the medians. A Kolmogorov-Smirnov test indicates clearly ($p\,\sim\,10^{-6}$) that the distributions are not drawn from same parent population.}\label{f:feepahs-agnfrac}
\vspace{.25cm}
\end{figure}

In Paper~I we studied how the FEE of the continuum at 13.2$\,\micron$ varies with the MIR AGN fraction of galaxies
% \citep[see][]{Petric2011}
and showed that the MIR continuum becomes more compact for AGN-dominated sources. In order to examine whether the presence of an AGN also influences the distribution of the PAH emission in (U)LIRGs, we plot in Figure~\ref{f:feepahs-agnfrac} the histogram of the ratio of the \PAHc\ FEE to the 13.2$\,\micron$ continuum FEE, grouping the galaxies as a function of their MIR AGN-fraction. Galaxies in which an AGN dominates the MIR emission display large PAH-to-continuum FEE ratios. The median \PAHc-to-continuum FEE ratio for MIR star formation-dominated galaxies, that is, sources for which \cite{Petric2011} find that the AGN contribution to the MIR emission is less than 50\%, is $1.08\,\pm\,0.10$, while for AGN-dominated sources, it is $1.86\,\pm\,0.41$. The corresponding median ratios for the \PAHd-to-continuum FEE ratio are $1.04\,\pm\,0.11$ and $1.74\,\pm\,0.51$ respectively. We note that a much larger difference is seen in the FEE$_{11.3\mu m}$/FEE$_{13.2\mu m}$ ratio. However, as discussed in Section~\ref{ss:extrole}, the FEE$_{11.3\mu m}$ is affected by the 9.7$\,\micron$ silicate absorption feature and, at the same time, AGN-dominated galaxies tend to show large differences between the extinction of their nuclei and extended regions, i.e., large $\Delta\SSi$ values. Both effects combined may artificially increase the \PAHa-to-continuum FEE ratio leading to a possible misinterpretation of the observed correlation. The 7.7 and \PAHd, on the other hand, do not suffer from severe obscuration and therefore the trend and ratios seen in Figure~\ref{f:feepahs-agnfrac} are more reliable.

Although the difference between the median PAH-to-continuum FEE ratios of MIR AGN and star formation-dominated galaxies is at the $\sim\,2\,\sigma$ level, a Kolmogorov-Smirnov test performed between the two samples indicates clearly ($p\,\sim\,10^{-6}$) that the distributions are not drawn from same parent population. This, combined with the fact that the FEE$_{13.2\mu m}$ decreases as the MIR AGN fraction increases (see Paper~I) suggest that as the MIR emission of a (U)LIRG begins to be dominated by an AGN, the hot dust continuum becomes more compact while the PAH emitting regions do not. That is, the presence of an AGN modifies the spatial extent of the hot dust continuum but has a negligible effect on the overall, kpc-scale distribution of the PAH emission.
% more rapidly than the PAH emission does (if it does whatsoever).

Indeed, this is consistent with the way in which the AGN fraction is calculated in \cite{Petric2011}. In this work the contribution of an AGN to the MIR emission of a (U)LIRG is estimated through the so-called Laurent diagram (\citealt{Laurent2000}; \citealt{Armus2007}) which, for the GOALS sample, relies mainly on the ratio between the \PAHd\ emission and the 5.5$\,\micron$ continuum (see Figure~3 in \citealt{Petric2011}). Therefore, as the compactness of the hot dust continuum increases, so does the surface brightness of the galaxy and, as a consequence, this pseudo-equivalent width (EW) of the \PAHd\ decreases, which implies that the AGN contribution to the MIR is larger and dilutes the PAH emission.

We have also investigated the influence of an AGN depending on the FEE$_\lambda$ type of the host.
For example, there are no galaxies with constant/featureless FEEs$_\lambda$ having a MIR AGN fraction above 0.5. This is an important result since, as we have explained, the galaxies were classified into the different FEE$_\lambda$ types independently of their MIR spectral characteristics. Thus, it appears that when an AGN dominates the MIR emission it also leaves a ``footprint'' in the FEE$_\lambda$ function of a galaxy in the sense that the MIR emission of (U)LIRGs with a constant FEE$_\lambda$ will never be dominated by an AGN or, in general, appear compact.

We would like to stress here that not only AGN-dominated galaxies show large PAH-to-continuum FEE ratios. Nearly 30\% of (U)LIRGs whose MIR emission is not dominated by an AGN (AGN-fraction $<\,0.5$) display large FEE$_{7.7\mu m}$/FEE$_{13.2\mu m}$ ratios. The nature of these sources will be addressed in the next Section.

\subsection{PAH-to-Continuum FEE Ratios and FIR Emission}\label{ss:pahcontfir}

\begin{figure}
\epsscale{1.1}
\plotone{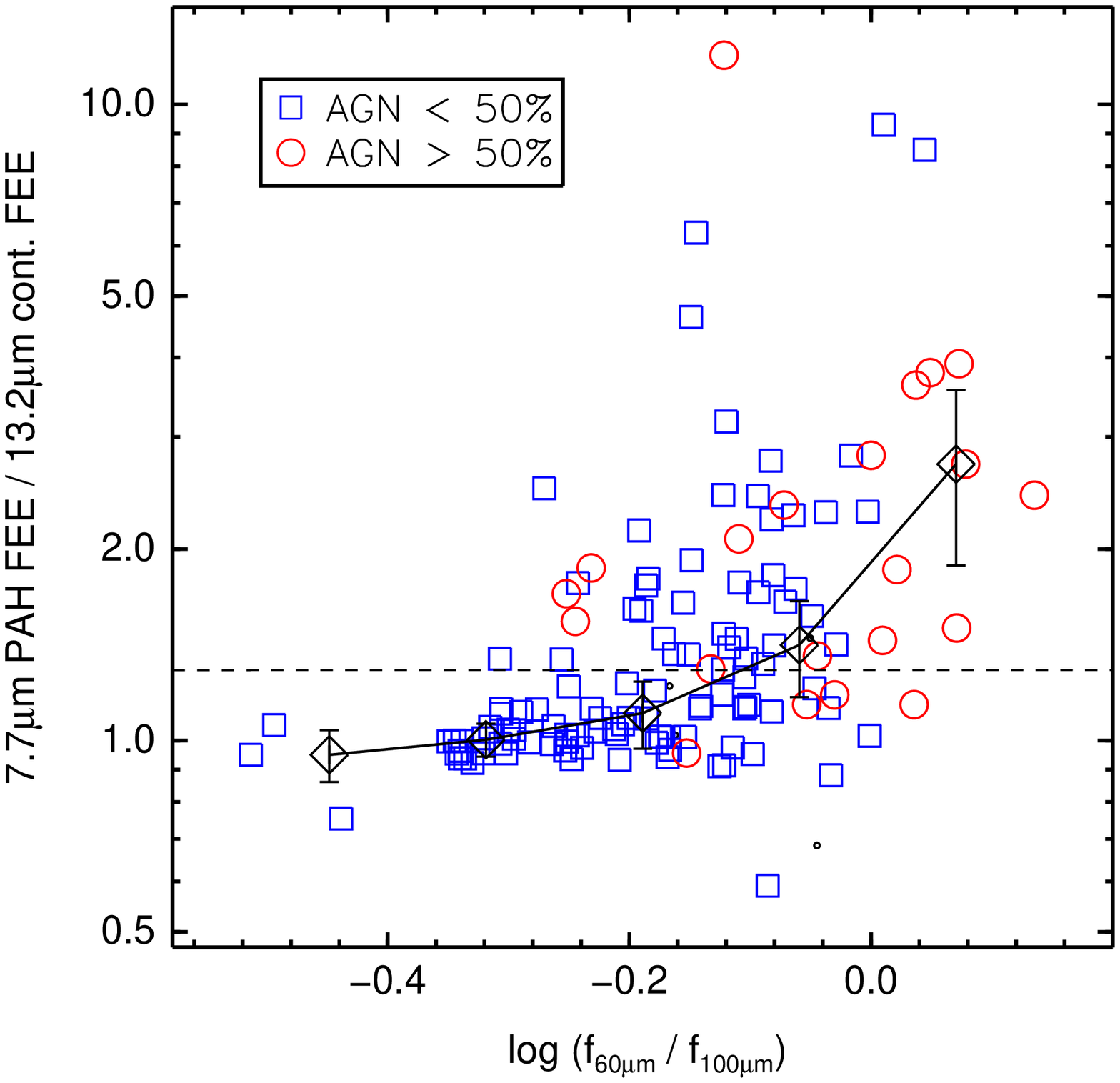}\plotone{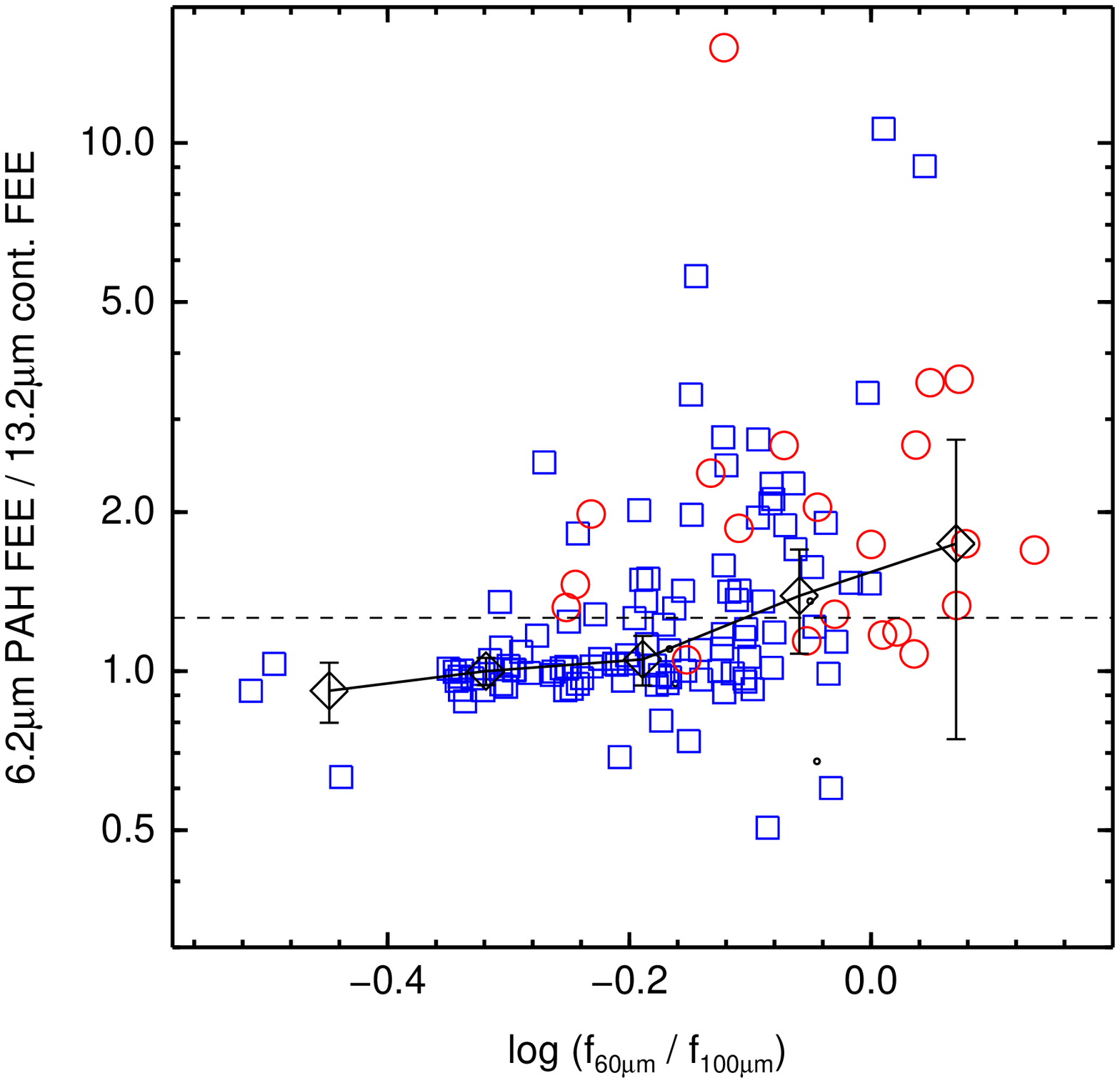}
%\vspace{.5cm}
\caption{\footnotesize Top: FEE$_{7.7\mu m}$/FEE$_{13.2\mu m}$ ratio as a function of the \textit{IRAS} log($f_{60\,\mu m}/f_{100\,\mu m}$) FIR color. The red circles and blue squares mark galaxies whose MIR AGN fraction is $\geq\,$0.5 and $<\,$0.5 respectively. The small black dots represent sources for which the MIR AGN fraction is not available. The dashed line is defined as the median PAH-to-continuum FEE ratio value + $2\,\sigma$ of star-formation dominated sources (see Figure~\ref{f:feepahs-agnfrac}). Bottom: Same plot but for the FEE$_{6.2\mu m}$/FEE$_{13.2\mu m}$ ratio.}\label{f:feepahcont-fir}
\vspace{.25cm}
\end{figure}

In this section we examine whether the PAH-to-continuum FEE ratios are related with the FIR colors of galaxies, that is, whether the differences in the compactness of the PAH and MIR continuum emission in (U)LIRGs are linked to their global dust temperatures traced by the FIR emission. We have already discussed that PAH emission probes time-scales of star formation larger than the MIR continuum emission, which is tightly associated to young, ionizing stellar populations (e.g., \citealt{Calzetti2007}; \citealt{DS2010a}). Likewise, results based on \textit{IRAS} as well as on recent high spatial resolution \textit{Herschel} observations have clearly demonstrated that the $100-500\,\micron$ emission of galaxies arises mainly from a diffuse cold dust component that extends smoothly along their disks and is heated by old stars, rather than being concentrated in knots of current star formation which are better traced by dust continuum emission at $<\,100\,\micron$ (\citealt{Sauvage1992}; \citealt{Walterbos1996}; \citealt{Bendo2010}; \citealt{Kramer2010}). Moreover, in a recent \textit{Herschel} study of local spiral galaxies it has been found that their cores show enhanced, warmer 70/160$\,\micron$ colors than their disks \citep{Sauvage2010}. Therefore, since in Paper~I we showed that the FEE$_{13.2\mu m}$ decreases when (U)LIRGs become warmer, i.e., when their \textit{IRAS} log($f_{60\,\mu m}/f_{100\,\mu m}$) FIR color increases, we would also expect to find a trend between the PAH-to-continuum FEE ratios and the FIR color of galaxies.

Figure~\ref{f:feepahcont-fir} shows that this is actually the case. Although the trend for the \PAHd-to-continuum FEE ratio is less significant, it is clear that the \PAHc-to-continuum FEE ratio and its dispersion increase as galaxies become warmer. Moreover, practically all (90\%) MIR star formation-dominated sources with large PAH-to-continuum FEE ratios (sources above the dashed lines, which are defined as the medians + $2\,\sigma$ of the distributions shown in Figure~\ref{f:feepahs-agnfrac}) have log($f_{60\,\mu m}/f_{100\,\mu m}$)$\,\gtrsim\,-0.2$. A plausible explanation for this is the following: as the nuclear starburst activity in a (U)LIRG increasingly dominates its integrated IR emission, the dust in the central regions would reach higher temperatures. This is because more and more of the atoms and molecules of the nuclear gas would be ionized and dissociated by the high energy photons of the stars formed in the \HII\ regions of their proximity, causing the ratio between the volume of the spherical shells of the PDRs, from where most of the PAH emission originate, and the volume of the dust emitting region to decrease.
%However, this would not affect the gas further away from the nucleus along with its PAH emission.
That is, a larger fraction of the ionizing photons would be absorbed by dust close to the stars increasing its temperature, and would not reach the PDRs. As a result, the MIR continuum emission due to this hot dust will be more compact (having a higher luminosity surface density), and the global galaxy colors will become warmer.

We could then consider that the ``normal'' behavior of a galaxy is to show the same spatial extent of the PAH and MIR continuum emission probably as a consequence of a sufficient mixing of its gas and stars. It is only when a strong, compact emitter starts to dominate the nuclear MIR continuum
%, which is seen when galaxies reach the LIRG or ULIRGs phrase,
that we see a deviation (i.e., PAH-to-continuum FEE ratios above the dashed lines in Figures~\ref{f:feepahs-agnfrac} and \ref{f:feepahcont-fir}). This is the case for 80\% of AGN-dominated GOALS galaxies and 60\% of warm sources with log($f_{60\,\mu m}/f_{100\,\mu m}$)$\,>\,-0.2$, excluding AGN-dominated galaxies. In these systems, the MIR continuum is more compact than the PAH emission, which is still as extended as in the ``normal'' population since it arises mostly from a diffuse component (the disk of the galaxies). Therefore, a more appropriate interpretation would not be that there is an excess of PAH emission or that the PAHs are more extended in absolute terms in these galaxies. It is the MIR continuum that is more compact due to the enhancement caused by the nuclear power-source that increases the PAH-to-continuum FEE ratio.
% In addition, it is important to stress that, as seen in Figure~\ref{f:feepahcont-lir} (Section~\ref{ss:feedist}), these compact galaxies are not only classified as ULIRGs but there are also included some systems within the IR luminosity range of LIRGs.
It is important to stress that these compact galaxy cores are not only observed in ULIRGs \citep[see, e.g.,][]{Charmandaris2002}; we also see them in LIRGs (see also Figure~\ref{f:feepahcont-lir}). In addition, Figures~\ref{f:feepahs-agnfrac} and \ref{f:feepahcont-fir} also suggests that irrespective of whether the nuclear MIR emission of (U)LIRGs is dominated by an AGN or a nuclear starburst, their impact on the compactness of the MIR continuum and PAH emission is the same. In other words, both processes are indistinguishable in terms of the relative spatial extent of the PAH and MIR continuum emission.
%, which is in agreement with the result we obtained in Section~\ref{ss:pahagnfrac} from different arguments.

We note that the MIR emission of 27\% of the warm compact sources, those with log($f_{60\,\mu m}/f_{100\,\mu m}$)$> -0.2$ and above the dashed lines in the Figures, is dominated by an AGN (MIR AGN fraction $\geq\,0.5$). It is known that AGN and star formation dominates the emission mainly at different wavelengths, i.e., the FIR colors of these galaxies are not affected by the existence of an AGN but they are associated principally to their star formation processes.
% which could weaken the previous argument since for these sources, the measurement of their 13.2$\,\micron$ FEE could be affected by the  of the AGN.
% {\bf I AM A BIT CONFUSED WITH THIS: And indeed it does, as we showed in \cite{DS2010b} and in the previous Section. However, we have to note that (1) the trend in Figure~\ref{f:feepahcont-fir} starts to be seen at colder FIR colors, and (2)}
This implies that in these (U)LIRGs, nuclear \textit{compact} star formation, traced by their warm FIR colors, must be coexisting with MIR-detected AGN at the same time (see, e.g., \citealt{Schweitzer2006}; \citealt{Netzer2007}; \citealt{Veilleux2009}), both contributing to the large PAH-to-continuum FEE ratios they show.

\subsection{Averaged Spectra of Unresolved and Extended Components}\label{ss:resspec}

The technique we use to calculate the FEE$_\lambda$ from the 2 dimensional \textit{Spitzer}/IRS images allows us to derive not only the spatial extent of specific spectral features but also to separate the MIR spectrum of the extended emission component of galaxies from that of the unresolved component in a simple manner. This can be done just by multiplying their total, flux-calibrated \textit{Spitzer}/IRS spectrum calculated by \cite{Petric2011}, by their FEE$_\lambda$ and $1-$FEE$_\lambda$ functions respectively. More technical details are available in the Appendix of Paper~I.

\begin{figure}
\epsscale{1.15}
%\plotone{/data/spitzer/goals/sample/results/feecomponentspecs.ps}
\plotone{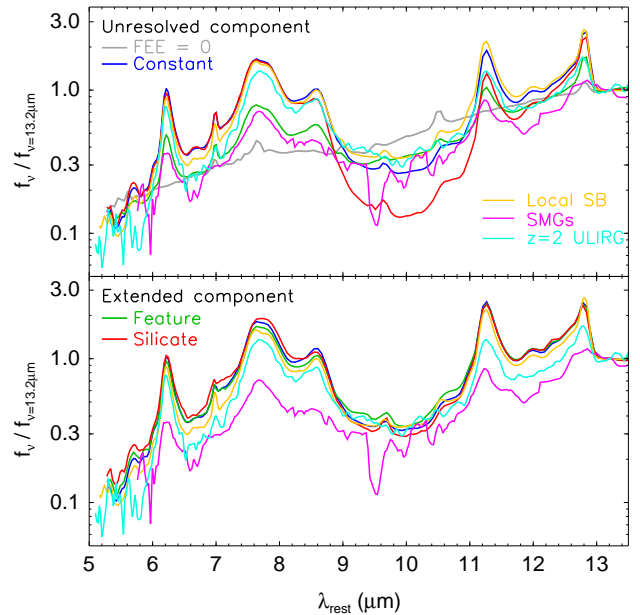}
\caption{\footnotesize Weighted averaged spectra of the different FEE$_\lambda$ types scaled to 13.2$\,\micron$ (unresolved: gray; constant/featureless: blue; PAH-dominated: green; silicate-dominated: red). The nuclear and extended emission components of each type are presented in the top and bottom panels respectively.
% The $1\,\sigma$ uncertainties on the weighted averages are shown as light shadows colored similarly to the corresponding FEE$_\lambda$ types.
On both panels we over-plot the local spectral template of \cite{Brandl2006} of normal star-forming galaxies in yellow, the averaged spectrum of the high-redshift sub-millimeter galaxies from \cite{MD2009} in pink, as well as the template of the z$\,\sim\,$2 ULIRGs from \cite{Farrah2008} in cyan. Unlike our GOALS galaxies, for which we show their nuclear and extended components, the comparison spectra are from the integrated emission of galaxies.
% The dotted lines represent the continua used to measure the \SSi\ in the spectra (see text for details on the fit).
}\label{f:feecompspecs}
\vspace{.25cm}
\end{figure}

\begin{figure}
\epsscale{1.15}
%\plotone{/data/spitzer/goals/sample/results/meanfeetypespecs.ps}
\plotone{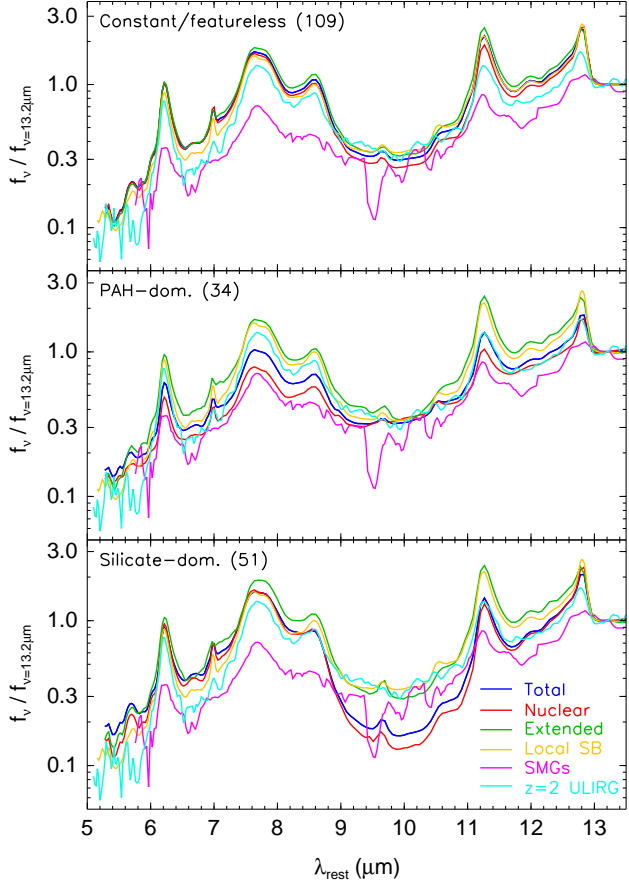}
\caption{\footnotesize Averaged spectra of the total (blue), unresolved (red), and extended (green) emission components of our sample, normalized at 13.2$\,\micron$, for the different types of FEE$_\lambda$. The integrated spectral templates used to compare our sample with low and high-redshift star-forming galaxies are as in Figure~\ref{f:feecompspecs}. A total of 194 sources, for which both their FEE$_{13.2\mu m}$ and $1-$FEE$_{13.2\mu m}$ are larger than 0.05 and therefore could be normalized at 13.2$\,\micron$, were used for this plot.
% The dotted lines represent the continua used to measure the \SSi\ in the spectra (see text for details on the fit).
}\label{f:feetypespecs}
\vspace{.25cm}
\end{figure}

\begin{deluxetable*}{lcccccc}
\tabletypesize{\scriptsize}
%\rotate
\tablewidth{0pc}
\tablecaption{\scriptsize Silicate Strengths and PAH EWs of the FEE$_\lambda$ Types}
\tablehead{\colhead{FEE$_\lambda$} & \multicolumn{2}{c}{Total} & \multicolumn{2}{c}{Unresolved} & \multicolumn{2}{c}{Extended} \\
\cline{2-7}
\colhead{Type} & \colhead{\SSi} & \colhead{\PAHd\ EW} & \colhead{\SSi} & \colhead{\PAHd\ EW} & \colhead{\SSi} & \colhead{\PAHd\ EW}  \\
\colhead{(1)} & \colhead{(2)} & \colhead{(3)} & \colhead{(2)} & \colhead{(3)} & \colhead{(2)} & \colhead{(3)}}
\startdata
Unresolved            &  -0.55 +/- 0.01  &  0.05 +/- 0.01  &  -0.31 +/- 0.01  &  0.00 +/- 0.01  &  \dots & \dots  \\
Constant/featureless  &  -0.89 +/- 0.01  &  0.50 +/- 0.01  &  -1.01 +/- 0.01  &  0.50 +/- 0.02  &  -0.81 +/- 0.02  &  0.54 +/- 0.01  \\
PAH-dominated         &  -0.74 +/- 0.01  &  0.29 +/- 0.01  &  -0.60 +/- 0.03  &  0.24 +/- 0.01  &  -0.77 +/- 0.01  &  0.48 +/- 0.01  \\
Silicate-dominated    &  -1.53 +/- 0.02  &  0.34 +/- 0.02  &  -1.72 +/- 0.02  &  0.43 +/- 0.02  &  -0.94 +/- 0.01  &  0.51 +/- 0.01
\enddata
\tablecomments{\scriptsize (1) FEE$_\lambda$ type; (2) Silicate strength; (3) \PAHd\ EW.}\label{t:sample}
\end{deluxetable*}

Using this method, we display together in Figure~\ref{f:feecompspecs} the averaged spectra of the different FEE$_\lambda$ types for the unresolved and extended components of (U)LIRGs. The component averages were weighted by the mean FEE$_\lambda$ and $1-$FEE$_\lambda$ of each galaxy, respectively. This was done in order to give proportionally more importance in the calculation of the extended component averages to (U)LIRGs showing larger values of FEE and, on the other hand, more importance to (U)LIRGs showing lower values of FEE when the unresolved component averages are calculated. In any case, we note that the results are the same irrespectively of this weighting.

Strikingly, we find that the intensity of the spectral features (PAHs, emission lines and 9.7$\,\micron$ silicate absorption) of the extended components are very similar in all FEE$_\lambda$ types, which suggests that the global physical properties of the
%physical mechanism responsible for this emission, that is, the 
star formation in the external parts (disks) of galaxies ($d\,\gtrsim\,1.5\,$kpc) are likely the same. Moreover, the averaged spectra of the extended emission of all FEE$_\lambda$ types resemble that of the template of lower IR luminosity starburst galaxies from \cite{Brandl2006}. On the other hand, the spectra of the unresolved component are substantially different among the three FEE$_\lambda$ types even though the $f_{13.2\mu m}/f_{5.5\mu m}$ continuum ratio is similar in all of them, ranging between $\sim\,5-6$. The nuclear MIR spectra of the GOALS galaxy sample will be analyzed in detail in a forthcoming work of Stierwalt et al. (2011, in prep.). From another point of view, these results also imply that, on average, the differences seen in the global/integrated MIR spectra of local ULIRGs and also LIRGs arise only from the processes that are taking place in their cores, while the extended star formation appear not to be affected by the nuclear activity and is similar to the more quiescent mode of star formation found in lower IR luminosity systems.

The difference between the MIR spectra of the unresolved emission of the constant and silicate-dominated FEE$_\lambda$ types can be explained in terms of dust obscuration, with the silicate-dominated FEE$_\lambda$ type having a larger 9.7$\,\micron$ optical depth. The relative emission of the PAHs with respect to that of the continuum in the unresolved spectrum of the PAH-dominated FEE$_\lambda$ type is lower than in the constant/featureless FEE$_\lambda$ type. This implies that a mechanism, such as an AGN or a compact nuclear starburst, is either destroying the PAH carriers in the nuclei of these galaxies, or probably simply diluting their strength (peak to continuum ratio) by heating the small dust grains to higher temperatures thus increasing the MIR continuum emission; a result that is in agreement with Sections~\ref{ss:pahagnfrac} and \ref{ss:pahcontfir}.
%Moreover, the spectrum of the unresolved emission of the PAH-dominated FEE$_\lambda$ type appears to be an intermediate case between those of the constant and the unresolved FEE$_\lambda$ types.

In Figure~\ref{f:feetypespecs} we show the averaged spectra of the total (blue), unresolved nuclear (red), and extended (green) emission for the galaxies in our sample corresponding to each FEE$_\lambda$ type. In Table~\ref{t:sample} we present the silicate strength (\SSi) and \PAHd\ EW of the different averaged spectra. The \SSi\ was calculated by fitting a power-law to the spectra, with anchors at 5.5, 6.6, and 13.2$\,\micron$, and evaluated at its maximum. The \PAHd\ EW was obtained by fitting the local continuum of the feature using a linear function with anchors at 5.8 and 6.6$\,\micron$ and integrating its emission from 5.9 to 6.5$\,\micron$.

By definition, the averaged spectra of the total as well as the unresolved and extended components for the constant/featureless FEE$_\lambda$ type galaxies are very similar. The silicate strengths vary between $-0.8$ and $-1.00$ (see Table~\ref{t:sample}), and the \PAHd\ EWs $\sim\,0.5\,\micron$, which are close to the values found for the starburst galaxy sample of \cite{Brandl2006}. The average spectrum of the extended emission component for the PAH-dominated FEE$_\lambda$ type, on the other hand, is very different from those of the unresolved and total components. Although the \SSi\ is similar, the \PAHd\ EW of the extended spectrum (0.48$\,\micron$) is around twice that of the unresolved and total spectra ($0.24-0.29\,\micron$; see Table~\ref{t:sample}). The average $f_{13.2\mu m}/f_{5.5\mu m}$ continuum ratios of the nuclear and extended spectra are also very similar $\sim\,7$. The averaged spectra of the components of the silicate-dominated FEE$_\lambda$ type are intrinsically different. The spectrum of the unresolved emission shows a deep silicate absorption feature ($\SSi\,=\,-1.72$) while the spectrum of the extended emission is significantly shallower ($\SSi\,=\,-0.94$) but similar to the values found for the extended components of the constant/featureless and PAH-dominated FEE$_\lambda$ types. If we define $\Delta$\SSi\ as the difference between the silicate strength of the nuclear and extended component of a galaxy, then $\Delta\SSi\,=\,-0.78$ for the silicate-dominated FEE$_\lambda$ type (see also Section~\ref{ss:extrole}). The averaged nuclear spectra of the unresolved FEE$_\lambda$ type (Figure~\ref{f:feecompspecs}) show an almost flat rising continuum, a moderate silicate strength ($\SSi\,=\,-0.31$), and almost no PAH emission (\PAHd\ EW $\sim\,0.00\,\micron$), all of them typical signatures of AGN-dominated galaxies. Their MIR spectrum is similar to sources in region 1A of \cite{Spoon2007} diagram, in which the AGN emission is the likely culprit for the dilution of the PAH features.

The classification of the GOALS galaxies into different stages of interaction carried out by \cite{Petric2011} and \cite{Haan2011} also allows us to explore whether the PAH-to-continuum FEE ratios or the MIR spectral shape of the nuclear and extended components of galaxies are related to the merger stage of the systems. Unfortunately, no such clear trend is found. Only at the final stage of interaction (stage 4), when the nuclei of the galaxies are already assembled, the averaged spectrum of their unresolved component shows a deeper silicate absorption feature than that displayed by the spectrum of the extended component. This is understood if we take into account that most of the ULIRGs in GOALS, which show the largest silicate strengths, are classified as being in this latest merger stage. On the other hand, the averaged spectra of both components for galaxies grouped in stages 0 to 3 are very similar, and resemble that of local starburst galaxies.

\subsection{High-redshift Implications}

In Figures~\ref{f:feecompspecs} and \ref{f:feetypespecs} we also compare the averaged spectra of the GOALS sample, both nuclear and extended components, with the composite, integrated spectra of the $z\,\sim\,2$ ULIRG sample from \cite{Farrah2008}, and the sub-millimeter galaxies (SMGs) studied in \cite{MD2009} which are located at a redshift range of $z\,\sim\,0.65-3.2$. As already mentioned in \cite{Farrah2008}, their composite spectrum of high-$z$ ULIRGs and HyperLIRGs ($\LIR\,=\,10^{12.9}-10^{13.8}\,\Lsun$) is very similar to that of the starburst galaxies from \cite{Brandl2006}. We also note that it is also very similar to the averaged spectra of the extended components of our galaxy sample.
%, although it has a slightly higher $f_{13.2\mu m}/f_{5.5\mu m}$ continuum ratio.
This suggests that the properties of star forming high-$z$ ULIRGs, often selected based on the characteristic \textit{Spitzer}/IRAC colors due to the $1.6\,\micron$ stellar bump and their \textit{Spitzer}/MIPS $24\,\micron$ fluxes (e.g., \citealt{Magdis2010}) is very similar to the \textit{extended disks} of local (U)LIRGs. However, the composite, integrated SMG spectrum is different from our average spectrum of the extended component, displaying PAHs with a considerably lower EW. \cite{MD2009} argued that an additional power-law-type source of hot dust emission was needed to partially enhance the MIR continuum of these galaxies. They suggested that this would arise either from an AGN torus or from optically thick dust emission around star-forming regions, in agreement with the scenario we propose for the compact nuclear sources in the previous Sections. Interestingly, the SMG composite spectrum is more similar to the averaged nuclear spectrum of (U)LIRGs belonging to the PAH-dominated FEE$_\lambda$ galaxy type (see Stierwalt et al. 2011, in prep.). We also note that while the \NeII\ emission line is clearly visible in all of our averaged spectra as well as in the high-$z$ ULIRG spectrum, it is almost absent from that of the SMGs. Nevertheless, we attribute this difference to the limited quality of the composite spectrum of the SMGs, which prevent us from performing meaningful comparisons regarding specific features in the spectra.

Summarizing, the spectra of the extended components of (U)LIRGs are very similar independently of their FEE$_{\lambda}$ type, suggesting that the extended disk emission of all LIRGs \textit{and} ULIRGs could be dominated by PAHs excited principally by non-ionizing stars and quiescent star-forming regions. The fact that recent \textit{Herschel} observations imply that the galactic disks are responsible for the bulk of the diffuse cold dust emission peaking at $100-500\,\micron$ (\citealt{Bendo2010}) leads us to speculate whether this would also hold in (U)LIRGs. Within this picture, their outer disks, traced by the PAH spatial profiles, would be responsible for the cold FIR emission. Meanwhile, the inner nuclei of the most compact sources, where most of their energy is produced, would dominate the MIR continuum and also contribute to show warmer FIR colors than galaxies without compact cores. Due to their distance and small physical size this can not be verified directly by \textit{Herschel} or even future missions, such as SPICA. However, modeling the FIR line emission (e.g., \citealt{vdW2010}; \citealt{GA2010}) will enable us to probe the excitation and density of gas and thus provide further constrains on its spatial distribution.

\section{Conclusions}\label{s:summary}

We analyzed the spatial profiles of low spectral resolution $5-14\,\micron$ \textit{Spitzer}/IRS spectra of the GOALS galaxy sample and quantified the spatial extent of their MIR emission, FEE$_\lambda$. Our work indicates that:

\begin{itemize}

\item The \NeII\ emission is as compact as the MIR continuum, but the PAHs are more extended in many galaxies. This is in agreement with studies showing that the \NeII\ and MIR continuum emissions trace the ionizing stars (i.e., clumpy, current star formation), while the PAH emission is more representative of older, colder, and diffuse stellar populations.

\item The \PAHa\ emission is more extended than that of the 6.2 and \PAHcs, which is consistent with the formers being enhanced in a more ionized medium while the \PAHa\ emission is similar for neutral and ionized molecules.

%\item For silicate-dominated FEE$_\lambda$ type galaxies, the log(FEE$_{11.3\mu m}$/FEE$_{13.2\mu m}$) ratio depends on the differential extinction between the unresolved and extended emission components of galaxies, $\Delta$\SSi, which is in turn related with the AGN contribution to their MIR emission. These sources show high log(FEE$_{11.3\mu m}$/FEE$_{13.2\mu m}$) ratios due to the absorption of the \PAHa\ emission in their nuclei. However, this is not the case for the PAH-dominated FEE$_\lambda$ type galaxies which present a rather constant $\Delta\,\SSi\,\sim\,0$ independently of the MIR AGN fraction, but still can have large log(FEE$_{11.3\mu m}$/FEE$_{13.2\mu m}$) ratios.

\item On average, the MIR continuum becomes more compact than the PAH emission as the MIR is increasingly dominated by an AGN. That is, the presence of an AGN modifies the spatial extent of the hot dust continuum but has a negligible effect on the overall, kpc-scale distribution of the PAH emission. None of the AGN-dominated (U)LIRGs shows a constant/featureless FEE$_\lambda$ type, implying that the AGN leaves a ``footprint'' in the FEE$_\lambda$ function of a galaxy. Therefore, the MIR emission of the galaxies showing a constant FEE$_\lambda$ will not be dominated by an AGN or, in general, show compact MIR continuum emission.

\item The PAH-to-continuum FEE ratio of (U)LIRGs increases as the global dust temperature, traced by their \textit{IRAS} log($f_{60\,\mu m}/f_{100\,\mu m}$) FIR color, becomes warmer. This can be attributed to the presence of a compact, powerful (circum-)nuclear starburst in these galaxies which would increase the temperature of the dust, and also the ratio between the volume of the shells of the photo-dissociation regions --from where PAH emission arises-- and the volume of the hot dust emitting region. ``Normal'' galaxies do not show enhanced PAH-to-continuum FEE ratios but a rather constant value. On the other hand, in (U)LIRGs with large PAH-to-continuum FEE ratios the MIR continuum is more compact than the PAH emission, which is still as extended as in the ``normal'' population since it arises mostly from a diffuse component (their disks). Nonetheless, in terms of the relative spatial distribution of the PAH and continuum emissions, this effect is indistinguishable from that caused by an AGN. Moreover, both processes, \textit{compact} star formation and AGN activity, coexist in 27\% of our (U)LIRGs sample.

\item The intensity of the spectral features and continuum emission of the extended emission component of all (U)LIRGs are very similar, indicating that the properties of the star formation in the external parts of galaxies (disks; $d\,\gtrsim\,1.5\,$kpc) are uniform. Moreover, they are similar to local starburst galaxies as well as to $z\,\sim\,2$ ULIRGs (selected by their \textit{Spitzer}/IRAC and MIPS colors), but different from sub-millimeter-selected high-redshift galaxies. On the other hand, the MIR spectra of the unresolved nuclear component of galaxies vary widely.

\item These results imply that, on average, the diversity seen in the properties of global/integrated MIR spectra of local ULIRGs as well as LIRGs arise only from the processes that are taking place in their cores. The extended star formation appears unaffected by the nuclear activity and it is similar to the more quiescent mode of star formation found in lower IR luminosity systems. We speculate that the outer disks/regions of (U)LIRGs, traced by their PAH spatial profiles, are responsible for most of the cold FIR emission as seen in galaxies with lower IR luminosities, while the nuclei of the most compact (U)LIRGs, as stated in \cite{DS2010b}, would dominate the MIR continuum and contribute to show overall warmer FIR colors.\\

\end{itemize}

%\section*{Acknowledgments}
\acknowledgments

We thank the referee for her/his useful comments that helped to improved the paper. TDS would like to thank D. Elbaz, E. Le Floc'h, V. Lebouteiller, E. Daddi, and G. Magdis for stimulating discussions, as well as all colleagues at CEA/Saclay (France), where part of this work was done, for their hospitality. TDS and VC would like to acknowledge partial support from the EU ToK grant 39965 and FP7-REGPOT 206469. This research has made use of the NASA/IPAC Extragalactic Database (NED), which is operated by the Jet Propulsion Laboratory, California Institute of Technology, under contract with the National Aeronautics and Space Administration, and of NASA's Astrophysics Data System (ADS) abstract service.\\

%\bibliographystyle{../../apj}%este estilo nombra en la lista de referencia solo a los tres primeros  autores
%\bibliography{../../bib}{}
%%\bibliographystyle{/Users/tanio/mypapers/apj}%este estilo nombra en la lista de referencia solo a los tres primeros  autores
%%\bibliography{/Users/tanio/mypapers/bib}{}

\begin{thebibliography}{52}
\expandafter\ifx\csname natexlab\endcsname\relax\def\natexlab#1{#1}\fi

\bibitem[{{Allamandola} {et~al.}(1999){Allamandola}, {Hudgins}, \&
  {Sandford}}]{Allamandola1999}
{Allamandola}, L.~J., {Hudgins}, D.~M., \& {Sandford}, S.~A. 1999, \apjl, 511,
  L115

\bibitem[{{Alonso-Herrero} {et~al.}(2002){Alonso-Herrero}, {Rieke}, {Rieke}, \&
  {Scoville}}]{AAH2002}
{Alonso-Herrero}, A., {Rieke}, G.~H., {Rieke}, M.~J., \& {Scoville}, N.~Z.
  2002, \aj, 124, 166

\bibitem[{{Armus} {et~al.}(2007){Armus}, {Charmandaris}, {Bernard-Salas},
  {Spoon}, {Marshall}, {Higdon}, {Desai}, {Teplitz}, {Hao}, {Devost}, {Brandl},
  {Wu}, {Sloan}, {Soifer}, {Houck}, \& {Herter}}]{Armus2007}
{Armus}, L., {et~al.} 2007, \apj, 656, 148

\bibitem[{{Armus} {et~al.}(2009){Armus}, {Mazzarella}, {Evans}, {Surace},
  {Sanders}, {Iwasawa}, {Frayer}, {Howell}, {Chan}, {Petric}, {Vavilkin},
  {Kim}, {Haan}, {Inami}, {Murphy}, {Appleton}, {Barnes}, {Bothun}, {Bridge},
  {Charmandaris}, {Jensen}, {Kewley}, {Lord}, {Madore}, {Marshall},
  {Melbourne}, {Rich}, {Satyapal}, {Schulz}, {Spoon}, {Sturm}, {U}, {Veilleux},
  \& {Xu}}]{Armus2009}
---. 2009, \pasp, 121, 559

\bibitem[{{Bendo} {et~al.}(2010){Bendo}, {Wilson}, {Pohlen}, {Sauvage}, {Auld},
  {Baes}, {Barlow}, {Bock}, {Boselli}, {Bradford}, {Buat}, {Castro-Rodriguez},
  {Chanial}, {Charlot}, {Ciesla}, {Clements}, {Cooray}, {Cormier}, {Cortese},
  {Davies}, {Dwek}, {Eales}, {Elbaz}, {Galametz}, {Galliano}, {Gear}, {Glenn},
  {Gomez}, {Griffin}, {Hony}, {Isaak}, {Levenson}, {Lu}, {Madden},
  {O'Halloran}, {Okumura}, {Oliver}, {Page}, {Panuzzo}, {Papageorgiou},
  {Parkin}, {Perez-Fournon}, {Rangwala}, {Rigby}, {Roussel}, {Rykala},
  {Sacchi}, {Schulz}, {Schirm}, {Smith}, {Spinoglio}, {Stevens}, {Sundar},
  {Symeonidis}, {Trichas}, {Vaccari}, {Vigroux}, {Wozniak}, {Wright}, \&
  {Zeilinger}}]{Bendo2010}
{Bendo}, G.~J., {et~al.} 2010, \aap, 518, L65+

\bibitem[{{Brandl} {et~al.}(2006){Brandl}, {Bernard-Salas}, {Spoon}, {Devost},
  {Sloan}, {Guilles}, {Wu}, {Houck}, {Weedman}, {Armus}, {Appleton}, {Soifer},
  {Charmandaris}, {Hao}, {Higdon}, \& {Herter}}]{Brandl2006}
{Brandl}, B.~R., {et~al.} 2006, \apj, 653, 1129

\bibitem[{{Calzetti} {et~al.}(2007){Calzetti}, {Kennicutt}, {Engelbracht},
  {Leitherer}, {Draine}, {Kewley}, {Moustakas}, {Sosey}, {Dale}, {Gordon},
  {Helou}, {Hollenbach}, {Armus}, {Bendo}, {Bot}, {Buckalew}, {Jarrett}, {Li},
  {Meyer}, {Murphy}, {Prescott}, {Regan}, {Rieke}, {Roussel}, {Sheth}, {Smith},
  {Thornley}, \& {Walter}}]{Calzetti2007}
{Calzetti}, D., {et~al.} 2007, \apj, 666, 870

\bibitem[{{Caputi} {et~al.}(2007){Caputi}, {Lagache}, {Yan}, {Dole},
  {Bavouzet}, {Le Floc'h}, {Choi}, {Helou}, \& {Reddy}}]{Caputi2007}
{Caputi}, K.~I., {et~al.} 2007, \apj, 660, 97

\bibitem[{{Charmandaris} {et~al.}(2004){Charmandaris}, {Le Floc'h}, \&
  {Mirabel}}]{Charmandaris2004}
{Charmandaris}, V., {Le Floc'h}, E., \& {Mirabel}, I.~F. 2004, \apjl, 600, L15

\bibitem[{{Charmandaris} {et~al.}(2002){Charmandaris}, {Laurent}, {Le Floc'h},
  {Mirabel}, {Sauvage}, {Madden}, {Gallais}, {Vigroux}, \&
  {Cesarsky}}]{Charmandaris2002}
{Charmandaris}, V., {et~al.} 2002, \aap, 391, 429

\bibitem[{{D{\'{\i}}az-Santos}
  {et~al.}(2010{\natexlab{a}}){D{\'{\i}}az-Santos}, {Alonso-Herrero}, {Colina},
  {Packham}, {Levenson}, {Pereira-Santaella}, {Roche}, \& {Telesco}}]{DS2010a}
{D{\'{\i}}az-Santos}, T., {Alonso-Herrero}, A., {Colina}, L., {Packham}, C.,
  {Levenson}, N.~A., {Pereira-Santaella}, M., {Roche}, P.~F., \& {Telesco},
  C.~M. 2010{\natexlab{a}}, \apj, 711, 328

\bibitem[{{D{\'{\i}}az-Santos} {et~al.}(2008){D{\'{\i}}az-Santos},
  {Alonso-Herrero}, {Colina}, {Packham}, {Radomski}, \& {Telesco}}]{DS2008}
{D{\'{\i}}az-Santos}, T., {Alonso-Herrero}, A., {Colina}, L., {Packham}, C.,
  {Radomski}, J.~T., \& {Telesco}, C.~M. 2008, \apj, 685, 211

\bibitem[{{D{\'{\i}}az-Santos}
  {et~al.}(2010{\natexlab{b}}){D{\'{\i}}az-Santos}, {Charmandaris}, {Armus},
  {Petric}, {Howell}, {Murphy}, {Mazzarella}, {Veilleux}, {Bothun}, {Inami},
  {Appleton}, {Evans}, {Haan}, {Marshall}, {Sanders}, {Stierwalt}, \&
  {Surace}}]{DS2010b}
{D{\'{\i}}az-Santos}, T., {et~al.} 2010{\natexlab{b}}, \apj, 723, 993

\bibitem[{{Draine} \& {Li}(2001)}]{Draine2001}
{Draine}, B.~T., \& {Li}, A. 2001, \apj, 551, 807

\bibitem[{{Farrah} {et~al.}(2008){Farrah}, {Lonsdale}, {Weedman}, {Spoon},
  {Rowan-Robinson}, {Polletta}, {Oliver}, {Houck}, \& {Smith}}]{Farrah2008}
{Farrah}, D., {et~al.} 2008, \apj, 677, 957

\bibitem[{{Galliano} {et~al.}(2008){Galliano}, {Madden}, {Tielens}, {Peeters},
  \& {Jones}}]{Galliano2008}
{Galliano}, F., {Madden}, S.~C., {Tielens}, A.~G.~G.~M., {Peeters}, E., \&
  {Jones}, A.~P. 2008, \apj, 679, 310

\bibitem[{{Garc{\'{\i}}a-Mar{\'{\i}}n}
  {et~al.}(2009){Garc{\'{\i}}a-Mar{\'{\i}}n}, {Colina}, \& {Arribas}}]{GM2009}
{Garc{\'{\i}}a-Mar{\'{\i}}n}, M., {Colina}, L., \& {Arribas}, S. 2009, \aap,
  505, 1017

\bibitem[{{Genzel} {et~al.}(1998){Genzel}, {Lutz}, {Sturm}, {Egami}, {Kunze},
  {Moorwood}, {Rigopoulou}, {Spoon}, {Sternberg}, {Tacconi-Garman}, {Tacconi},
  \& {Thatte}}]{Genzel1998}
{Genzel}, R., {et~al.} 1998, \apj, 498, 579

\bibitem[{{Gonz{\'a}lez-Alfonso} {et~al.}(2010){Gonz{\'a}lez-Alfonso},
  {Fischer}, {Isaak}, {Rykala}, {Savini}, {Spaans}, {van der Werf},
  {Meijerink}, {Israel}, {Loenen}, {Vlahakis}, {Smith}, {Charmandaris},
  {Aalto}, {Henkel}, {Wei{\ss}}, {Walter}, {Greve}, {Mart{\'{\i}}n-Pintado},
  {Naylor}, {Spinoglio}, {Veilleux}, {Harris}, {Armus}, {Lord}, {Mazzarella},
  {Xilouris}, {Sanders}, {Dasyra}, {Wiedner}, {Kramer}, {Papadopoulos},
  {Stacey}, {Evans}, \& {Gao}}]{GA2010}
{Gonz{\'a}lez-Alfonso}, E., {et~al.} 2010, \aap, 518, L43+

\bibitem[{{Haan} {et~al.}(2011){Haan}, {Surace}, {Armus}, {Evans}, {Howell},
  {Mazzarella}, {Kim}, {Vavilkin}, {Inami}, {Sanders}, {Petric}, {Bridge},
  {Melbourne}, {Charmandaris}, {Diaz-Santos}, {Murphy}, {U}, {Stierwalt}, \&
  {Marshall}}]{Haan2011}
{Haan}, S., {et~al.} 2011, \aj, 141, 100

\bibitem[{{Howell} {et~al.}(2010){Howell}, {Armus}, {Mazzarella}, {Evans},
  {Surace}, {Sanders}, {Petric}, {Appleton}, {Bothun}, {Bridge}, {Chan},
  {Charmandaris}, {Frayer}, {Haan}, {Inami}, {Kim}, {Lord}, {Madore},
  {Melbourne}, {Schulz}, {U}, {Vavilkin}, {Veilleux}, \& {Xu}}]{Howell2010}
{Howell}, J.~H., {et~al.} 2010, \apj, 715, 572

\bibitem[{{Imanishi}(2009)}]{Imanishi2009}
{Imanishi}, M. 2009, \apj, 694, 751

\bibitem[{{Imanishi} {et~al.}(2007){Imanishi}, {Dudley}, {Maiolino}, {Maloney},
  {Nakagawa}, \& {Risaliti}}]{Imanishi2007}
{Imanishi}, M., {Dudley}, C.~C., {Maiolino}, R., {Maloney}, P.~R., {Nakagawa},
  T., \& {Risaliti}, G. 2007, \apjs, 171, 72

\bibitem[{{Imanishi} {et~al.}(2010){Imanishi}, {Maiolino}, \&
  {Nakagawa}}]{Imanishi2010}
{Imanishi}, M., {Maiolino}, R., \& {Nakagawa}, T. 2010, \apj, 709, 801

\bibitem[{{Kramer} {et~al.}(2010){Kramer}, {Buchbender}, {Xilouris}, {Boquien},
  {Braine}, {Calzetti}, {Lord}, {Mookerjea}, {Quintana-Lacaci}, {Rela{\~n}o},
  {Stacey}, {Tabatabaei}, {Verley}, {Aalto}, {Akras}, {Albrecht}, {Anderl},
  {Beck}, {Bertoldi}, {Combes}, {Dumke}, {Garcia-Burillo}, {Gonzalez},
  {Gratier}, {G{\"u}sten}, {Henkel}, {Israel}, {Koribalski}, {Lundgren},
  {Martin-Pintado}, {R{\"o}llig}, {Rosolowsky}, {Schuster}, {Sheth}, {Sievers},
  {Stutzki}, {Tilanus}, {van der Tak}, {van der Werf}, \&
  {Wiedner}}]{Kramer2010}
{Kramer}, C., {et~al.} 2010, \aap, 518, L67+

\bibitem[{{Laurent} {et~al.}(2000){Laurent}, {Mirabel}, {Charmandaris},
  {Gallais}, {Madden}, {Sauvage}, {Vigroux}, \& {Cesarsky}}]{Laurent2000}
{Laurent}, O., {Mirabel}, I.~F., {Charmandaris}, V., {Gallais}, P., {Madden},
  S.~C., {Sauvage}, M., {Vigroux}, L., \& {Cesarsky}, C. 2000, \aap, 359, 887

\bibitem[{{Le Floc'h} {et~al.}(2005){Le Floc'h}, {Papovich}, {Dole}, {Bell},
  {Lagache}, {Rieke}, {Egami}, {P{\'e}rez-Gonz{\'a}lez}, {Alonso-Herrero},
  {Rieke}, {Blaylock}, {Engelbracht}, {Gordon}, {Hines}, {Misselt}, {Morrison},
  \& {Mould}}]{LeFloch2005}
{Le Floc'h}, E., {et~al.} 2005, \apj, 632, 169

\bibitem[{{Lutz} {et~al.}(1998b){Lutz}, {Spoon}, {Rigopoulou}, {Moorwood}, \&
  {Genzel}}]{Lutz1998b}
{Lutz}, D., {Spoon}, H.~W.~W., {Rigopoulou}, D., {Moorwood}, A.~F.~M., \&
  {Genzel}, R. 1998b, \apjl, 505, L103

\bibitem[{{Magdis} {et~al.}(2010){Magdis}, {Elbaz}, {Hwang}, {Amblard},
  {Arumugam}, {Aussel}, {Blain}, {Bock}, {Boselli}, {Buat},
  {Castro-Rodr{\'{\i}}guez}, {Cava}, {Chanial}, {Clements}, {Conley},
  {Conversi}, {Cooray}, {Dowell}, {Dwek}, {Eales}, {Farrah}, {Franceschini},
  {Glenn}, {Griffin}, {Halpern}, {Hatziminaoglou}, {Huang}, {Ibar}, {Isaak},
  {Le Floc'h}, {Lagache}, {Levenson}, {Lonsdale}, {Lu}, {Madden}, {Maffei},
  {Mainetti}, {Marchetti}, {Morrison}, {Nguyen}, {O'Halloran}, {Oliver},
  {Omont}, {Owen}, {Page}, {Pannella}, {Panuzzo}, {Papageorgiou}, {Pearson},
  {P{\'e}rez-Fournon}, {Pohlen}, {Rigopoulou}, {Rizzo}, {Roseboom},
  {Rowan-Robinson}, {Schulz}, {Scott}, {Seymour}, {Shupe}, {Smith}, {Stevens},
  {Strazzullo}, {Symeonidis}, {Trichas}, {Tugwell}, {Vaccari}, {Valtchanov},
  {Vigroux}, {Wang}, {Wright}, {Xu}, \& {Zemcov}}]{Magdis2010}
{Magdis}, G.~E., {et~al.} 2010, \mnras, 409, 22

\bibitem[{{Magnelli} {et~al.}(2011){Magnelli}, {Elbaz}, {Chary}, {Dickinson},
  {Le Borgne}, {Frayer}, \& {Willmer}}]{Magnelli2011}
{Magnelli}, B., {Elbaz}, D., {Chary}, R.~R., {Dickinson}, M., {Le Borgne}, D.,
  {Frayer}, D.~T., \& {Willmer}, C.~N.~A. 2011, \aap, 528, A35+

\bibitem[{{Men{\'e}ndez-Delmestre} {et~al.}(2009){Men{\'e}ndez-Delmestre},
  {Blain}, {Smail}, {Alexander}, {Chapman}, {Armus}, {Frayer}, {Ivison}, \&
  {Teplitz}}]{MD2009}
{Men{\'e}ndez-Delmestre}, K., {et~al.} 2009, \apj, 699, 667

\bibitem[{{Murphy} {et~al.}(2011){Murphy}, {Chary}, {Dickinson}, {Pope},
  {Frayer}, \& {Lin}}]{Murphy2011}
{Murphy}, E.~J., {Chary}, R., {Dickinson}, M., {Pope}, A., {Frayer}, D.~T., \&
  {Lin}, L. 2011, ArXiv e-prints

\bibitem[{{Netzer} {et~al.}(2007){Netzer}, {Lutz}, {Schweitzer}, {Contursi},
  {Sturm}, {Tacconi}, {Veilleux}, {Kim}, {Rupke}, {Baker}, {Dasyra},
  {Mazzarella}, \& {Lord}}]{Netzer2007}
{Netzer}, H., {et~al.} 2007, \apj, 666, 806

\bibitem[{{Pereira-Santaella} {et~al.}(2010){Pereira-Santaella},
  {Alonso-Herrero}, {Rieke}, {Colina}, {D{\'{\i}}az-Santos}, \& {et
  al}}]{PS2010}
{Pereira-Santaella}, M., {Alonso-Herrero}, A., {Rieke}, G.~H., {Colina}, L.,
  {D{\'{\i}}az-Santos}, T., \& {et al}. 2010, \apj, in press

\bibitem[{{P{\'e}rez-Gonz{\'a}lez} {et~al.}(2005){P{\'e}rez-Gonz{\'a}lez},
  {Rieke}, {Egami}, {Alonso-Herrero}, {Dole}, {Papovich}, {Blaylock}, {Jones},
  {Rieke}, {Rigby}, {Barmby}, {Fazio}, {Huang}, \& {Martin}}]{PG2005}
{P{\'e}rez-Gonz{\'a}lez}, P.~G., {et~al.} 2005, \apj, 630, 82

\bibitem[{{Petric} {et~al.}(2011){Petric}, {Armus}, {Howell}, {Chan},
  {Mazzarella}, {Evans}, {Surace}, {Sanders}, {Appleton}, {Charmandaris},
  {D{\'{\i}}az-Santos}, {Frayer}, {Haan}, {Inami}, {Iwasawa}, {Kim}, {Madore},
  {Marshall}, {Spoon}, {Stierwalt}, {Sturm}, {U}, {Vavilkin}, \&
  {Veilleux}}]{Petric2011}
{Petric}, A.~O., {et~al.} 2011, \apj, 730, 28

\bibitem[{{Sales} {et~al.}(2010){Sales}, {Pastoriza}, \& {Riffel}}]{Sales2010}
{Sales}, D.~A., {Pastoriza}, M.~G., \& {Riffel}, R. 2010, \apj, 725, 605

\bibitem[{{Sanders} {et~al.}(2003){Sanders}, {Mazzarella}, {Kim}, {Surace}, \&
  {Soifer}}]{Sanders2003}
{Sanders}, D.~B., {Mazzarella}, J.~M., {Kim}, D.-C., {Surace}, J.~A., \&
  {Soifer}, B.~T. 2003, \aj, 126, 1607

\bibitem[{{Sanders} \& {Mirabel}(1996)}]{Sanders1996}
{Sanders}, D.~B., \& {Mirabel}, I.~F. 1996, \araa, 34, 749

\bibitem[{{Sauvage} \& {Thuan}(1992)}]{Sauvage1992}
{Sauvage}, M., \& {Thuan}, T.~X. 1992, \apjl, 396, L69

\bibitem[{{Sauvage} {et~al.}(2010){Sauvage}, {Sacchi}, {Bendo}, {Boselli},
  {Pohlen}, {Wilson}, {Auld}, {Baes}, {Barlow}, {Bock}, {Bradford}, {Buat},
  {Castro-Rodriguez}, {Chanial}, {Charlot}, {Ciesla}, {Clements}, {Cooray},
  {Cormier}, {Cortese}, {Davies}, {Dwek}, {Eales}, {Elbaz}, {Galametz},
  {Galliano}, {Gear}, {Glenn}, {Gomez}, {Griffin}, {Hony}, {Isaak}, {Levenson},
  {Lu}, {Madden}, {O'Halloran}, {Okumura}, {Oliver}, {Page}, {Panuzzo},
  {Papageorgiou}, {Parkin}, {Perez-Fournon}, {Rangwala}, {Rigby}, {Roussel},
  {Rykala}, {Schulz}, {Schirm}, {Smith}, {Spinoglio}, {Stevens}, {Srinivasan},
  {Symeonidis}, {Trichas}, {Vaccari}, {Vigroux}, {Wozniak}, {Wright}, \&
  {Zeilinger}}]{Sauvage2010}
{Sauvage}, M., {et~al.} 2010, \aap, 518, L64+

\bibitem[{{Schweitzer} {et~al.}(2006){Schweitzer}, {Lutz}, {Sturm}, {Contursi},
  {Tacconi}, {Lehnert}, {Dasyra}, {Genzel}, {Veilleux}, {Rupke}, {Kim},
  {Baker}, {Netzer}, {Sternberg}, {Mazzarella}, \& {Lord}}]{Schweitzer2006}
{Schweitzer}, M., {et~al.} 2006, \apj, 649, 79

\bibitem[{{Smith} {et~al.}(2004){Smith}, {Dale}, {Armus}, {Draine},
  {Hollenbach}, {Roussel}, {Helou}, {Kennicutt}, {Li}, {Bendo}, {Calzetti},
  {Engelbracht}, {Gordon}, {Jarrett}, {Kewley}, {Leitherer}, {Malhotra},
  {Meyer}, {Murphy}, {Regan}, {Rieke}, {Rieke}, {Thornley}, {Walter}, \&
  {Wolfire}}]{Smith2004}
{Smith}, J.~D.~T., {et~al.} 2004, \apjs, 154, 199

\bibitem[{{Smith} {et~al.}(2007){Smith}, {Draine}, {Dale}, {Moustakas},
  {Kennicutt}, {Helou}, {Armus}, {Roussel}, {Sheth}, {Bendo}, {Buckalew},
  {Calzetti}, {Engelbracht}, {Gordon}, {Hollenbach}, {Li}, {Malhotra},
  {Murphy}, \& {Walter}}]{Smith2007}
---. 2007, \apj, 656, 770

\bibitem[{{Soifer} {et~al.}(2003){Soifer}, {Bock}, {Marsh}, {Neugebauer},
  {Matthews}, {Egami}, \& {Armus}}]{Soifer2003}
{Soifer}, B.~T., {Bock}, J.~J., {Marsh}, K., {Neugebauer}, G., {Matthews}, K.,
  {Egami}, E., \& {Armus}, L. 2003, \aj, 126, 143

\bibitem[{{Soifer} {et~al.}(2002){Soifer}, {Neugebauer}, {Matthews}, {Egami},
  \& {Weinberger}}]{Soifer2002}
{Soifer}, B.~T., {Neugebauer}, G., {Matthews}, K., {Egami}, E., \&
  {Weinberger}, A.~J. 2002, \aj, 124, 2980

\bibitem[{{Spoon} {et~al.}(2007){Spoon}, {Marshall}, {Houck}, {Elitzur}, {Hao},
  {Armus}, {Brandl}, \& {Charmandaris}}]{Spoon2007}
{Spoon}, H.~W.~W., {Marshall}, J.~A., {Houck}, J.~R., {Elitzur}, M., {Hao}, L.,
  {Armus}, L., {Brandl}, B.~R., \& {Charmandaris}, V. 2007, \apjl, 654, L49

\bibitem[{{van der Werf} {et~al.}(2010){van der Werf}, {Isaak}, {Meijerink},
  {Spaans}, {Rykala}, {Fulton}, {Loenen}, {Walter}, {Wei{\ss}}, {Armus},
  {Fischer}, {Israel}, {Harris}, {Veilleux}, {Henkel}, {Savini}, {Lord},
  {Smith}, {Gonz{\'a}lez-Alfonso}, {Naylor}, {Aalto}, {Charmandaris}, {Dasyra},
  {Evans}, {Gao}, {Greve}, {G{\"u}sten}, {Kramer}, {Mart{\'{\i}}n-Pintado},
  {Mazzarella}, {Papadopoulos}, {Sanders}, {Spinoglio}, {Stacey}, {Vlahakis},
  {Wiedner}, \& {Xilouris}}]{vdW2010}
{van der Werf}, P.~P., {et~al.} 2010, \aap, 518, L42+

\bibitem[{{Veilleux} {et~al.}(2002){Veilleux}, {Kim}, \&
  {Sanders}}]{Veilleux2002}
{Veilleux}, S., {Kim}, D., \& {Sanders}, D.~B. 2002, \apjs, 143, 315

\bibitem[{{Veilleux} {et~al.}(2009){Veilleux}, {Rupke}, {Kim}, {Genzel},
  {Sturm}, {Lutz}, {Contursi}, {Schweitzer}, {Tacconi}, {Netzer}, {Sternberg},
  {Mihos}, {Baker}, {Mazzarella}, {Lord}, {Sanders}, {Stockton}, {Joseph}, \&
  {Barnes}}]{Veilleux2009}
{Veilleux}, S., {et~al.} 2009, \apjs, 182, 628

\bibitem[{{Walterbos} \& {Greenawalt}(1996)}]{Walterbos1996}
{Walterbos}, R.~A.~M., \& {Greenawalt}, B. 1996, \apj, 460, 696

\bibitem[{{Wu} {et~al.}(2009){Wu}, {Charmandaris}, {Huang}, {Spinoglio}, \&
  {Tommasin}}]{Wu2009}
{Wu}, Y., {Charmandaris}, V., {Huang}, J., {Spinoglio}, L., \& {Tommasin}, S.
  2009, \apj, 701, 658

\end{thebibliography}

\clearpage
%\LongTables

%\begin{landscape}

\setcounter{table}{1}
\begin{deluxetable}{lccccccccc}
\tabletypesize{\scriptsize}
\tablewidth{0pc}
\tablecaption{\scriptsize FEEs of Various MIR Features}
\tablehead{\colhead{Galaxy} & \colhead{R.A.} & \colhead{Declination} & \colhead{FEE$_{6.2\mu m}$} & \colhead{FEE$_{6.7\mu m}$} & \colhead{FEE$_{7.7\mu m}$} & \colhead{FEE$_{9.7\mu m}$} & \colhead{FEE$_{11.3\mu m}$} & \colhead{FEE$_{12.8\mu m}$} & FEE$_{\lambda}$ \\
\colhead{name} & \colhead{(J2000)} & \colhead{(J2000)} & \colhead{(PAH)} & \colhead{(Cont.)} & \colhead{(PAH)} & \colhead{(PAH)} & \colhead{(Si abs.)} & \colhead{(\NeIIno)} &  \colhead{Type} \\
\colhead{(1)} & \colhead{(2)} & \colhead{(3)} & \colhead{(4)} & \colhead{(5)} & \colhead{(6)} & \colhead{(7)} & \colhead{(8)} & \colhead{(9)} & \colhead{(10)}}
\startdata
             NGC0023 & 00h 09m 53.35s & +25\deg\ 55m 27.8s & 0.66$\,\pm\,$0.02 & 0.65$\,\pm\,$0.01 & 0.68$\,\pm\,$0.02 & 0.66$\,\pm\,$0.01 & 0.67$\,\pm\,$0.03 & 0.68$\,\pm\,$0.02 & 1 \\
       MCG-02-01-051 & 00h 18m 50.90s & $-$10\deg\ 22m 36.8s & 0.37$\,\pm\,$0.04 & 0.38$\,\pm\,$0.02 & 0.36$\,\pm\,$0.05 & 0.40$\,\pm\,$0.02 & 0.29$\,\pm\,$0.06 & 0.43$\,\pm\,$0.03 & 1 \\
        ESO350-IG038 & 00h 36m 52.49s & $-$33\deg\ 33m 17.3s & 0.08$\,\pm\,$0.04 & 0.12$\,\pm\,$0.02 & 0.13$\,\pm\,$0.04 & 0.17$\,\pm\,$0.04 & 0.10$\,\pm\,$0.06 & 0.17$\,\pm\,$0.04 & 2 \\
             NGC0232 & 00h 42m 45.84s & $-$23\deg\ 33m 41.0s & 0.36$\,\pm\,$0.07 & 0.40$\,\pm\,$0.01 & 0.49$\,\pm\,$0.04 & 0.41$\,\pm\,$0.01 & 0.38$\,\pm\,$0.05 & 0.47$\,\pm\,$0.04 & 3 \\
             NGC0232 & 00h 42m 52.82s & $-$23\deg\ 32m 28.0s & 0.28$\,\pm\,$0.09 & 0.35$\,\pm\,$0.02 & 0.29$\,\pm\,$0.07 & 0.35$\,\pm\,$0.02 & 0.22$\,\pm\,$0.06 & 0.37$\,\pm\,$0.04 & 2 \\
       MCG+12-02-001 & 00h 54m 03.89s & +73\deg\ 05m 06.0s & 0.47$\,\pm\,$0.02 & 0.47$\,\pm\,$0.04 & 0.32$\,\pm\,$0.04 & 0.50$\,\pm\,$0.01 & 0.44$\,\pm\,$0.05 & 0.49$\,\pm\,$0.04 & 2 \\
            NGC0317B & 00h 57m 40.42s & +43\deg\ 47m 32.6s & 0.25$\,\pm\,$0.06 & 0.24$\,\pm\,$0.01 & 0.36$\,\pm\,$0.06 & 0.29$\,\pm\,$0.07 & 0.19$\,\pm\,$0.07 & 0.33$\,\pm\,$0.04 & 3 \\
       MCG-03-04-014 & 01h 10m 08.93s & $-$16\deg\ 51m 10.1s & 0.50$\,\pm\,$0.04 & 0.53$\,\pm\,$0.01 & 0.58$\,\pm\,$0.03 & 0.54$\,\pm\,$0.01 & 0.53$\,\pm\,$0.04 & 0.57$\,\pm\,$0.03 & 1 \\
         ESO244-G012 & 01h 18m 08.23s & $-$44\deg\ 28m 00.5s & 0.16$\,\pm\,$0.05 & 0.18$\,\pm\,$0.02 & 0.20$\,\pm\,$0.03 & 0.20$\,\pm\,$0.01 & 0.12$\,\pm\,$0.05 & 0.20$\,\pm\,$0.04 & 1 \\
         CGCG436-030 & 01h 20m 02.64s & +14\deg\ 21m 42.1s & 0.17$\,\pm\,$0.06 & 0.26$\,\pm\,$0.03 & 0.36$\,\pm\,$0.06 & 0.24$\,\pm\,$0.05 & 0.07$\,\pm\,$0.07 & 0.35$\,\pm\,$0.05 & 3 \\
         ESO353-G020 & 01h 34m 51.26s & $-$36\deg\ 08m 14.3s & 0.36$\,\pm\,$0.04 & 0.41$\,\pm\,$0.02 & 0.53$\,\pm\,$0.04 & 0.44$\,\pm\,$0.01 & 0.44$\,\pm\,$0.05 & 0.50$\,\pm\,$0.04 & 3 \\
         ESO297-G011 & 01h 36m 23.40s & $-$37\deg\ 19m 18.1s & 0.52$\,\pm\,$0.04 & 0.53$\,\pm\,$0.02 & 0.51$\,\pm\,$0.04 & 0.56$\,\pm\,$0.01 & 0.53$\,\pm\,$0.05 & 0.58$\,\pm\,$0.04 & 1 \\
         ESO297-G011 & 01h 36m 24.14s & $-$37\deg\ 20m 25.8s & 0.16$\,\pm\,$0.07 & 0.16$\,\pm\,$0.01 & 0.21$\,\pm\,$0.07 & 0.19$\,\pm\,$0.03 & 0.09$\,\pm\,$0.06 & 0.21$\,\pm\,$0.05 & 1 \\
     IRASF01364-1042 & 01h 38m 52.80s & $-$10\deg\ 27m 12.2s & 0.09$\,\pm\,$0.10 & 0.08$\,\pm\,$0.04 & 0.24$\,\pm\,$0.06 & 0.16$\,\pm\,$0.10 & 0.06$\,\pm\,$0.08 & 0.23$\,\pm\,$0.07 & 2 \\
     IRASF01417+1651 & 01h 44m 30.55s & +17\deg\ 06m 09.0s & 0.35$\,\pm\,$0.06 & 0.28$\,\pm\,$0.09 & 0.57$\,\pm\,$0.13 & 0.42$\,\pm\,$0.05 & 0.41$\,\pm\,$0.07 & 0.49$\,\pm\,$0.03 & 3 \\
             NGC0695 & 01h 51m 14.35s & +22\deg\ 34m 55.9s & 0.71$\,\pm\,$0.01 & 0.74$\,\pm\,$0.01 & 0.74$\,\pm\,$0.02 & 0.74$\,\pm\,$0.01 & 0.73$\,\pm\,$0.03 & 0.75$\,\pm\,$0.02 & 1 \\
            UGC01385 & 01h 54m 53.83s & +36\deg\ 55m 04.4s & 0.29$\,\pm\,$0.07 & 0.25$\,\pm\,$0.07 & 0.29$\,\pm\,$0.03 & 0.28$\,\pm\,$0.07 & 0.23$\,\pm\,$0.05 & 0.29$\,\pm\,$0.04 & 1 \\
             NGC0838 & 02h 09m 42.82s & $-$10\deg\ 11m 02.0s & 0.11$\,\pm\,$0.05 & 0.19$\,\pm\,$0.01 & 0.30$\,\pm\,$0.05 & 0.19$\,\pm\,$0.02 & 0.11$\,\pm\,$0.06 & 0.31$\,\pm\,$0.04 & 2 \\
             NGC0838 & 02h 09m 38.66s & $-$10\deg\ 08m 47.0s & 0.54$\,\pm\,$0.04 & 0.54$\,\pm\,$0.01 & 0.59$\,\pm\,$0.03 & 0.55$\,\pm\,$0.01 & 0.58$\,\pm\,$0.04 & 0.61$\,\pm\,$0.03 & 1 \\
             NGC0838 & 02h 09m 20.88s & $-$10\deg\ 07m 59.5s & 0.73$\,\pm\,$0.01 & 0.75$\,\pm\,$0.02 & 0.72$\,\pm\,$0.02 & 0.74$\,\pm\,$0.01 & 0.74$\,\pm\,$0.03 & 0.74$\,\pm\,$0.02 & 1 \\
             NGC0828 & 02h 10m 09.53s & +39\deg\ 11m 24.7s & 0.70$\,\pm\,$0.01 & 0.71$\,\pm\,$0.02 & 0.74$\,\pm\,$0.02 & 0.72$\,\pm\,$0.01 & 0.72$\,\pm\,$0.03 & 0.73$\,\pm\,$0.02 & 1 \\
              IC0214 & 02h 14m 05.57s & +05\deg\ 10m 23.9s & 0.72$\,\pm\,$0.02 & 0.72$\,\pm\,$0.01 & 0.72$\,\pm\,$0.02 & 0.72$\,\pm\,$0.01 & 0.72$\,\pm\,$0.03 & 0.73$\,\pm\,$0.03 & 1 \\
             NGC0877 & 02h 17m 59.69s & +14\deg\ 32m 38.0s & 0.80$\,\pm\,$0.05 & 0.79$\,\pm\,$0.06 & 0.79$\,\pm\,$0.07 & 0.76$\,\pm\,$0.04 & 0.80$\,\pm\,$0.06 & 0.80$\,\pm\,$0.06 & 1 \\
             NGC0877 & 02h 17m 53.26s & +14\deg\ 31m 18.5s & 0.41$\,\pm\,$0.04 & 0.39$\,\pm\,$0.03 & 0.58$\,\pm\,$0.04 & 0.34$\,\pm\,$0.05 & 0.33$\,\pm\,$0.05 & 0.50$\,\pm\,$0.03 & 3 \\
       MCG+05-06-036 & 02h 23m 21.98s & +32\deg\ 11m 48.8s & 0.29$\,\pm\,$0.05 & 0.33$\,\pm\,$0.03 & 0.35$\,\pm\,$0.04 & 0.36$\,\pm\,$0.05 & 0.26$\,\pm\,$0.06 & 0.35$\,\pm\,$0.04 & 1 \\
       MCG+05-06-036 & 02h 23m 18.96s & +32\deg\ 11m 18.6s & 0.50$\,\pm\,$0.03 & 0.47$\,\pm\,$0.02 & 0.54$\,\pm\,$0.03 & 0.51$\,\pm\,$0.01 & 0.49$\,\pm\,$0.05 & 0.55$\,\pm\,$0.03 & 1 \\
            UGC01845 & 02h 24m 07.97s & +47\deg\ 58m 12.0s & 0.49$\,\pm\,$0.03 & 0.50$\,\pm\,$0.04 & 0.56$\,\pm\,$0.02 & 0.53$\,\pm\,$0.01 & 0.53$\,\pm\,$0.04 & 0.56$\,\pm\,$0.03 & 1 \\
             NGC0992 & 02h 37m 25.46s & +21\deg\ 06m 02.9s & 0.75$\,\pm\,$0.02 & 0.74$\,\pm\,$0.01 & 0.76$\,\pm\,$0.02 & 0.74$\,\pm\,$0.01 & 0.76$\,\pm\,$0.02 & 0.76$\,\pm\,$0.02 & 1 \\
            UGC02238 & 02h 46m 17.45s & +13\deg\ 05m 44.5s & 0.67$\,\pm\,$0.02 & 0.67$\,\pm\,$0.01 & 0.74$\,\pm\,$0.02 & 0.67$\,\pm\,$0.01 & 0.67$\,\pm\,$0.03 & 0.72$\,\pm\,$0.02 & 3 \\
     IRASF02437+2122 & 02h 46m 39.12s & +21\deg\ 35m 10.3s & 0.14$\,\pm\,$0.05 & 0.11$\,\pm\,$0.02 & 0.37$\,\pm\,$0.11 & 0.11$\,\pm\,$0.02 & 0.04$\,\pm\,$0.06 & 0.15$\,\pm\,$0.07 & 3 \\
            UGC02369 & 02h 54m 01.75s & +14\deg\ 58m 36.5s & 0.49$\,\pm\,$0.04 & 0.52$\,\pm\,$0.02 & 0.37$\,\pm\,$0.04 & 0.53$\,\pm\,$0.01 & 0.50$\,\pm\,$0.04 & 0.53$\,\pm\,$0.03 & 2 \\
            UGC02608 & 03h 15m 01.46s & +42\deg\ 02m 08.5s & 0.17$\,\pm\,$0.03 & 0.21$\,\pm\,$0.05 & 0.26$\,\pm\,$0.03 & 0.28$\,\pm\,$0.05 & 0.22$\,\pm\,$0.05 & 0.29$\,\pm\,$0.04 & 2 \\
            UGC02608 & 03h 15m 14.57s & +41\deg\ 58m 50.2s & 0.41$\,\pm\,$0.01 & 0.55$\,\pm\,$0.18 & 0.47$\,\pm\,$0.01 & 0.57$\,\pm\,$0.07 & 0.70$\,\pm\,$0.06 & 0.72$\,\pm\,$0.04 & 2 \\
             NGC1275 & 03h 19m 48.17s & +41\deg\ 30m 42.1s & 0.09$\,\pm\,$0.07 & 0.14$\,\pm\,$0.01 & 0.04$\,\pm\,$0.04 & 0.14$\,\pm\,$0.06 & 0.01$\,\pm\,$0.01 & 0.02$\,\pm\,$0.06 & 0 \\
     IRASF03217+4022 & 03h 25m 05.38s & +40\deg\ 33m 32.0s & 0.46$\,\pm\,$0.02 & 0.49$\,\pm\,$0.03 & 0.52$\,\pm\,$0.03 & 0.51$\,\pm\,$0.01 & 0.42$\,\pm\,$0.05 & 0.53$\,\pm\,$0.04 & 1 \\
             NGC1365 & 03h 33m 36.41s & $-$36\deg\ 08m 25.8s & 0.47$\,\pm\,$0.01 & 0.70$\,\pm\,$0.03 & 0.30$\,\pm\,$0.10 & 0.74$\,\pm\,$0.01 & 0.65$\,\pm\,$0.03 & 0.67$\,\pm\,$0.02 & 2 \\
     IRASF03359+1523 & 03h 38m 47.06s & +15\deg\ 32m 54.2s & \dots$\,\pm\,$\dots & 0.33$\,\pm\,$0.11 & 0.41$\,\pm\,$0.09 & 0.38$\,\pm\,$0.02 & 0.63$\,\pm\,$0.04 & 0.59$\,\pm\,$0.03 & 4 \\
         CGCG465-012 & 03h 54m 07.68s & +15\deg\ 59m 24.4s & 0.72$\,\pm\,$0.03 & 0.73$\,\pm\,$0.01 & 0.72$\,\pm\,$0.05 & 0.71$\,\pm\,$0.02 & 0.72$\,\pm\,$0.04 & 0.72$\,\pm\,$0.04 & 1 \\
         CGCG465-012 & 03h 54m 15.96s & +15\deg\ 55m 43.3s & 0.58$\,\pm\,$0.03 & 0.59$\,\pm\,$0.01 & 0.61$\,\pm\,$0.03 & 0.59$\,\pm\,$0.01 & 0.57$\,\pm\,$0.04 & 0.64$\,\pm\,$0.03 & 1 \\
      IRAS03582+6012 & 04h 02m 33.00s & +60\deg\ 20m 41.6s & 0.02$\,\pm\,$0.01 & 0.10$\,\pm\,$0.08 & 0.49$\,\pm\,$0.05 & 0.07$\,\pm\,$0.04 & 0.12$\,\pm\,$0.07 & 0.42$\,\pm\,$0.04 & 3 \\
      IRAS03582+6012 & 04h 02m 31.97s & +60\deg\ 20m 38.4s & 0.50$\,\pm\,$0.04 & 0.49$\,\pm\,$0.01 & 0.53$\,\pm\,$0.03 & 0.49$\,\pm\,$0.01 & 0.47$\,\pm\,$0.05 & 0.52$\,\pm\,$0.04 & 1 \\
            UGC02982 & 04h 12m 22.68s & +05\deg\ 32m 49.2s & 0.66$\,\pm\,$0.01 & 0.66$\,\pm\,$0.02 & 0.72$\,\pm\,$0.01 & 0.67$\,\pm\,$0.02 & 0.69$\,\pm\,$0.03 & 0.72$\,\pm\,$0.02 & 1 \\
         ESO420-G013 & 04h 13m 49.70s & $-$32\deg\ 00m 25.2s & 0.25$\,\pm\,$0.04 & 0.32$\,\pm\,$0.05 & 0.23$\,\pm\,$0.07 & 0.38$\,\pm\,$0.01 & 0.29$\,\pm\,$0.05 & 0.38$\,\pm\,$0.04 & 2 \\
             NGC1572 & 04h 22m 42.82s & $-$40\deg\ 36m 03.2s & 0.31$\,\pm\,$0.05 & 0.38$\,\pm\,$0.01 & 0.44$\,\pm\,$0.03 & 0.39$\,\pm\,$0.01 & 0.39$\,\pm\,$0.05 & 0.43$\,\pm\,$0.04 & 1 \\
      IRAS04271+3849 & 04h 30m 33.10s & +38\deg\ 55m 47.6s & 0.32$\,\pm\,$0.05 & 0.37$\,\pm\,$0.01 & 0.46$\,\pm\,$0.04 & 0.39$\,\pm\,$0.01 & 0.35$\,\pm\,$0.05 & 0.42$\,\pm\,$0.05 & 3 \\
             NGC1614 & 04h 33m 59.95s & $-$08\deg\ 34m 46.6s & 0.33$\,\pm\,$0.07 & 0.34$\,\pm\,$0.10 & 0.40$\,\pm\,$0.04 & 0.43$\,\pm\,$0.01 & 0.36$\,\pm\,$0.05 & 0.45$\,\pm\,$0.04 & 1 \\
            UGC03094 & 04h 35m 33.82s & +19\deg\ 10m 18.1s & 0.73$\,\pm\,$0.01 & 0.78$\,\pm\,$0.02 & 0.72$\,\pm\,$0.02 & 0.78$\,\pm\,$0.01 & 0.76$\,\pm\,$0.02 & 0.78$\,\pm\,$0.02 & 2
\enddata
\tablecomments{\footnotesize (1) Galaxy name. Multiple systems are indicated with the same name but providing the right ascension and declination of the individual galaxies. An asterisk next to a name indicates that multiple nuclei, not resolved by \textit{Spitzer}/IRS, are detected using higher angular resolution \textit{HST} NIR continuum imaging \citep{Haan2011}. (2) Right Ascension (J2000). (3) Declination (J2000). (4)$-$(9) Fraction of extended emission (FEE) at: 6.2, 6.7, 7.7, 9.7, 11.3, and 12.8$\,\micron$ respectively. (10) FEE$_\lambda$ function type: 0, FEE$_\lambda\,\sim\,0$; 1, constant/featureless; 2, PAH-dominated; 3, silicate-dominated; 4, undefined.}\label{t:fees}
%\vspace{2cm}
\end{deluxetable}

\clearpage

\setcounter{table}{1}
\begin{deluxetable}{lccccccccc}
\tabletypesize{\scriptsize}
\tablewidth{0pc}
\tablecaption{\scriptsize FEEs of Various MIR Features}
\tablehead{\colhead{Galaxy} & \colhead{R.A.} & \colhead{Declination} & \colhead{FEE$_{6.2\mu m}$} & \colhead{FEE$_{6.7\mu m}$} & \colhead{FEE$_{7.7\mu m}$} & \colhead{FEE$_{9.7\mu m}$} & \colhead{FEE$_{11.3\mu m}$} & \colhead{FEE$_{12.8\mu m}$} & FEE$_{\lambda}$ \\
\colhead{name} & \colhead{(J2000)} & \colhead{(J2000)} & \colhead{(PAH)} & \colhead{(Cont.)} & \colhead{(PAH)} & \colhead{(PAH)} & \colhead{(Si abs.)} & \colhead{(\NeIIno)} &  \colhead{Type} \\
\colhead{(1)} & \colhead{(2)} & \colhead{(3)} & \colhead{(4)} & \colhead{(5)} & \colhead{(6)} & \colhead{(7)} & \colhead{(8)} & \colhead{(9)} & \colhead{(10)}}
\startdata
        ESO203-IG001$^*$ & 04h 46m 49.56s & $-$48\deg\ 33m 30.6s & 0.12$\,\pm\,$0.07 & 0.16$\,\pm\,$0.02 & 0.39$\,\pm\,$0.14 & 0.22$\,\pm\,$0.04 & 0.05$\,\pm\,$0.01 & 0.30$\,\pm\,$0.07 & 3 \\
       MCG-05-12-006 & 04h 52m 04.97s & $-$32\deg\ 59m 26.2s & 0.08$\,\pm\,$0.08 & 0.20$\,\pm\,$0.01 & 0.18$\,\pm\,$0.05 & 0.20$\,\pm\,$0.01 & 0.10$\,\pm\,$0.06 & 0.21$\,\pm\,$0.05 & 1 \\
             NGC1797 & 05h 07m 44.83s & $-$08\deg\ 01m 08.8s & 0.25$\,\pm\,$0.06 & 0.29$\,\pm\,$0.02 & 0.30$\,\pm\,$0.04 & 0.26$\,\pm\,$0.08 & 0.29$\,\pm\,$0.05 & 0.33$\,\pm\,$0.03 & 1 \\
         CGCG468-002 & 05h 08m 19.70s & +17\deg\ 21m 47.9s & 0.14$\,\pm\,$0.09 & 0.24$\,\pm\,$0.02 & 0.10$\,\pm\,$0.06 & 0.26$\,\pm\,$0.03 & 0.14$\,\pm\,$0.06 & 0.20$\,\pm\,$0.04 & 2 \\
         CGCG468-002 & 05h 08m 21.22s & +17\deg\ 22m 08.0s & 0.39$\,\pm\,$0.06 & 0.40$\,\pm\,$0.01 & 0.52$\,\pm\,$0.04 & 0.41$\,\pm\,$0.02 & 0.28$\,\pm\,$0.06 & 0.49$\,\pm\,$0.04 & 3 \\
      IRAS05083+2441 & 05h 11m 25.87s & +24\deg\ 45m 18.4s & 0.39$\,\pm\,$0.06 & 0.39$\,\pm\,$0.03 & 0.45$\,\pm\,$0.04 & 0.42$\,\pm\,$0.01 & 0.37$\,\pm\,$0.05 & 0.44$\,\pm\,$0.04 & 1 \\
     IRASF05081+7936 & 05h 16m 46.39s & +79\deg\ 40m 13.1s & 0.42$\,\pm\,$0.04 & 0.43$\,\pm\,$0.02 & 0.49$\,\pm\,$0.04 & 0.45$\,\pm\,$0.01 & 0.41$\,\pm\,$0.07 & 0.48$\,\pm\,$0.04 & 1 \\
      IRAS05129+5128 & 05h 16m 55.97s & +51\deg\ 31m 57.0s & 0.18$\,\pm\,$0.07 & 0.34$\,\pm\,$0.01 & 0.39$\,\pm\,$0.06 & 0.34$\,\pm\,$0.03 & 0.25$\,\pm\,$0.06 & 0.37$\,\pm\,$0.05 & 1 \\
     IRASF05189-2524 & 05h 21m 01.44s & $-$25\deg\ 21m 46.1s & \dots$\,\pm\,$\dots & 0.17$\,\pm\,$0.04 & 0.07$\,\pm\,$0.01 & 0.18$\,\pm\,$0.04 & \dots$\,\pm\,$\dots & \dots$\,\pm\,$\dots & 0 \\
     IRASF05187-1017 & 05h 21m 06.53s & $-$10\deg\ 14m 46.3s & 0.15$\,\pm\,$0.03 & 0.20$\,\pm\,$0.07 & 0.47$\,\pm\,$0.12 & 0.25$\,\pm\,$0.05 & 0.15$\,\pm\,$0.07 & 0.31$\,\pm\,$0.06 & 3 \\
      IRAS05223+1908 & 05h 25m 16.68s & +19\deg\ 10m 47.6s & 0.05$\,\pm\,$0.06 & 0.04$\,\pm\,$0.01 & 0.09$\,\pm\,$0.09 & 0.10$\,\pm\,$0.05 & 0.02$\,\pm\,$0.01 & 0.03$\,\pm\,$0.01 & 0 \\
       MCG+08-11-002 & 05h 40m 43.70s & +49\deg\ 41m 41.6s & 0.26$\,\pm\,$0.05 & 0.27$\,\pm\,$0.01 & 0.56$\,\pm\,$0.03 & 0.27$\,\pm\,$0.08 & 0.28$\,\pm\,$0.06 & 0.47$\,\pm\,$0.03 & 3 \\
             NGC1961 & 05h 42m 04.56s & +69\deg\ 22m 43.0s & 0.66$\,\pm\,$0.05 & 0.70$\,\pm\,$0.03 & 0.71$\,\pm\,$0.02 & 0.72$\,\pm\,$0.01 & 0.71$\,\pm\,$0.02 & 0.74$\,\pm\,$0.02 & 1 \\
            UGC03351 & 05h 45m 48.02s & +58\deg\ 42m 03.6s & 0.74$\,\pm\,$0.01 & 0.78$\,\pm\,$0.02 & 0.81$\,\pm\,$0.02 & 0.78$\,\pm\,$0.02 & 0.75$\,\pm\,$0.03 & 0.79$\,\pm\,$0.02 & 1 \\
      IRAS05442+1732 & 05h 47m 11.21s & +17\deg\ 33m 46.4s & 0.28$\,\pm\,$0.03 & 0.24$\,\pm\,$0.07 & 0.29$\,\pm\,$0.03 & 0.29$\,\pm\,$0.06 & 0.25$\,\pm\,$0.06 & 0.34$\,\pm\,$0.04 & 1 \\
     IRASF06076-2139 & 06h 09m 45.74s & $-$21\deg\ 40m 24.6s & 0.16$\,\pm\,$0.05 & 0.17$\,\pm\,$0.02 & 0.27$\,\pm\,$0.05 & 0.22$\,\pm\,$0.05 & 0.09$\,\pm\,$0.10 & 0.22$\,\pm\,$0.05 & 3 \\
            UGC03410 & 06h 14m 29.62s & +80\deg\ 26m 59.6s & 0.75$\,\pm\,$0.01 & 0.75$\,\pm\,$0.01 & 0.77$\,\pm\,$0.02 & 0.75$\,\pm\,$0.01 & 0.76$\,\pm\,$0.03 & 0.77$\,\pm\,$0.02 & 1 \\
            UGC03410 & 06h 13m 57.89s & +80\deg\ 28m 34.7s & 0.68$\,\pm\,$0.01 & 0.67$\,\pm\,$0.01 & 0.73$\,\pm\,$0.02 & 0.72$\,\pm\,$0.02 & 0.73$\,\pm\,$0.02 & 0.74$\,\pm\,$0.01 & 1 \\
             NGC2146 & 06h 18m 37.82s & +78\deg\ 21m 24.1s & 0.60$\,\pm\,$0.03 & 0.60$\,\pm\,$0.01 & 0.70$\,\pm\,$0.02 & 0.60$\,\pm\,$0.02 & 0.65$\,\pm\,$0.03 & 0.68$\,\pm\,$0.02 & 3 \\
        ESO255-IG007 & 06h 27m 21.70s & $-$47\deg\ 10m 36.1s & 0.17$\,\pm\,$0.08 & 0.24$\,\pm\,$0.01 & 0.24$\,\pm\,$0.05 & 0.26$\,\pm\,$0.01 & 0.16$\,\pm\,$0.06 & 0.27$\,\pm\,$0.04 & 1 \\
        ESO255-IG007 & 06h 27m 22.56s & $-$47\deg\ 10m 47.3s & 0.54$\,\pm\,$0.07 & 0.49$\,\pm\,$0.02 & 0.48$\,\pm\,$0.04 & 0.55$\,\pm\,$0.01 & 0.54$\,\pm\,$0.04 & 0.56$\,\pm\,$0.04 & 1 \\
        ESO255-IG007 & 06h 27m 23.09s & $-$47\deg\ 11m 02.8s & 0.27$\,\pm\,$0.17 & 0.26$\,\pm\,$0.03 & 0.25$\,\pm\,$0.13 & 0.30$\,\pm\,$0.01 & 0.24$\,\pm\,$0.06 & 0.36$\,\pm\,$0.04 & 1 \\
         ESO557-G002 & 06h 31m 47.21s & $-$17\deg\ 37m 16.7s & 0.37$\,\pm\,$0.21 & 0.26$\,\pm\,$0.20 & 0.58$\,\pm\,$0.03 & 0.48$\,\pm\,$0.04 & 0.43$\,\pm\,$0.05 & 0.56$\,\pm\,$0.03 & 3 \\
            UGC03608 & 06h 57m 34.42s & +46\deg\ 24m 10.4s & 0.57$\,\pm\,$0.05 & 0.56$\,\pm\,$0.04 & 0.43$\,\pm\,$0.04 & 0.54$\,\pm\,$0.06 & 0.37$\,\pm\,$0.05 & 0.54$\,\pm\,$0.02 & 2 \\
     IRASF06592-6313 & 06h 59m 40.25s & $-$63\deg\ 17m 52.4s & 0.10$\,\pm\,$0.04 & 0.17$\,\pm\,$0.06 & 0.20$\,\pm\,$0.02 & 0.17$\,\pm\,$0.07 & 0.11$\,\pm\,$0.05 & 0.18$\,\pm\,$0.04 & 1 \\
     IRASF07027-6011$^*$ & 07h 03m 28.51s & $-$60\deg\ 16m 43.7s & 0.36$\,\pm\,$0.04 & 0.34$\,\pm\,$0.04 & 0.40$\,\pm\,$0.04 & 0.40$\,\pm\,$0.01 & 0.37$\,\pm\,$0.06 & 0.43$\,\pm\,$0.05 & 1 \\
             NGC2342 & 07h 09m 18.07s & +20\deg\ 38m 10.3s & 0.31$\,\pm\,$0.03 & 0.33$\,\pm\,$0.02 & 0.39$\,\pm\,$0.03 & 0.33$\,\pm\,$0.04 & 0.29$\,\pm\,$0.05 & 0.38$\,\pm\,$0.04 & 1 \\
      IRAS07251-0248 & 07h 27m 37.63s & $-$02\deg\ 54m 54.7s & 0.14$\,\pm\,$0.05 & 0.10$\,\pm\,$0.07 & 0.32$\,\pm\,$0.10 & 0.13$\,\pm\,$0.06 & 0.09$\,\pm\,$0.01 & 0.14$\,\pm\,$0.06 & 3 \\
             NGC2388 & 07h 29m 04.58s & +33\deg\ 51m 38.2s & 0.33$\,\pm\,$0.03 & 0.32$\,\pm\,$0.03 & 0.32$\,\pm\,$0.03 & 0.36$\,\pm\,$0.02 & 0.30$\,\pm\,$0.05 & 0.39$\,\pm\,$0.04 & 1 \\
       MCG+02-20-003 & 07h 35m 43.44s & +11\deg\ 42m 34.9s & 0.12$\,\pm\,$0.06 & 0.30$\,\pm\,$0.01 & 0.51$\,\pm\,$0.02 & 0.28$\,\pm\,$0.08 & 0.31$\,\pm\,$0.05 & 0.47$\,\pm\,$0.03 & 3 \\
      IRAS08355-4944 & 08h 37m 01.87s & $-$49\deg\ 54m 29.9s & 0.10$\,\pm\,$0.06 & 0.26$\,\pm\,$0.01 & 0.18$\,\pm\,$0.07 & 0.30$\,\pm\,$0.09 & 0.24$\,\pm\,$0.06 & 0.30$\,\pm\,$0.03 & 2 \\
             NGC2623 & 08h 38m 24.12s & +25\deg\ 45m 16.6s & 0.21$\,\pm\,$0.03 & 0.22$\,\pm\,$0.04 & 0.42$\,\pm\,$0.03 & 0.26$\,\pm\,$0.06 & 0.13$\,\pm\,$0.08 & 0.30$\,\pm\,$0.05 & 3 \\
        ESO432-IG006 & 08h 44m 27.22s & $-$31\deg\ 41m 50.6s & 0.37$\,\pm\,$0.06 & 0.42$\,\pm\,$0.01 & 0.45$\,\pm\,$0.04 & 0.42$\,\pm\,$0.01 & 0.42$\,\pm\,$0.05 & 0.45$\,\pm\,$0.04 & 1 \\
        ESO432-IG006 & 08h 44m 28.92s & $-$31\deg\ 41m 30.1s & 0.20$\,\pm\,$0.06 & 0.20$\,\pm\,$0.06 & 0.28$\,\pm\,$0.05 & 0.16$\,\pm\,$0.09 & 0.13$\,\pm\,$0.06 & 0.19$\,\pm\,$0.04 & 1 \\
         ESO60-IG016 & 08h 52m 30.50s & $-$69\deg\ 01m 59.2s & 0.11$\,\pm\,$0.01 & 0.24$\,\pm\,$0.07 & 0.36$\,\pm\,$0.05 & 0.22$\,\pm\,$0.06 & 0.18$\,\pm\,$0.06 & 0.33$\,\pm\,$0.05 & 3 \\
     IRASF08572+3915 & 09h 00m 25.34s & +39\deg\ 03m 54.0s & 0.04$\,\pm\,$0.02 & 0.16$\,\pm\,$0.06 & 0.17$\,\pm\,$0.17 & 0.10$\,\pm\,$0.09 & \dots$\,\pm\,$\dots & 0.06$\,\pm\,$0.05 & 0 \\
      IRAS09022-3615 & 09h 04m 12.70s & $-$36\deg\ 27m 01.4s & 0.10$\,\pm\,$0.03 & 0.13$\,\pm\,$0.03 & 0.17$\,\pm\,$0.04 & 0.20$\,\pm\,$0.02 & 0.10$\,\pm\,$0.08 & 0.20$\,\pm\,$0.05 & 2 \\
     IRASF09111-1007 & 09h 13m 36.50s & $-$10\deg\ 19m 29.6s & 0.24$\,\pm\,$0.08 & 0.28$\,\pm\,$0.02 & 0.35$\,\pm\,$0.04 & 0.27$\,\pm\,$0.04 & 0.18$\,\pm\,$0.06 & 0.30$\,\pm\,$0.04 & 1 \\
     IRASF09111-1007 & 09h 13m 38.88s & $-$10\deg\ 19m 19.6s & 0.43$\,\pm\,$0.12 & 0.42$\,\pm\,$0.04 & 0.39$\,\pm\,$0.04 & 0.42$\,\pm\,$0.01 & 0.32$\,\pm\,$0.05 & 0.43$\,\pm\,$0.04 & 1 \\
            UGC04881 & 09h 15m 55.51s & +44\deg\ 19m 57.4s & 0.37$\,\pm\,$0.03 & 0.52$\,\pm\,$0.03 & 0.59$\,\pm\,$0.05 & 0.49$\,\pm\,$0.03 & 0.35$\,\pm\,$0.06 & 0.57$\,\pm\,$0.04 & 3 \\
            UGC04881 & 09h 15m 54.70s & +44\deg\ 19m 50.9s & 0.33$\,\pm\,$0.05 & 0.38$\,\pm\,$0.02 & 0.48$\,\pm\,$0.03 & 0.37$\,\pm\,$0.03 & 0.40$\,\pm\,$0.06 & 0.44$\,\pm\,$0.04 & 3 \\
            UGC05101 & 09h 35m 51.60s & +61\deg\ 21m 11.9s & 0.21$\,\pm\,$0.07 & 0.30$\,\pm\,$0.02 & 0.35$\,\pm\,$0.03 & 0.28$\,\pm\,$0.02 & 0.20$\,\pm\,$0.05 & 0.36$\,\pm\,$0.04 & 2 \\
       MCG+08-18-013 & 09h 36m 37.20s & +48\deg\ 28m 27.8s & \dots$\,\pm\,$\dots & 0.42$\,\pm\,$0.01 & \dots$\,\pm\,$\dots & 0.47$\,\pm\,$0.07 & 0.50$\,\pm\,$0.01 & 0.56$\,\pm\,$0.11 & 4 \\
     IRASF09437+0317 & 09h 46m 21.10s & +03\deg\ 04m 16.3s & 0.64$\,\pm\,$0.02 & 0.60$\,\pm\,$0.04 & 0.67$\,\pm\,$0.07 & 0.62$\,\pm\,$0.03 & 0.64$\,\pm\,$0.06 & 0.66$\,\pm\,$0.06 & 1 \\
     IRASF09437+0317 & 09h 46m 20.30s & +03\deg\ 02m 44.5s & 0.66$\,\pm\,$0.05 & 0.66$\,\pm\,$0.01 & 0.64$\,\pm\,$0.09 & 0.59$\,\pm\,$0.06 & 0.57$\,\pm\,$0.10 & 0.64$\,\pm\,$0.08 & 1 \\
             NGC3110 & 10h 04m 02.11s & $-$06\deg\ 28m 29.6s & 0.65$\,\pm\,$0.03 & 0.66$\,\pm\,$0.01 & 0.68$\,\pm\,$0.03 & 0.66$\,\pm\,$0.01 & 0.65$\,\pm\,$0.03 & 0.68$\,\pm\,$0.02 & 1 \\
     IRASF10038-3338 & 10h 06m 04.66s & $-$33\deg\ 53m 06.0s & 0.05$\,\pm\,$0.04 & 0.08$\,\pm\,$0.03 & 0.39$\,\pm\,$0.03 & 0.09$\,\pm\,$0.09 & 0.06$\,\pm\,$0.09 & 0.33$\,\pm\,$0.04 & 3 \\
     IRASF10173+0828 & 10h 20m 00.24s & +08\deg\ 13m 32.9s & 0.14$\,\pm\,$0.09 & 0.14$\,\pm\,$0.02 & 0.14$\,\pm\,$0.01 & 0.11$\,\pm\,$0.03 & \dots$\,\pm\,$\dots & 0.07$\,\pm\,$0.09 & 0 \\
             NGC3221 & 10h 22m 19.97s & +21\deg\ 34m 10.6s & \dots$\,\pm\,$\dots & 0.77$\,\pm\,$0.02 & \dots$\,\pm\,$\dots & 0.80$\,\pm\,$0.01 & 0.76$\,\pm\,$0.04 & \dots$\,\pm\,$\dots & 1
\enddata
\tablecomments{\footnotesize Continued.}
%\vspace{2cm}
\end{deluxetable}

\clearpage

\setcounter{table}{1}
\begin{deluxetable}{lccccccccc}
\tabletypesize{\scriptsize}
\tablewidth{0pc}
\tablecaption{\scriptsize FEEs of Various MIR Features}
\tablehead{\colhead{Galaxy} & \colhead{R.A.} & \colhead{Declination} & \colhead{FEE$_{6.2\mu m}$} & \colhead{FEE$_{6.7\mu m}$} & \colhead{FEE$_{7.7\mu m}$} & \colhead{FEE$_{9.7\mu m}$} & \colhead{FEE$_{11.3\mu m}$} & \colhead{FEE$_{12.8\mu m}$} & FEE$_{\lambda}$ \\
\colhead{name} & \colhead{(J2000)} & \colhead{(J2000)} & \colhead{(PAH)} & \colhead{(Cont.)} & \colhead{(PAH)} & \colhead{(PAH)} & \colhead{(Si abs.)} & \colhead{(\NeIIno)} &  \colhead{Type} \\
\colhead{(1)} & \colhead{(2)} & \colhead{(3)} & \colhead{(4)} & \colhead{(5)} & \colhead{(6)} & \colhead{(7)} & \colhead{(8)} & \colhead{(9)} & \colhead{(10)}}
\startdata
             NGC3256 & 10h 27m 51.31s & $-$43\deg\ 54m 14.0s & 0.69$\,\pm\,$0.03 & 0.66$\,\pm\,$0.01 & 0.47$\,\pm\,$0.05 & 0.66$\,\pm\,$0.01 & 0.57$\,\pm\,$0.04 & 0.62$\,\pm\,$0.03 & 2 \\
         ESO264-G036 & 10h 43m 07.51s & $-$46\deg\ 12m 44.3s & 0.65$\,\pm\,$0.02 & 0.65$\,\pm\,$0.01 & 0.69$\,\pm\,$0.02 & 0.65$\,\pm\,$0.01 & 0.67$\,\pm\,$0.03 & 0.70$\,\pm\,$0.02 & 1 \\
         ESO264-G057 & 10h 59m 01.70s & $-$43\deg\ 26m 25.1s & 0.33$\,\pm\,$0.05 & 0.39$\,\pm\,$0.02 & 0.44$\,\pm\,$0.05 & 0.37$\,\pm\,$0.06 & 0.38$\,\pm\,$0.06 & 0.43$\,\pm\,$0.04 & 1 \\
     IRASF10565+2448 & 10h 59m 18.14s & +24\deg\ 32m 34.1s & 0.11$\,\pm\,$0.07 & 0.15$\,\pm\,$0.01 & 0.17$\,\pm\,$0.05 & 0.19$\,\pm\,$0.03 & 0.05$\,\pm\,$0.08 & 0.16$\,\pm\,$0.05 & 1 \\
       MCG+07-23-019 & 11h 03m 53.98s & +40\deg\ 51m 00.4s & 0.50$\,\pm\,$0.12 & 0.42$\,\pm\,$0.21 & 0.57$\,\pm\,$0.10 & 0.57$\,\pm\,$0.09 & 0.62$\,\pm\,$0.08 & 0.61$\,\pm\,$0.09 & 4 \\
         CGCG011-076 & 11h 21m 12.24s & $-$02\deg\ 59m 02.4s & 0.32$\,\pm\,$0.03 & 0.32$\,\pm\,$0.04 & 0.32$\,\pm\,$0.04 & 0.41$\,\pm\,$0.01 & 0.33$\,\pm\,$0.06 & 0.39$\,\pm\,$0.05 & 2 \\
              IC2810 & 11h 25m 45.07s & +14\deg\ 40m 36.1s & 0.45$\,\pm\,$0.03 & 0.49$\,\pm\,$0.02 & 0.55$\,\pm\,$0.03 & 0.51$\,\pm\,$0.02 & 0.48$\,\pm\,$0.04 & 0.56$\,\pm\,$0.03 & 1 \\
         ESO319-G022 & 11h 27m 54.19s & $-$41\deg\ 36m 51.8s & 0.17$\,\pm\,$0.03 & 0.20$\,\pm\,$0.07 & 0.18$\,\pm\,$0.09 & 0.13$\,\pm\,$0.05 & 0.10$\,\pm\,$0.09 & 0.18$\,\pm\,$0.06 & 1 \\
        ESO440-IG058 & 12h 06m 51.86s & $-$31\deg\ 56m 59.3s & 0.56$\,\pm\,$0.04 & 0.56$\,\pm\,$0.01 & 0.61$\,\pm\,$0.03 & 0.56$\,\pm\,$0.01 & 0.54$\,\pm\,$0.04 & 0.59$\,\pm\,$0.03 & 1 \\
        ESO440-IG058 & 12h 06m 51.70s & $-$31\deg\ 56m 46.3s & 0.52$\,\pm\,$0.08 & 0.45$\,\pm\,$0.09 & 0.50$\,\pm\,$0.06 & 0.47$\,\pm\,$0.01 & 0.50$\,\pm\,$0.07 & 0.52$\,\pm\,$0.03 & 1 \\
     IRASF12112+0305 & 12h 13m 46.03s & +02\deg\ 48m 42.1s & 0.51$\,\pm\,$0.08 & 0.46$\,\pm\,$0.02 & 0.56$\,\pm\,$0.05 & 0.49$\,\pm\,$0.02 & 0.49$\,\pm\,$0.06 & 0.52$\,\pm\,$0.04 & 1 \\
             NGC4194 & 12h 14m 09.72s & +54\deg\ 31m 35.4s & 0.48$\,\pm\,$0.01 & 0.45$\,\pm\,$0.04 & 0.50$\,\pm\,$0.03 & 0.51$\,\pm\,$0.01 & 0.47$\,\pm\,$0.04 & 0.55$\,\pm\,$0.04 & 1 \\
         ESO267-G030 & 12h 14m 12.82s & $-$47\deg\ 13m 42.6s & 0.60$\,\pm\,$0.02 & 0.60$\,\pm\,$0.03 & 0.56$\,\pm\,$0.03 & 0.61$\,\pm\,$0.01 & 0.60$\,\pm\,$0.03 & 0.62$\,\pm\,$0.03 & 2 \\
         ESO267-G030 & 12h 13m 52.27s & $-$47\deg\ 16m 25.3s & 0.33$\,\pm\,$0.04 & 0.32$\,\pm\,$0.04 & 0.36$\,\pm\,$0.04 & 0.32$\,\pm\,$0.05 & 0.27$\,\pm\,$0.05 & 0.38$\,\pm\,$0.03 & 1 \\
      IRAS12116-5615 & 12h 14m 22.08s & $-$56\deg\ 32m 32.6s & 0.15$\,\pm\,$0.07 & 0.27$\,\pm\,$0.01 & 0.11$\,\pm\,$0.06 & 0.25$\,\pm\,$0.01 & 0.18$\,\pm\,$0.05 & 0.28$\,\pm\,$0.05 & 2 \\
     IRASF12224-0624 & 12h 25m 03.91s & $-$06\deg\ 40m 52.0s & 0.20$\,\pm\,$0.05 & 0.21$\,\pm\,$0.10 & 0.37$\,\pm\,$0.06 & 0.12$\,\pm\,$0.08 & 0.10$\,\pm\,$0.01 & 0.22$\,\pm\,$0.05 & 3 \\
              Mrk231 & 12h 56m 14.26s & +56\deg\ 52m 25.0s & 0.03$\,\pm\,$0.01 & 0.18$\,\pm\,$0.04 & 0.04$\,\pm\,$0.01 & 0.13$\,\pm\,$0.09 & \dots$\,\pm\,$\dots & 0.04$\,\pm\,$0.01 & 0 \\
             NGC4922 & 13h 01m 25.27s & +29\deg\ 18m 49.7s & 0.11$\,\pm\,$0.04 & 0.13$\,\pm\,$0.07 & 0.13$\,\pm\,$0.06 & 0.11$\,\pm\,$0.04 & 0.06$\,\pm\,$0.08 & 0.10$\,\pm\,$0.06 & 4 \\
         CGCG043-099 & 13h 01m 50.28s & +04\deg\ 20m 01.0s & 0.13$\,\pm\,$0.05 & 0.18$\,\pm\,$0.03 & 0.37$\,\pm\,$0.04 & 0.23$\,\pm\,$0.07 & 0.10$\,\pm\,$0.08 & 0.25$\,\pm\,$0.05 & 3 \\
       MCG-02-33-098 & 13h 02m 19.66s & $-$15\deg\ 46m 04.1s & 0.15$\,\pm\,$0.09 & 0.19$\,\pm\,$0.01 & 0.11$\,\pm\,$0.04 & 0.25$\,\pm\,$0.03 & 0.07$\,\pm\,$0.08 & 0.20$\,\pm\,$0.05 & 2 \\
       MCG-02-33-098 & 13h 02m 20.38s & $-$15\deg\ 45m 59.8s & 0.49$\,\pm\,$0.05 & 0.40$\,\pm\,$0.06 & 0.45$\,\pm\,$0.04 & 0.47$\,\pm\,$0.02 & 0.43$\,\pm\,$0.05 & 0.50$\,\pm\,$0.04 & 1 \\
         ESO507-G070 & 13h 02m 52.42s & $-$23\deg\ 55m 17.8s & 0.13$\,\pm\,$0.07 & 0.17$\,\pm\,$0.01 & 0.43$\,\pm\,$0.04 & 0.15$\,\pm\,$0.13 & 0.11$\,\pm\,$0.08 & 0.22$\,\pm\,$0.05 & 3 \\
      IRAS13052-5711 & 13h 08m 18.72s & $-$57\deg\ 27m 30.2s & 0.37$\,\pm\,$0.07 & 0.43$\,\pm\,$0.03 & 0.52$\,\pm\,$0.04 & 0.42$\,\pm\,$0.01 & 0.40$\,\pm\,$0.05 & 0.47$\,\pm\,$0.05 & 3 \\
              IC0860 & 13h 15m 03.48s & +24\deg\ 37m 07.7s & 0.29$\,\pm\,$0.04 & 0.24$\,\pm\,$0.06 & 0.51$\,\pm\,$0.04 & 0.17$\,\pm\,$0.08 & 0.21$\,\pm\,$0.05 & 0.39$\,\pm\,$0.03 & 3 \\
      IRAS13120-5453 & 13h 15m 06.36s & $-$55\deg\ 09m 22.3s & 0.22$\,\pm\,$0.04 & 0.17$\,\pm\,$0.04 & 0.28$\,\pm\,$0.05 & 0.23$\,\pm\,$0.04 & 0.15$\,\pm\,$0.06 & 0.26$\,\pm\,$0.04 & 1 \\
              VV250a & 13h 15m 34.97s & +62\deg\ 07m 29.3s & 0.12$\,\pm\,$0.06 & 0.14$\,\pm\,$0.02 & 0.15$\,\pm\,$0.04 & 0.16$\,\pm\,$0.01 & 0.07$\,\pm\,$0.06 & 0.17$\,\pm\,$0.04 & 1 \\
              VV250a & 13h 15m 30.70s & +62\deg\ 07m 45.8s & 0.20$\,\pm\,$0.08 & 0.24$\,\pm\,$0.08 & 0.32$\,\pm\,$0.13 & 0.27$\,\pm\,$0.04 & 0.20$\,\pm\,$0.06 & 0.28$\,\pm\,$0.05 & 1 \\
            UGC08387 & 13h 20m 35.38s & +34\deg\ 08m 22.2s & 0.17$\,\pm\,$0.06 & 0.23$\,\pm\,$0.01 & 0.36$\,\pm\,$0.05 & 0.27$\,\pm\,$0.01 & 0.16$\,\pm\,$0.06 & 0.31$\,\pm\,$0.05 & 3 \\
             NGC5104 & 13h 21m 23.09s & +00\deg\ 20m 33.4s & 0.45$\,\pm\,$0.04 & 0.45$\,\pm\,$0.02 & 0.58$\,\pm\,$0.03 & 0.46$\,\pm\,$0.01 & 0.42$\,\pm\,$0.06 & 0.53$\,\pm\,$0.04 & 3 \\
       MCG-03-34-064 & 13h 22m 24.46s & $-$16\deg\ 43m 42.2s & 0.02$\,\pm\,$0.05 & 0.05$\,\pm\,$0.02 & 0.05$\,\pm\,$0.03 & 0.08$\,\pm\,$0.03 & 0.04$\,\pm\,$0.01 & 0.07$\,\pm\,$0.05 & 0 \\
             NGC5135 & 13h 25m 44.02s & $-$29\deg\ 50m 00.2s & 0.55$\,\pm\,$0.03 & 0.54$\,\pm\,$0.03 & 0.55$\,\pm\,$0.03 & 0.55$\,\pm\,$0.01 & 0.57$\,\pm\,$0.04 & 0.59$\,\pm\,$0.03 & 1 \\
         ESO173-G015 & 13h 27m 23.78s & $-$57\deg\ 29m 21.8s & 0.35$\,\pm\,$0.01 & 0.34$\,\pm\,$0.05 & 0.56$\,\pm\,$0.02 & 0.33$\,\pm\,$0.07 & 0.29$\,\pm\,$0.05 & 0.51$\,\pm\,$0.04 & 3 \\
              IC4280 & 13h 32m 53.40s & $-$24\deg\ 12m 25.6s & 0.76$\,\pm\,$0.02 & 0.76$\,\pm\,$0.01 & 0.78$\,\pm\,$0.02 & 0.75$\,\pm\,$0.01 & 0.76$\,\pm\,$0.02 & 0.77$\,\pm\,$0.02 & 1 \\
             NGC5256 & 13h 38m 17.26s & +48\deg\ 16m 32.9s & 0.33$\,\pm\,$0.04 & 0.39$\,\pm\,$0.01 & 0.47$\,\pm\,$0.04 & 0.39$\,\pm\,$0.01 & 0.39$\,\pm\,$0.06 & 0.42$\,\pm\,$0.05 & 1 \\
             NGC5256 & 13h 38m 17.78s & +48\deg\ 16m 41.5s & 0.49$\,\pm\,$0.01 & 0.41$\,\pm\,$0.01 & 0.44$\,\pm\,$0.16 & 0.34$\,\pm\,$0.10 & 0.29$\,\pm\,$0.21 & 0.37$\,\pm\,$0.19 & 1 \\
             NGC5257 & 13h 39m 57.72s & +00\deg\ 49m 53.0s & 0.79$\,\pm\,$0.03 & 0.78$\,\pm\,$0.02 & 0.77$\,\pm\,$0.03 & 0.77$\,\pm\,$0.02 & 0.79$\,\pm\,$0.03 & 0.78$\,\pm\,$0.03 & 1 \\
             NGC5257 & 13h 39m 52.94s & +00\deg\ 50m 25.8s & 0.76$\,\pm\,$0.04 & 0.82$\,\pm\,$0.03 & 0.83$\,\pm\,$0.01 & 0.81$\,\pm\,$0.03 & 0.82$\,\pm\,$0.03 & 0.82$\,\pm\,$0.02 & 1 \\
              Mrk273 & 13h 44m 42.12s & +55\deg\ 53m 13.2s & 0.08$\,\pm\,$0.03 & 0.24$\,\pm\,$0.02 & 0.24$\,\pm\,$0.08 & 0.18$\,\pm\,$0.08 & 0.03$\,\pm\,$0.08 & 0.18$\,\pm\,$0.05 & 3 \\
            UGC08739 & 13h 49m 13.94s & +35\deg\ 15m 26.3s & 0.37$\,\pm\,$0.02 & 0.30$\,\pm\,$0.07 & 0.61$\,\pm\,$0.04 & 0.36$\,\pm\,$0.07 & 0.45$\,\pm\,$0.04 & 0.54$\,\pm\,$0.02 & 3 \\
        ESO221-IG010 & 13h 50m 56.93s & $-$49\deg\ 03m 18.7s & 0.44$\,\pm\,$0.03 & 0.43$\,\pm\,$0.04 & 0.45$\,\pm\,$0.04 & 0.48$\,\pm\,$0.01 & 0.43$\,\pm\,$0.05 & 0.52$\,\pm\,$0.04 & 1 \\
             NGC5331 & 13h 52m 16.20s & +02\deg\ 06m 05.0s & 0.58$\,\pm\,$0.03 & 0.57$\,\pm\,$0.01 & 0.66$\,\pm\,$0.02 & 0.58$\,\pm\,$0.01 & 0.60$\,\pm\,$0.03 & 0.64$\,\pm\,$0.03 & 1 \\
             NGC5331 & 13h 52m 16.44s & +02\deg\ 06m 31.0s & 0.59$\,\pm\,$0.09 & 0.58$\,\pm\,$0.07 & 0.65$\,\pm\,$0.04 & 0.59$\,\pm\,$0.04 & 0.62$\,\pm\,$0.06 & 0.64$\,\pm\,$0.05 & 1 \\
             NGC5395 & 13h 58m 37.97s & +37\deg\ 25m 28.2s & 0.62$\,\pm\,$0.12 & 0.67$\,\pm\,$0.07 & 0.63$\,\pm\,$0.04 & 0.71$\,\pm\,$0.01 & 0.71$\,\pm\,$0.01 & 0.71$\,\pm\,$0.01 & 1 \\
             NGC5395 & 13h 58m 33.65s & +37\deg\ 27m 13.0s & 0.31$\,\pm\,$0.03 & 0.31$\,\pm\,$0.02 & 0.33$\,\pm\,$0.03 & 0.35$\,\pm\,$0.01 & 0.30$\,\pm\,$0.05 & 0.34$\,\pm\,$0.04 & 1 \\
         CGCG247-020 & 14h 19m 43.27s & +49\deg\ 14m 11.8s & 0.03$\,\pm\,$0.07 & 0.14$\,\pm\,$0.01 & 0.11$\,\pm\,$0.06 & 0.19$\,\pm\,$0.05 & 0.06$\,\pm\,$0.08 & 0.12$\,\pm\,$0.06 & 1 \\
             NGC5653 & 14h 30m 10.44s & +31\deg\ 12m 55.8s & 0.75$\,\pm\,$0.01 & 0.76$\,\pm\,$0.02 & 0.76$\,\pm\,$0.02 & 0.76$\,\pm\,$0.01 & 0.76$\,\pm\,$0.02 & 0.77$\,\pm\,$0.02 & 1 \\
     IRASF14348-1447$^*$ & 14h 37m 38.28s & $-$15\deg\ 00m 24.1s & 0.41$\,\pm\,$0.03 & 0.50$\,\pm\,$0.04 & 0.58$\,\pm\,$0.03 & 0.53$\,\pm\,$0.03 & 0.56$\,\pm\,$0.05 & 0.60$\,\pm\,$0.03 & 4 \\
     IRASF14378-3651 & 14h 40m 59.04s & $-$37\deg\ 04m 32.2s & 0.16$\,\pm\,$0.07 & 0.11$\,\pm\,$0.07 & 0.29$\,\pm\,$0.12 & 0.13$\,\pm\,$0.03 & 0.05$\,\pm\,$0.01 & 0.08$\,\pm\,$0.07 & 4 \\
             NGC5734 & 14h 45m 09.05s & $-$20\deg\ 52m 13.1s & 0.77$\,\pm\,$0.01 & 0.79$\,\pm\,$0.02 & 0.77$\,\pm\,$0.02 & 0.78$\,\pm\,$0.01 & 0.77$\,\pm\,$0.03 & 0.78$\,\pm\,$0.03 & 1
\enddata
\tablecomments{\footnotesize Continued.}
%\vspace{2cm}
\end{deluxetable}

\clearpage

\setcounter{table}{1}
\begin{deluxetable}{lccccccccc}
\tabletypesize{\scriptsize}
\tablewidth{0pc}
\tablecaption{\scriptsize FEEs of Various MIR Features}
\tablehead{\colhead{Galaxy} & \colhead{R.A.} & \colhead{Declination} & \colhead{FEE$_{6.2\mu m}$} & \colhead{FEE$_{6.7\mu m}$} & \colhead{FEE$_{7.7\mu m}$} & \colhead{FEE$_{9.7\mu m}$} & \colhead{FEE$_{11.3\mu m}$} & \colhead{FEE$_{12.8\mu m}$} & FEE$_{\lambda}$ \\
\colhead{name} & \colhead{(J2000)} & \colhead{(J2000)} & \colhead{(PAH)} & \colhead{(Cont.)} & \colhead{(PAH)} & \colhead{(PAH)} & \colhead{(Si abs.)} & \colhead{(\NeIIno)} &  \colhead{Type} \\
\colhead{(1)} & \colhead{(2)} & \colhead{(3)} & \colhead{(4)} & \colhead{(5)} & \colhead{(6)} & \colhead{(7)} & \colhead{(8)} & \colhead{(9)} & \colhead{(10)}}
\startdata
             NGC5734 & 14h 45m 11.02s & $-$20\deg\ 54m 48.6s & 0.67$\,\pm\,$0.02 & 0.70$\,\pm\,$0.02 & 0.68$\,\pm\,$0.06 & 0.67$\,\pm\,$0.04 & 0.68$\,\pm\,$0.06 & 0.69$\,\pm\,$0.05 & 1 \\
              VV340a & 14h 57m 00.70s & +24\deg\ 37m 05.9s & 0.78$\,\pm\,$0.02 & 0.79$\,\pm\,$0.01 & 0.81$\,\pm\,$0.02 & 0.78$\,\pm\,$0.01 & 0.77$\,\pm\,$0.02 & 0.80$\,\pm\,$0.02 & 1 \\
              VV340a & 14h 57m 00.31s & +24\deg\ 36m 24.5s & 0.70$\,\pm\,$0.04 & 0.67$\,\pm\,$0.03 & 0.63$\,\pm\,$0.06 & 0.61$\,\pm\,$0.05 & 0.64$\,\pm\,$0.08 & 0.65$\,\pm\,$0.07 & 1 \\
               VV705 & 15h 18m 06.14s & +42\deg\ 44m 44.9s & 0.52$\,\pm\,$0.23 & 0.41$\,\pm\,$0.07 & 0.65$\,\pm\,$0.10 & 0.41$\,\pm\,$0.02 & 0.52$\,\pm\,$0.10 & 0.50$\,\pm\,$0.12 & 4 \\
         ESO099-G004 & 15h 24m 57.98s & $-$63\deg\ 07m 29.3s & 0.50$\,\pm\,$0.05 & 0.55$\,\pm\,$0.07 & 0.41$\,\pm\,$0.03 & 0.48$\,\pm\,$0.02 & 0.24$\,\pm\,$0.08 & 0.46$\,\pm\,$0.04 & 2 \\
     IRASF15250+3608 & 15h 26m 59.42s & +35\deg\ 58m 37.9s & 0.05$\,\pm\,$0.03 & 0.05$\,\pm\,$0.04 & 0.18$\,\pm\,$0.10 & 0.08$\,\pm\,$0.06 & \dots$\,\pm\,$\dots & 0.17$\,\pm\,$0.05 & 3 \\
             NGC5936 & 15h 30m 00.84s & +12\deg\ 59m 22.2s & 0.39$\,\pm\,$0.01 & 0.32$\,\pm\,$0.06 & 0.39$\,\pm\,$0.05 & 0.38$\,\pm\,$0.07 & 0.35$\,\pm\,$0.04 & 0.42$\,\pm\,$0.02 & 1 \\
              Arp220 & 15h 34m 57.24s & +23\deg\ 30m 11.2s & 0.19$\,\pm\,$0.06 & 0.24$\,\pm\,$0.01 & 0.48$\,\pm\,$0.04 & 0.16$\,\pm\,$0.08 & 0.10$\,\pm\,$0.09 & 0.43$\,\pm\,$0.03 & 3 \\
             NGC5990 & 15h 46m 16.42s & +02\deg\ 24m 55.4s & 0.27$\,\pm\,$0.05 & 0.34$\,\pm\,$0.01 & 0.30$\,\pm\,$0.05 & 0.44$\,\pm\,$0.02 & 0.36$\,\pm\,$0.05 & 0.48$\,\pm\,$0.04 & 2 \\
             NGC6090 & 16h 11m 40.85s & +52\deg\ 27m 27.4s & 0.50$\,\pm\,$0.03 & 0.50$\,\pm\,$0.02 & 0.53$\,\pm\,$0.03 & 0.54$\,\pm\,$0.01 & 0.53$\,\pm\,$0.04 & 0.55$\,\pm\,$0.03 & 1 \\
     IRASF16164-0746 & 16h 19m 11.76s & $-$07\deg\ 54m 02.9s & 0.17$\,\pm\,$0.06 & 0.20$\,\pm\,$0.02 & 0.52$\,\pm\,$0.05 & 0.25$\,\pm\,$0.03 & 0.12$\,\pm\,$0.08 & 0.37$\,\pm\,$0.05 & 3 \\
         CGCG052-037 & 16h 30m 53.26s & +04\deg\ 04m 23.9s & 0.35$\,\pm\,$0.03 & 0.35$\,\pm\,$0.04 & 0.41$\,\pm\,$0.03 & 0.38$\,\pm\,$0.04 & 0.35$\,\pm\,$0.04 & 0.41$\,\pm\,$0.03 & 1 \\
             NGC6156 & 16h 34m 52.56s & $-$60\deg\ 37m 08.0s & 0.41$\,\pm\,$0.02 & 0.46$\,\pm\,$0.03 & 0.14$\,\pm\,$0.04 & 0.46$\,\pm\,$0.01 & 0.23$\,\pm\,$0.06 & 0.32$\,\pm\,$0.04 & 2 \\
        ESO069-IG006 & 16h 38m 11.86s & $-$68\deg\ 26m 08.2s & 0.39$\,\pm\,$0.07 & 0.40$\,\pm\,$0.03 & 0.51$\,\pm\,$0.04 & 0.42$\,\pm\,$0.01 & 0.41$\,\pm\,$0.06 & 0.46$\,\pm\,$0.04 & 1 \\
     IRASF16399-0937$^*$ & 16h 42m 40.10s & $-$09\deg\ 43m 13.8s & 0.72$\,\pm\,$0.02 & 0.73$\,\pm\,$0.01 & 0.74$\,\pm\,$0.02 & 0.71$\,\pm\,$0.01 & 0.61$\,\pm\,$0.04 & 0.73$\,\pm\,$0.02 & 4 \\
         ESO453-G005 & 16h 47m 31.08s & $-$29\deg\ 21m 21.6s & 0.70$\,\pm\,$0.02 & 0.73$\,\pm\,$0.01 & 0.73$\,\pm\,$0.01 & 0.70$\,\pm\,$0.02 & 0.69$\,\pm\,$0.03 & 0.71$\,\pm\,$0.02 & 1 \\
             NGC6240$^*$ & 16h 52m 58.90s & +02\deg\ 24m 03.2s & 0.32$\,\pm\,$0.06 & 0.34$\,\pm\,$0.01 & 0.48$\,\pm\,$0.04 & 0.35$\,\pm\,$0.02 & 0.31$\,\pm\,$0.05 & 0.45$\,\pm\,$0.04 & 3 \\
     IRASF16516-0948 & 16h 54m 23.71s & $-$09\deg\ 53m 20.8s & 0.71$\,\pm\,$0.02 & 0.72$\,\pm\,$0.02 & 0.75$\,\pm\,$0.04 & 0.71$\,\pm\,$0.03 & 0.73$\,\pm\,$0.05 & 0.74$\,\pm\,$0.04 & 1 \\
             NGC6286 & 16h 58m 31.63s & +58\deg\ 56m 13.2s & 0.80$\,\pm\,$0.04 & 0.80$\,\pm\,$0.03 & 0.80$\,\pm\,$0.01 & 0.80$\,\pm\,$0.01 & 0.80$\,\pm\,$0.02 & 0.80$\,\pm\,$0.01 & 1 \\
             NGC6286 & 16h 58m 24.00s & +58\deg\ 57m 21.6s & 0.50$\,\pm\,$0.10 & 0.47$\,\pm\,$0.01 & 0.59$\,\pm\,$0.03 & 0.48$\,\pm\,$0.02 & 0.49$\,\pm\,$0.05 & 0.55$\,\pm\,$0.03 & 3 \\
     IRASF17132+5313$^*$ & 17h 14m 20.45s & +53\deg\ 10m 31.4s & 0.66$\,\pm\,$0.03 & 0.67$\,\pm\,$0.02 & 0.68$\,\pm\,$0.03 & 0.66$\,\pm\,$0.01 & 0.70$\,\pm\,$0.03 & 0.70$\,\pm\,$0.03 & 1 \\
     IRASF17138-1017 & 17h 16m 35.69s & $-$10\deg\ 20m 40.6s & 0.65$\,\pm\,$0.03 & 0.64$\,\pm\,$0.03 & 0.69$\,\pm\,$0.02 & 0.66$\,\pm\,$0.01 & 0.70$\,\pm\,$0.03 & 0.71$\,\pm\,$0.03 & 1 \\
     IRASF17207-0014 & 17h 23m 21.96s & +00\deg\ 17m 00.6s & 0.17$\,\pm\,$0.06 & 0.24$\,\pm\,$0.01 & 0.45$\,\pm\,$0.04 & 0.25$\,\pm\,$0.03 & 0.19$\,\pm\,$0.05 & 0.40$\,\pm\,$0.04 & 3 \\
         ESO138-G027 & 17h 26m 43.34s & $-$59\deg\ 55m 55.2s & 0.48$\,\pm\,$0.03 & 0.55$\,\pm\,$0.03 & 0.36$\,\pm\,$0.04 & 0.55$\,\pm\,$0.01 & 0.36$\,\pm\,$0.06 & 0.54$\,\pm\,$0.04 & 2 \\
            UGC11041 & 17h 54m 51.82s & +34\deg\ 46m 34.3s & 0.56$\,\pm\,$0.02 & 0.56$\,\pm\,$0.03 & 0.61$\,\pm\,$0.02 & 0.57$\,\pm\,$0.02 & 0.60$\,\pm\,$0.03 & 0.62$\,\pm\,$0.02 & 1 \\
         CGCG141-034 & 17h 56m 56.64s & +24\deg\ 01m 01.9s & 0.23$\,\pm\,$0.05 & 0.25$\,\pm\,$0.02 & 0.42$\,\pm\,$0.06 & 0.20$\,\pm\,$0.05 & 0.20$\,\pm\,$0.07 & 0.30$\,\pm\,$0.04 & 3 \\
      IRAS17578-0400 & 18h 00m 31.85s & $-$04\deg\ 00m 53.3s & 0.39$\,\pm\,$0.06 & 0.40$\,\pm\,$0.07 & 0.55$\,\pm\,$0.04 & 0.44$\,\pm\,$0.01 & 0.48$\,\pm\,$0.05 & 0.54$\,\pm\,$0.03 & 4 \\
      IRAS17578-0400 & 18h 00m 34.08s & $-$04\deg\ 01m 44.0s & 0.43$\,\pm\,$0.05 & 0.49$\,\pm\,$0.02 & 0.54$\,\pm\,$0.04 & 0.50$\,\pm\,$0.01 & 0.50$\,\pm\,$0.04 & 0.56$\,\pm\,$0.03 & 1 \\
      IRAS17578-0400 & 18h 00m 24.29s & $-$04\deg\ 01m 03.7s & 0.60$\,\pm\,$0.05 & 0.60$\,\pm\,$0.02 & 0.68$\,\pm\,$0.04 & 0.64$\,\pm\,$0.01 & 0.67$\,\pm\,$0.03 & 0.68$\,\pm\,$0.03 & 1 \\
      IRAS18090+0130 & 18h 11m 38.42s & +01\deg\ 31m 40.4s & 0.52$\,\pm\,$0.05 & 0.55$\,\pm\,$0.02 & 0.63$\,\pm\,$0.04 & 0.55$\,\pm\,$0.01 & 0.51$\,\pm\,$0.04 & 0.59$\,\pm\,$0.03 & 1 \\
      IRAS18090+0130 & 18h 11m 33.41s & +01\deg\ 31m 42.2s & 0.35$\,\pm\,$0.07 & 0.34$\,\pm\,$0.03 & 0.54$\,\pm\,$0.09 & 0.38$\,\pm\,$0.06 & 0.32$\,\pm\,$0.05 & 0.46$\,\pm\,$0.04 & 3 \\
         CGCG142-034 & 18h 16m 40.68s & +22\deg\ 06m 46.4s & 0.24$\,\pm\,$0.03 & 0.23$\,\pm\,$0.04 & 0.46$\,\pm\,$0.03 & 0.21$\,\pm\,$0.11 & 0.33$\,\pm\,$0.04 & 0.40$\,\pm\,$0.03 & 3 \\
         CGCG142-034 & 18h 16m 33.84s & +22\deg\ 06m 38.9s & 0.47$\,\pm\,$0.04 & 0.49$\,\pm\,$0.02 & 0.55$\,\pm\,$0.05 & 0.47$\,\pm\,$0.01 & 0.44$\,\pm\,$0.05 & 0.49$\,\pm\,$0.04 & 1 \\
     IRASF18293-3413 & 18h 32m 41.09s & $-$34\deg\ 11m 26.9s & 0.45$\,\pm\,$0.04 & 0.49$\,\pm\,$0.01 & 0.54$\,\pm\,$0.03 & 0.51$\,\pm\,$0.01 & 0.51$\,\pm\,$0.04 & 0.54$\,\pm\,$0.03 & 1 \\
           NGC6670AB & 18h 33m 34.25s & +59\deg\ 53m 17.9s & 0.25$\,\pm\,$0.06 & 0.27$\,\pm\,$0.01 & 0.40$\,\pm\,$0.03 & 0.26$\,\pm\,$0.05 & 0.32$\,\pm\,$0.06 & 0.39$\,\pm\,$0.04 & 3 \\
           NGC6670AB & 18h 33m 37.73s & +59\deg\ 53m 22.9s & 0.24$\,\pm\,$0.06 & 0.25$\,\pm\,$0.01 & 0.32$\,\pm\,$0.03 & 0.28$\,\pm\,$0.05 & 0.23$\,\pm\,$0.07 & 0.35$\,\pm\,$0.03 & 1 \\
              IC4734 & 18h 38m 25.75s & $-$57\deg\ 29m 25.4s & 0.21$\,\pm\,$0.05 & 0.23$\,\pm\,$0.01 & 0.40$\,\pm\,$0.04 & 0.30$\,\pm\,$0.06 & 0.22$\,\pm\,$0.07 & 0.35$\,\pm\,$0.04 & 3 \\
             NGC6701 & 18h 43m 12.53s & +60\deg\ 39m 11.5s & 0.41$\,\pm\,$0.05 & 0.46$\,\pm\,$0.01 & 0.48$\,\pm\,$0.03 & 0.47$\,\pm\,$0.02 & 0.45$\,\pm\,$0.05 & 0.48$\,\pm\,$0.04 & 1 \\
             NGC6786 & 19h 10m 54.00s & +73\deg\ 24m 36.0s & 0.54$\,\pm\,$0.04 & 0.53$\,\pm\,$0.02 & 0.56$\,\pm\,$0.04 & 0.54$\,\pm\,$0.01 & 0.53$\,\pm\,$0.04 & 0.59$\,\pm\,$0.03 & 1 \\
             NGC6786 & 19h 11m 04.37s & +73\deg\ 25m 32.5s & 0.36$\,\pm\,$0.05 & 0.47$\,\pm\,$0.04 & 0.24$\,\pm\,$0.08 & 0.45$\,\pm\,$0.04 & 0.21$\,\pm\,$0.08 & 0.34$\,\pm\,$0.04 & 2 \\
        ESO593-IG008 & 19h 14m 31.15s & $-$21\deg\ 19m 06.2s & 0.56$\,\pm\,$0.05 & 0.58$\,\pm\,$0.01 & 0.63$\,\pm\,$0.03 & 0.53$\,\pm\,$0.01 & 0.55$\,\pm\,$0.05 & 0.61$\,\pm\,$0.03 & 1 \\
     IRASF19297-0406 & 19h 32m 22.30s & $-$04\deg\ 00m 01.1s & 0.16$\,\pm\,$0.04 & 0.18$\,\pm\,$0.04 & 0.25$\,\pm\,$0.12 & 0.15$\,\pm\,$0.04 & 0.13$\,\pm\,$0.08 & 0.18$\,\pm\,$0.06 & 1 \\
      IRAS19542+1110 & 19h 56m 35.78s & +11\deg\ 19m 04.8s & 0.11$\,\pm\,$0.03 & 0.09$\,\pm\,$0.06 & 0.10$\,\pm\,$0.07 & 0.06$\,\pm\,$0.03 & 0.03$\,\pm\,$0.08 & 0.05$\,\pm\,$0.06 & 0 \\
         ESO339-G011 & 19h 57m 37.61s & $-$37\deg\ 56m 08.5s & 0.29$\,\pm\,$0.03 & 0.36$\,\pm\,$0.04 & 0.25$\,\pm\,$0.04 & 0.38$\,\pm\,$0.04 & 0.23$\,\pm\,$0.06 & 0.38$\,\pm\,$0.03 & 2 \\
             NGC6907 & 20h 25m 06.58s & $-$24\deg\ 48m 32.8s & 0.64$\,\pm\,$0.04 & 0.62$\,\pm\,$0.01 & 0.64$\,\pm\,$0.02 & 0.64$\,\pm\,$0.01 & 0.60$\,\pm\,$0.04 & 0.66$\,\pm\,$0.03 & 1 \\
       MCG+04-48-002 & 20h 28m 35.06s & +25\deg\ 44m 00.2s & 0.41$\,\pm\,$0.03 & 0.43$\,\pm\,$0.04 & 0.52$\,\pm\,$0.03 & 0.46$\,\pm\,$0.01 & 0.48$\,\pm\,$0.04 & 0.52$\,\pm\,$0.04 & 1 \\
             NGC6926 & 20h 33m 06.12s & $-$02\deg\ 01m 39.0s & 0.73$\,\pm\,$0.08 & 0.85$\,\pm\,$0.01 & 0.85$\,\pm\,$0.01 & 0.85$\,\pm\,$0.01 & 0.79$\,\pm\,$0.02 & 0.85$\,\pm\,$0.01 & 4 \\
      IRAS20351+2521 & 20h 37m 17.74s & +25\deg\ 31m 37.6s & 0.40$\,\pm\,$0.07 & 0.45$\,\pm\,$0.02 & 0.54$\,\pm\,$0.04 & 0.53$\,\pm\,$0.02 & 0.44$\,\pm\,$0.06 & 0.56$\,\pm\,$0.04 & 1 \\
         CGCG448-020 & 20h 57m 24.10s & +17\deg\ 07m 35.0s & 0.49$\,\pm\,$0.08 & 0.49$\,\pm\,$0.02 & 0.52$\,\pm\,$0.05 & 0.55$\,\pm\,$0.02 & 0.60$\,\pm\,$0.05 & 0.56$\,\pm\,$0.03 & 1
\enddata
\tablecomments{\footnotesize Continued.}
%\vspace{2cm}
\end{deluxetable}

\clearpage

\setcounter{table}{1}
\begin{deluxetable}{lccccccccc}
\tabletypesize{\scriptsize}
\tablewidth{0pc}
\tablecaption{\scriptsize FEEs of Various MIR Features}
\tablehead{\colhead{Galaxy} & \colhead{R.A.} & \colhead{Declination} & \colhead{FEE$_{6.2\mu m}$} & \colhead{FEE$_{6.7\mu m}$} & \colhead{FEE$_{7.7\mu m}$} & \colhead{FEE$_{9.7\mu m}$} & \colhead{FEE$_{11.3\mu m}$} & \colhead{FEE$_{12.8\mu m}$} & FEE$_{\lambda}$ \\
\colhead{name} & \colhead{(J2000)} & \colhead{(J2000)} & \colhead{(PAH)} & \colhead{(Cont.)} & \colhead{(PAH)} & \colhead{(PAH)} & \colhead{(Si abs.)} & \colhead{(\NeIIno)} &  \colhead{Type} \\
\colhead{(1)} & \colhead{(2)} & \colhead{(3)} & \colhead{(4)} & \colhead{(5)} & \colhead{(6)} & \colhead{(7)} & \colhead{(8)} & \colhead{(9)} & \colhead{(10)}}
\startdata
         CGCG448-020 & 20h 57m 24.38s & +17\deg\ 07m 39.4s & 0.13$\,\pm\,$0.09 & 0.20$\,\pm\,$0.02 & 0.20$\,\pm\,$0.06 & 0.23$\,\pm\,$0.03 & 0.13$\,\pm\,$0.06 & 0.31$\,\pm\,$0.04 & 2 \\
        ESO286-IG019 & 20h 58m 26.78s & $-$42\deg\ 39m 00.4s & 0.12$\,\pm\,$0.09 & 0.20$\,\pm\,$0.04 & 0.41$\,\pm\,$0.04 & 0.22$\,\pm\,$0.08 & 0.05$\,\pm\,$0.08 & 0.26$\,\pm\,$0.05 & 3 \\
         ESO286-G035 & 21h 04m 11.11s & $-$43\deg\ 35m 36.2s & 0.57$\,\pm\,$0.02 & 0.56$\,\pm\,$0.02 & 0.62$\,\pm\,$0.02 & 0.56$\,\pm\,$0.01 & 0.59$\,\pm\,$0.03 & 0.61$\,\pm\,$0.03 & 1 \\
      IRAS21101+5810 & 21h 11m 29.28s & +58\deg\ 23m 07.8s & 0.27$\,\pm\,$0.07 & 0.26$\,\pm\,$0.03 & 0.25$\,\pm\,$0.08 & 0.29$\,\pm\,$0.02 & 0.11$\,\pm\,$0.08 & 0.25$\,\pm\,$0.06 & 4 \\
        ESO343-IG013 & 21h 36m 10.54s & $-$38\deg\ 32m 42.7s & 0.40$\,\pm\,$0.06 & 0.41$\,\pm\,$0.01 & 0.61$\,\pm\,$0.03 & 0.42$\,\pm\,$0.01 & 0.47$\,\pm\,$0.04 & 0.53$\,\pm\,$0.04 & 3 \\
        ESO343-IG013 & 21h 36m 10.92s & $-$38\deg\ 32m 33.0s & 0.18$\,\pm\,$0.06 & 0.28$\,\pm\,$0.02 & 0.19$\,\pm\,$0.05 & 0.25$\,\pm\,$0.06 & 0.04$\,\pm\,$0.06 & 0.25$\,\pm\,$0.05 & 2 \\
             NGC7130 & 21h 48m 19.54s & $-$34\deg\ 57m 04.7s & 0.46$\,\pm\,$0.03 & 0.53$\,\pm\,$0.05 & 0.47$\,\pm\,$0.07 & 0.57$\,\pm\,$0.05 & 0.48$\,\pm\,$0.08 & 0.59$\,\pm\,$0.06 & 2 \\
         ESO467-G027 & 22h 14m 39.96s & $-$27\deg\ 27m 50.4s & 0.71$\,\pm\,$0.02 & 0.73$\,\pm\,$0.02 & 0.73$\,\pm\,$0.02 & 0.72$\,\pm\,$0.01 & 0.72$\,\pm\,$0.02 & 0.74$\,\pm\,$0.02 & 1 \\
         ESO602-G025 & 22h 31m 25.49s & $-$19\deg\ 02m 03.8s & 0.35$\,\pm\,$0.06 & 0.44$\,\pm\,$0.01 & 0.52$\,\pm\,$0.04 & 0.43$\,\pm\,$0.01 & 0.39$\,\pm\,$0.05 & 0.47$\,\pm\,$0.04 & 3 \\
            UGC12150 & 22h 41m 12.22s & +34\deg\ 14m 56.8s & 0.33$\,\pm\,$0.06 & 0.39$\,\pm\,$0.02 & 0.43$\,\pm\,$0.04 & 0.39$\,\pm\,$0.03 & 0.35$\,\pm\,$0.04 & 0.43$\,\pm\,$0.04 & 1 \\
     IRASF22491-1808 & 22h 51m 49.34s & $-$17\deg\ 52m 25.0s & 0.19$\,\pm\,$0.03 & 0.27$\,\pm\,$0.06 & 0.35$\,\pm\,$0.07 & 0.35$\,\pm\,$0.02 & 0.27$\,\pm\,$0.08 & 0.39$\,\pm\,$0.06 & 2 \\
             NGC7469 & 23h 03m 15.65s & +08\deg\ 52m 25.7s & 0.25$\,\pm\,$0.06 & 0.37$\,\pm\,$0.01 & 0.15$\,\pm\,$0.06 & 0.41$\,\pm\,$0.01 & 0.30$\,\pm\,$0.05 & 0.35$\,\pm\,$0.04 & 2 \\
         CGCG453-062 & 23h 04m 56.54s & +19\deg\ 33m 07.2s & 0.69$\,\pm\,$0.02 & 0.72$\,\pm\,$0.03 & 0.77$\,\pm\,$0.03 & 0.72$\,\pm\,$0.02 & 0.71$\,\pm\,$0.03 & 0.75$\,\pm\,$0.02 & 3 \\
        ESO148-IG002 & 23h 15m 46.75s & $-$59\deg\ 03m 15.8s & 0.13$\,\pm\,$0.02 & 0.35$\,\pm\,$0.05 & 0.17$\,\pm\,$0.10 & 0.31$\,\pm\,$0.04 & 0.06$\,\pm\,$0.09 & 0.27$\,\pm\,$0.06 & 2 \\
              IC5298 & 23h 16m 00.67s & +25\deg\ 33m 24.5s & 0.15$\,\pm\,$0.05 & 0.26$\,\pm\,$0.04 & 0.07$\,\pm\,$0.04 & 0.20$\,\pm\,$0.02 & 0.09$\,\pm\,$0.05 & 0.18$\,\pm\,$0.05 & 2 \\
             NGC7591 & 23h 18m 16.25s & +06\deg\ 35m 09.2s & 0.44$\,\pm\,$0.05 & 0.39$\,\pm\,$0.04 & 0.49$\,\pm\,$0.04 & 0.38$\,\pm\,$0.03 & 0.31$\,\pm\,$0.04 & 0.45$\,\pm\,$0.03 & 3 \\
        ESO077-IG014 & 23h 21m 05.45s & $-$69\deg\ 12m 47.2s & 0.14$\,\pm\,$0.09 & 0.16$\,\pm\,$0.01 & 0.32$\,\pm\,$0.04 & 0.20$\,\pm\,$0.05 & 0.09$\,\pm\,$0.09 & 0.21$\,\pm\,$0.04 & 3 \\
        ESO077-IG014 & 23h 21m 03.72s & $-$69\deg\ 13m 00.8s & 0.18$\,\pm\,$0.07 & 0.20$\,\pm\,$0.02 & 0.37$\,\pm\,$0.05 & 0.24$\,\pm\,$0.03 & 0.15$\,\pm\,$0.09 & 0.26$\,\pm\,$0.05 & 3 \\
     IRASF23365+3604 & 23h 39m 01.32s & +36\deg\ 21m 08.3s & 0.37$\,\pm\,$0.08 & 0.28$\,\pm\,$0.03 & 0.49$\,\pm\,$0.17 & 0.38$\,\pm\,$0.02 & 0.16$\,\pm\,$0.12 & 0.43$\,\pm\,$0.06 & 3 \\
       MCG-01-60-022 & 23h 42m 00.91s & $-$03\deg\ 36m 54.4s & 0.43$\,\pm\,$0.07 & 0.48$\,\pm\,$0.01 & 0.60$\,\pm\,$0.04 & 0.47$\,\pm\,$0.02 & 0.42$\,\pm\,$0.06 & 0.59$\,\pm\,$0.04 & 3 \\
      IRAS23436+5257 & 23h 46m 05.45s & +53\deg\ 14m 01.7s & 0.35$\,\pm\,$0.05 & 0.39$\,\pm\,$0.05 & 0.31$\,\pm\,$0.04 & 0.45$\,\pm\,$0.04 & 0.36$\,\pm\,$0.08 & 0.41$\,\pm\,$0.05 & 2 \\
             NGC7753 & 23h 47m 04.85s & +29\deg\ 29m 00.6s & 0.39$\,\pm\,$0.06 & 0.37$\,\pm\,$0.03 & 0.41$\,\pm\,$0.06 & 0.37$\,\pm\,$0.02 & 0.32$\,\pm\,$0.04 & 0.41$\,\pm\,$0.04 & 1 \\
             NGC7753 & 23h 46m 58.63s & +29\deg\ 27m 32.0s & 0.62$\,\pm\,$0.03 & 0.62$\,\pm\,$0.02 & 0.67$\,\pm\,$0.02 & 0.63$\,\pm\,$0.01 & 0.63$\,\pm\,$0.03 & 0.68$\,\pm\,$0.03 & 1 \\
             NGC7771 & 23h 51m 03.91s & +20\deg\ 09m 01.8s & 0.55$\,\pm\,$0.07 & 0.56$\,\pm\,$0.02 & 0.51$\,\pm\,$0.04 & 0.57$\,\pm\,$0.05 & 0.58$\,\pm\,$0.05 & 0.65$\,\pm\,$0.05 & 2 \\
             NGC7771 & 23h 51m 24.79s & +20\deg\ 06m 42.1s & 0.46$\,\pm\,$0.01 & 0.41$\,\pm\,$0.05 & 0.52$\,\pm\,$0.03 & 0.48$\,\pm\,$0.01 & 0.50$\,\pm\,$0.04 & 0.53$\,\pm\,$0.03 & 1 \\
             NGC7771 & 23h 51m 22.56s & +20\deg\ 05m 49.2s & 0.48$\,\pm\,$0.05 & 0.56$\,\pm\,$0.05 & 0.52$\,\pm\,$0.04 & 0.59$\,\pm\,$0.01 & 0.58$\,\pm\,$0.04 & 0.64$\,\pm\,$0.03 & 2 \\
             Mrk0331 & 23h 51m 26.76s & +20\deg\ 35m 10.3s & 0.38$\,\pm\,$0.03 & 0.40$\,\pm\,$0.04 & 0.46$\,\pm\,$0.04 & 0.43$\,\pm\,$0.01 & 0.38$\,\pm\,$0.05 & 0.44$\,\pm\,$0.04 & 1
\enddata
\tablecomments{\footnotesize Continued.}
%\vspace{2cm}
\end{deluxetable}

\clearpage

\end{document}